\documentclass{PBCreport}
\usepackage{wrapfig}
\usepackage{booktabs}
\usepackage{afterpage}
\usepackage[T1]{fontenc}
\usepackage{textcomp}
\usepackage{color, colortbl}
\usepackage{booktabs}
\usepackage{afterpage}
\usepackage{tabularx}
\usepackage{siunitx}
\usepackage{hyperref}
\usepackage{graphicx,mathptmx,amsmath,amsfonts,amscd,amsthm,amssymb,eucal,psfrag,color,subfig,url,multirow,gensymb}
\usepackage{rotating}
\usepackage{soul}
\usepackage{appendix}
\usepackage{placeins}
\hypersetup{
     colorlinks = true,
     linkcolor = blue,
     anchorcolor = blue,
     citecolor = blue,
     filecolor = blue,
     urlcolor = blue
     }

\definecolor{LightCyan}{rgb}{0.88,1,1}
\definecolor{LightYellow}{rgb}{1,0.97,0.9}

\documentlabel{CERN-PBC Report-2023-002}

\begin{document}

\title{A Long-Baseline Atom Interferometer at CERN: Conceptual Feasibility Study}

\author{\parbox{\textwidth}{\it 
G.~Arduini$^{1,*}$, L.~Badurina$^2$, K.~Balazs$^1$, C.~Baynham$^3$, O.~Buchmueller$^{3,4,*}$, M.~Buzio$^1$,
S.~Calatroni$^{1,*}$, J.-P.~Corso$^1$, J.~Ellis$^{1,2,*}$, Ch.~Gaignant$^1$, M.~Guinchard$^1$, T.~Hakulinen$^1$, R.~Hobson$^3$, A.~Infantino$^1$,
D.~Lafarge$^1$, R.~Langlois$^1$, C.~Marcel$^1$, J.~Mitchell$^5$, M.~Parodi$^1$, M.~Pentella$^1$,
D.~Valuch$^1$, H.~Vincke$^1$}\\
~~\\
%\email{author.email@cern.ch}
%\affiliation{CERN, CH 1211 Geneva 23, Switzerland}
\small{$^1$ CERN, 
$^2$ King's College London, $^3$ Imperial College London, $^4$ University of Oxford,}\\
\small{$^5$ University of Cambridge}\\
%~~\\
\small{$^*$ Editors}}

\abstract{

We present results from exploratory studies, supported by the Physics Beyond Colliders~(PBC) Study Group, of the suitability of a CERN site and its infrastructure for hosting a vertical atom interferometer (AI) with a baseline of about $100$~m. We first review the scientific motivations for such an experiment to search for ultralight dark matter and measure gravitational waves, and then outline the general technical requirements for such an atom interferometer, using the AION-100 project as an example. We present a possible CERN site in the PX46 access shaft to the Large Hadron Collider~(LHC), including the motivations for this choice and a description of its infrastructure. We then assess its compliance with the technical requirements of such an experiment and what upgrades may be needed. We analyse issues related to the proximity of the LHC machine and its ancillary hardware and present a preliminary safety analysis and the required mitigation measures and infrastructure modifications. In conclusion, we identify primary cost drivers and describe constraints on the experimental installation and operation schedules arising from LHC operation. We find no technical obstacles: the CERN site is a very promising location for an AI experiment with a vertical baseline of about $100$~m.

%(The goal of this feasibility document is to present to the atom interferometer physics community the site and the infrastructure that can be made available at CERN for the installation of a long-baseline vertical interferometer. We first present the physics motivations and the general technical requirement of a long-baseline vertical ion interferometer such as AION-100. We then present the site and the motivations for its choice, its infrastructure, we assess its compliance with the technical requirements of an atom interferometer and discuss any upgrade needed. We analyse the safety aspects due to the proximity of the LHC machine and its ancillary hardware and present a preliminary risk analysis and the required mitigation measures and infrastructure. A preliminary cost estimate and the constraints on the schedule arising from LHC operation conclude the document)  

}

% \begin{figure}[t]
% 	\flushleft
% 	\hspace*{+5mm}
% 	\includegraphics[width=4cm]{Figures/QTI.png}
%  	\vspace*{-6mm}
% \end{figure}

\maketitle

\tableofcontents

\newpage

%\Section{Introduction (G. Arduini, S. Calatroni, O. Buchmuller, J. Ellis)}
\section{Introduction} %(G. Arduini, S. Calatroni, O. Buchmuller, J. Ellis)}
\label{sec:Intro}

Atom interferometer (AI) experiments offer interesting prospects for
searches for the interactions of ultralight bosonic dark matter with
Standard Model particles as well as groundbreaking detections of gravitational waves in
a frequency band inaccessible to experiments that are operating and under construction, via high-precision quantum interference measurements.
Ideal locations for the next generation of such experiments are provided by vertical shafts
of $\sim 100$~m, such as those providing access to the LHC.
PBC has supported exploratory studies of the PX46 LHC access shaft as a potential
location for an atom interferometer experiment such as the AION-100 proposal~\cite{Badurina:2019hst}, 
aimed at demonstrating its {\it prima facie} suitability,
analysing the principal technical requirements and their implications,
identifying potential engineering solutions and discussing principal cost and schedule drivers. In this document we summarize our findings.

Explanations of the acronyms used in this document are listed in the Appendix~\ref{sec:Acronyms}.

%\Section{Executive summary (G. Arduini, S. Calatroni, O. Buchmuller, J. Ellis)}
\section{Executive summary} 
\label{sec:Exec}

Quantum sensors are attracting increasing attention for their potential
to make precise measurements within the Standard Model (SM) and search
for possible new physics beyond the Standard Model (BSM). Among the proposed
applications of Quantum Technology to Fundamental Physics, one of the most
interesting is Atom Interferometry (AI), which offers interesting prospects for
searches for Dark Matter (DM) and Gravitational Waves (GWs)~\cite{Dimopoulos:2008sv} that are largely
complementary to established techniques.
AI is an established quantum sensor concept
based on the superposition and interference of atomic wave packets. AI experimental designs take advantage of features used by state-of-the-art atomic clocks in combination with established techniques for building inertial sensors.

Several current AI experiments and projects are designed to measure quantum
interference effects over distances ${\cal O}(10)$~m and can be sited in 
university laboratories (e.g., Stanford~\cite{Overstreet:2021hea} and Oxford~\cite{Badurina:2019hst}) or research institutes (e.g., Hannover~\cite{schlippert2020matter} and Wuhan~\cite{zhou2011development}). The proposed next
generation of AI experiments will be on a larger scale of ${\cal O}(100)$~m,
and may be sited in national facilities, e.g., MAGIS~(Matter-wave Atomic Gradiometer Interferometric Sensor experiment) at
Fermilab~\cite{MAGIS-100:2021etm} in the US, MIGA~(Matter wave-laser based Interferometer Gravitation Antenna) in the Laboratoire Souterrain 
{\` a} Bas Bruit (LSBB) in France~\cite{Canuel:2017rrp}, ZAIGA
(Zhaoshan Long-baseline Atom Interferometer Gravitation Antenna)~\cite{Zhan:2019quq} in China, and the Boulby Underground Laboratory is being considered as a UK possible site for AION-100~\cite{Badurina:2019hst}. 

CERN is also a compelling possible site
for a vertical AI of length ${\cal O}(100)$~m,
in view of its physical and technical infrastructure including LHC access shafts, and its experience in hosting international experimental collaborations. CERN's PBC and Quantum Technology Initiative (QTI) programmes provide suitable frameworks for exploring the feasibility of installing such an AI at CERN, which would extend the scope of
the CERN experimental programme in exciting new directions with no impact on the exploitation
of the LHC.

\begin{figure}[!]
%\centering
%\begin{wrapfigure}[41]{lt}{0.5\textwidth}
%\begin{figure}[t!]
\centering
%~~\\
%\vspace{-0.2cm}
%\includegraphics[width=0.7\textwidth]{Figures/3-PhysicsMotivations/PX46situation.png}
\includegraphics[width=1.0\textwidth]{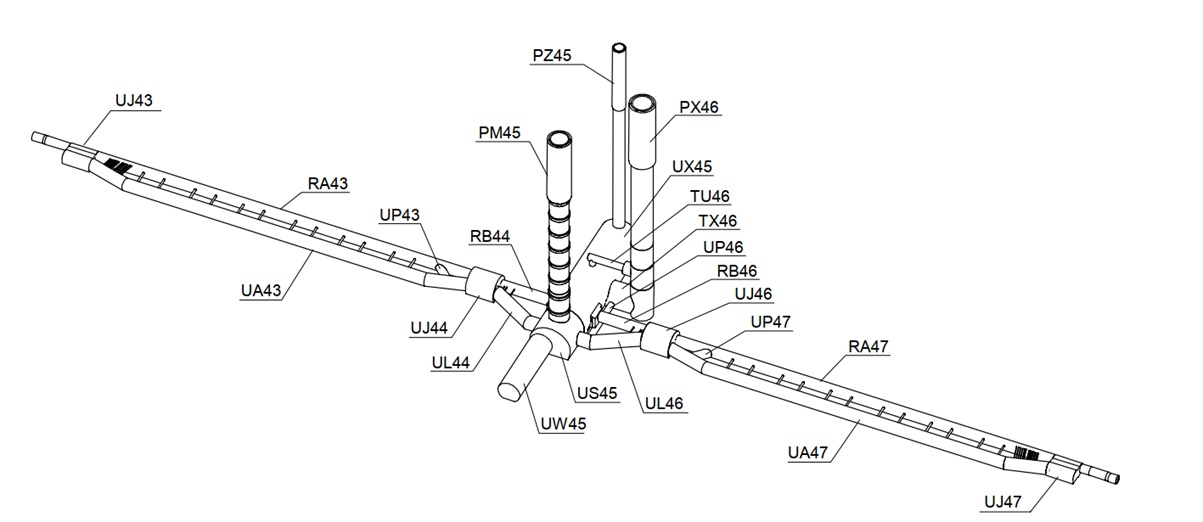}
%\hspace{5mm}
\includegraphics[width=0.45\textwidth]{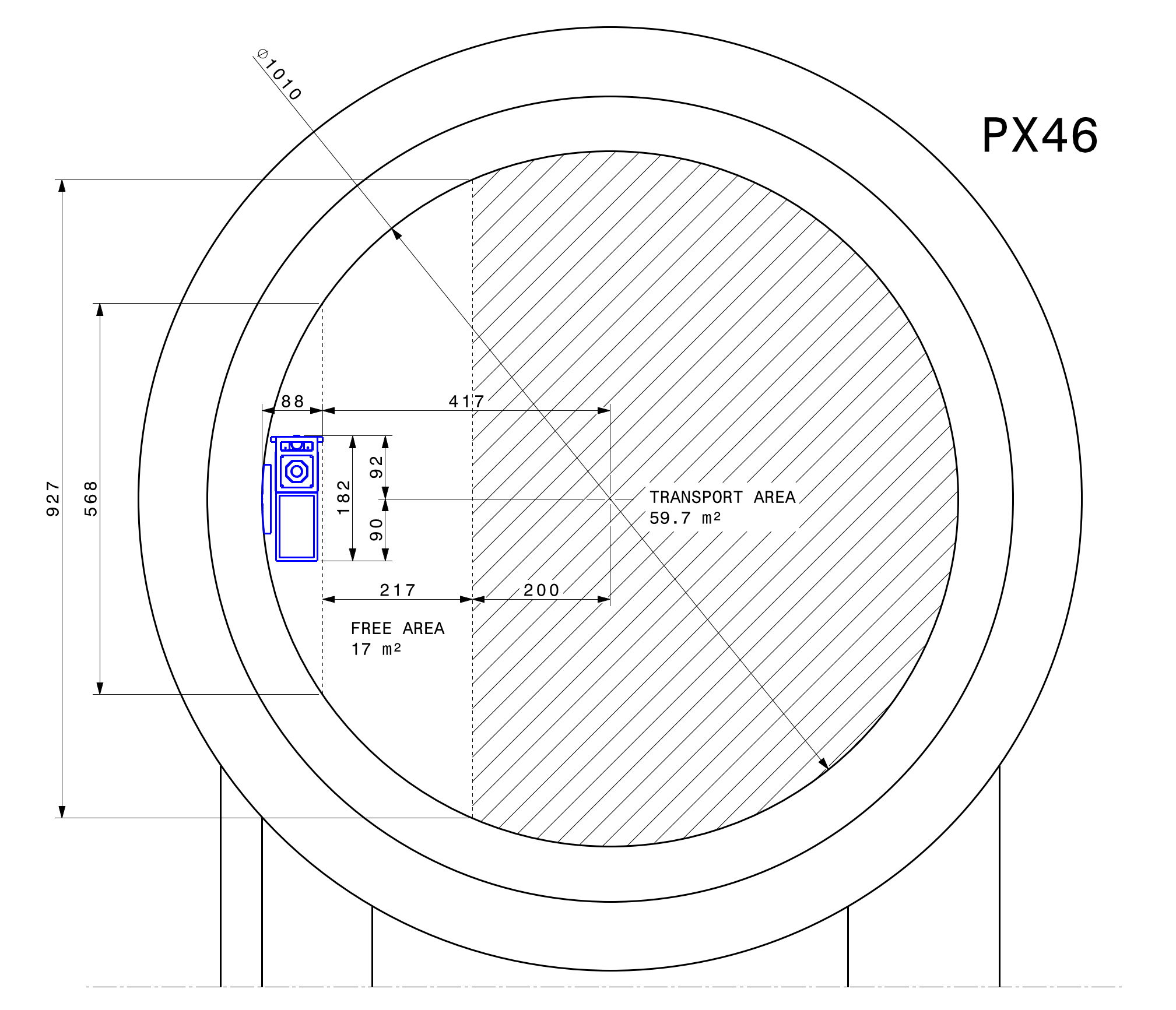}
%\quad
\includegraphics[width=0.45\textwidth]{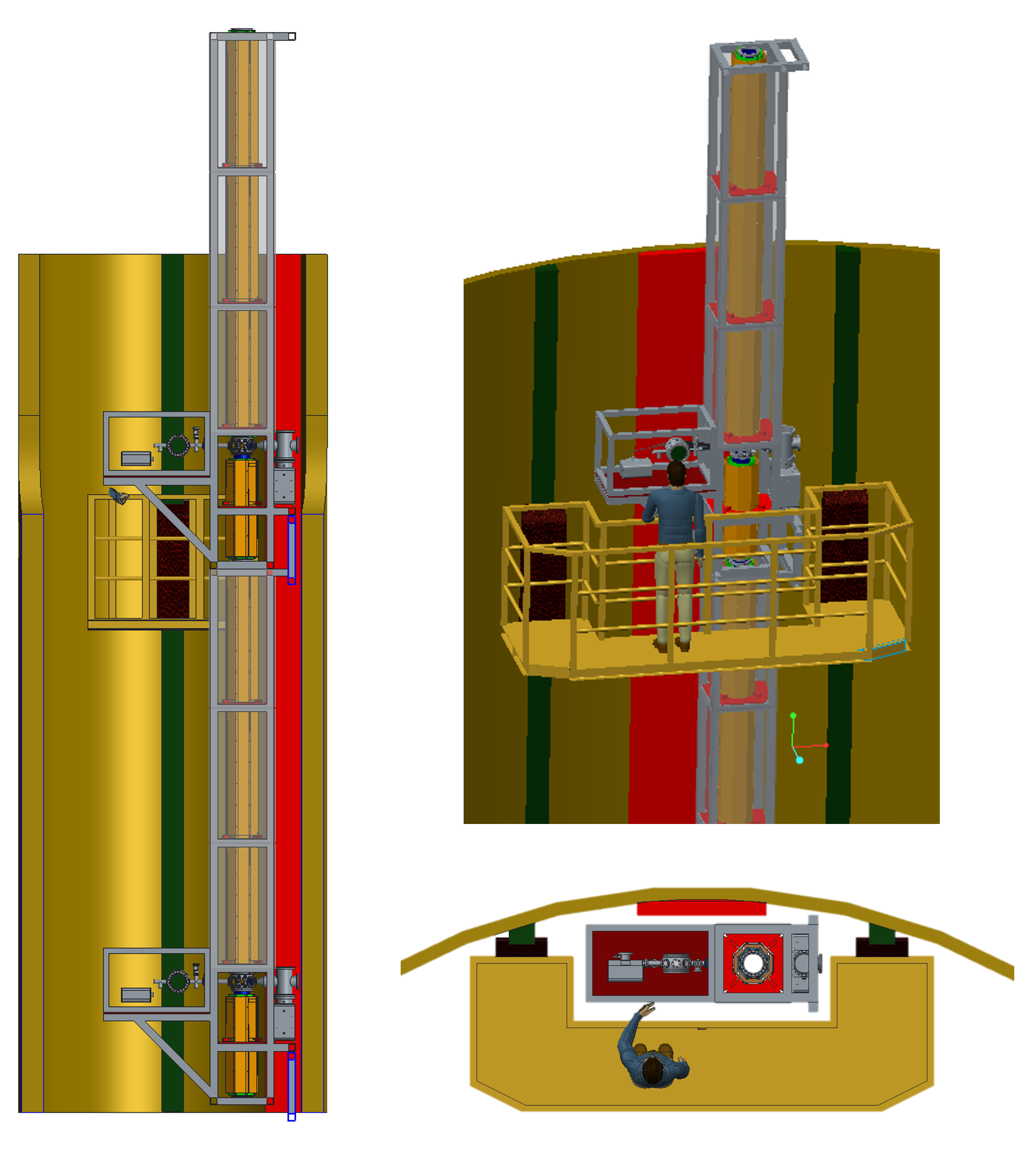}
\caption{%The left plot shows a s
{\it Top panel}:
Schematic drawing of the civil engineering infrastructure at Point~4 on the LHC ring,
showing the LHC tunnel, the PX46 shaft
%(in colour)
, UX45, where the LHC RF power system is located, and the horizontal TX46 gallery. The LHC machine components are installed in the nearby tunnel (RA43-RB44-RB46-RA47).
{\it Bottom left panel}: Horizontal cross section of the PX46 shaft.
{\it Bottom right panel}: Illustration how a 100~m vertical AI similar to AION-100
could be accommodated in PX46. The experiment and its magnetic shielding
would be contained within a cylinder with a diameter $\sim 1$~m, with
provision for external access via a mobile platform with approximate dimensions
2~m$\times$5~m.}
%For clarity, the sizes of the atom interferometers are shown on an exaggerated scale.}
\label{fig:PX46situation}
%\end{wrapfigure}
\end{figure}

An initial survey of the LHC access shafts has
identified the PX46 access shaft to the LHC as the most promising CERN site for
a ${\cal O}(100)$~m vertical AI. Figure~\ref{fig:PX46situation} shows a
schematic view of PX46 and the layout of the civil engineering 
infrastructure at Point~4 on the LHC ring.
The vertical height from the SX4 surface building 
down to the level of the LHC is $\sim 143$~m and the internal diameter of the shaft is 10.1~m. 
PX46 provides access to the main LHC radiofrequency (RF) 
system and its primary use is for raising and lowering technical equipment.

However, as seen in the bottom left  panel of Fig.~\ref{fig:PX46situation},
a substantial fraction of the horizontal cross-section of the
PX46 shaft is not required for LHC access, and would be large enough to 
accommodate an AI experiment such as AION-100, as illustrated in the bottom right panel
of Fig.~\ref{fig:PX46situation}.

We describe in a later Section of this report the general civil infrastructure 
on the surface and below ground at LHC Point 4, and outline the considerations that led to
the identification of PX46 as an interesting potential location for an AI
experiment. We summarise the results of a core sample taken at the location of
PX46, which revealed a thin layer of topsoil above a 48-m thick layer of glacial
till (moraine), below which there is a substrate of sandstone (molasse).
We also report exploratory seismic measurements at the top and bottom of
the PX46, which can be used to estimate the possible level of Gravity Gradient
Noise (GGN), an important source of background, which we find to be similar to other sites under consideration.  

%In addition, we report temperature measurements at the top and bottom of PX46, which are quite satisfactory.

One issue that is specific to PX46 is that of electromagnetic (EM) noise
associated with the LHC RF system and other electrical equipment at depth and on the surface.
The EM noise levels have been measured and found not to be of concern: we recall 
that operating frequency of the LHC RF system is many orders of magnitude different
from the range in which AI measurements would be made. We have also measured the variation in the ambient
magnetic field during a ramp of the LHC magnets: this is small and slow, and
can be dealt with by the magnetic shield of the AI. Another issue specific to
Point 4 of the LHC is the possibility of a major helium release from the LHC
cryogenic system. This will require some modification of the safety systems in the area that have been conceptually evaluated and costed.

An issue at any LHC access shaft is the unlikely possibility of a catastrophic LHC beam
loss near the base of the shaft, in view of which radioprotection measures must
be foreseen. These would include in particular the installation of a protective
shielding wall in TX46 at the base of PX46, with an access door and provision
for temporary opening when LHC equipment must be installed or removed.

General safety considerations imply the need for extending the LHC access control
system to PX46 and provision for rapid evacuation of experimental personnel when necessary, such as in the event of a fire in the nearby UX45 cavern, which will
require an {\it ad hoc} fast cage system. Preliminary consultations indicate that
a technical solution is feasible at a reasonable cost, and a detailed design 
can be obtained if requested.

These preliminary studies indicate that siting a $\sim 100$~m vertical AI in
PX46 appears feasible: no showstoppers have been found. We have therefore
proceeded to identify the main cost drivers for installing such an experiment,
which include the costs of the radioprotection shielding wall and the rapid
evacuation system. Initial estimates indicate that these costs should be small 
compared to the cost of the experiment itself.

%\Section{Physics motivations (O. Buchmuller, J. Ellis)}
\section{Physics motivations}
\label{sec:Phys}

%{\it 2-3 pages: short description of the physics goals and motivations. Why a vertical interferometer? }

The nature of Dark Matter (DM) is one of the greatest puzzles in fundamental physics, and lies beyond the scope of the Standard Model. The favoured hypothesis is that it is composed of non-relativistic particles such as weakly-interacting massive particles (WIMPs) or coherent waves of ultralight bosons. Experiments at the LHC and elsewhere have not yet found any evidence for WIMPs, though searches will continue during Run 3 of the LHC and at the high-luminosity LHC (HL-LHC). However, in the meantime the search for Ultra-Light Dark Matter (ULDM) is attracting growing interest, and this is one of the principal scientific objectives of atom interferometry (AI) experiments such as AION.

The other principal objective of such experiments is the search for GWs in the range of frequencies around 1~Hz that is intermediate between the peak sensitivities of present terrestrial experiments such as LIGO~\cite{LIGOScientific:2014pky}, Virgo~\cite{VIRGO:2014yos} and KAGRA~\cite{Aso:2013eba}, and the approved space-borne experiment LISA~\cite{LISA:2017pwj}. Among the targets of experiments in this frequency range are mergers of black holes with masses intermediate between those whose mergers have been detected by LIGO and Virgo and the supermassive black holes detected in the centres of galaxies~\cite{EventHorizonTelescope:2019dse,EventHorizonTelescope:2022wkp}. Detectors in the intermediate frequency range may also be sensitive to a background of GWs produced by fundamental physics processes such as first-order phase transitions in the early Universe or the evolution of a network of cosmic strings~\cite{Badurina:2019hst}. 

\begin{figure}[t!]
\centering
%~~\\
\vspace{-0.7cm}
\includegraphics[width=0.43\textwidth]{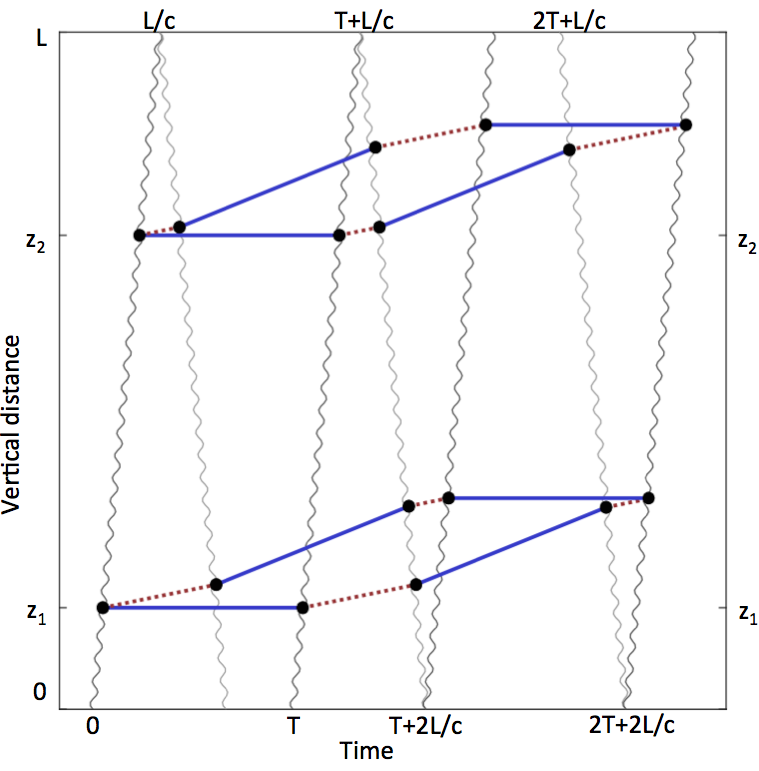}
\caption{%The left plot shows a s
Space-time diagram of the operation of a pair of cold-atom interferometers based on single-photon transitions between the ground state (blue) and the excited state (red dashed). Height is shown on the vertical axis and the time axis is horizontal.
The laser pulses (wavy lines) travelling across the baseline from opposite sides are used to divide, redirect, and recombine the atomic de Broglie waves, yielding interference patterns that are sensitive to the modulation of the atomic transition frequency caused by coupling to ULDM, or the space-time distortions caused by
GWs.}
\label{fig:space-time}
%\end{wrapfigure}
\end{figure}

The basic principle of AI is illustrated in Figure~\ref{fig:space-time}, and is analogous to that of laser interferometry as employed by current GW detectors. A cloud of cold atoms is split by a laser pulse into populations of ground-state and excited atoms, a second `mirror' pulse interchanges the populations of excited and ground-state atoms, which are finally recombined using another laser pulse, and the wave functions of the atomic populations interfere. Each laser interaction with an atom imparts momentum, so the two populations follow different space-time trajectories, as seen in Figure~\ref{fig:space-time}. Interactions of the atoms with a coherent wave of ULDM may alter the excited atomic energy level, modifying the atomic phase and hence the interference pattern. In practice, AI experiments use two or more sources of atoms that are exposed to the same laser beam, as also shown in Figure~\ref{fig:space-time}, thereby minimising the effects of laser noise, and measure the differences between the interference patterns they exhibit, which are sensitive to the space-time dependence of the ULDM field density. Such differential measurements are also sensitive to the distortions of space-time caused by the passage of a GW.

Two types of configurations have been considered for AI experiments, using either vertical shafts~\cite{Badurina:2019hst,MAGIS-100:2021etm} or horizontal galleries~\cite{Canuel:2017rrp,Canuel:2019abg,Zhan:2019quq}. In the option considered here the cold atom clouds are launched vertically into a vacuum tube and follow ballistic trajectories modified by laser pulses that are also directed vertically. This configuration has the advantage that the atom clouds may have a relatively long flight time, $T$, that is limited only by the length, $L$, of the tube. During this time, the atoms may be subjected to multiple laser pulses, $n$, enabling large momentum transfers (LMTs) that enlarge the `hysteresis diamonds' illustrated in Figure~\ref{fig:space-time}. Also, separating the sources by relatively large vertical distances, $\Delta z$, enables atom clouds to be projected over large vertical distances and helps suppress the gravity gradient noise (GGN) due to seismic movements in the surrounding rock, which decays exponentially with depth~\cite{Badurina:2022ngn}.
The indicative values of the experimental parameters assumed here for a two-interferometer design are listed in Table~\ref{tab:AION100parameters}, including also the repetition rate, $\Delta z$, the laser phase noise and the experimental duration, $T_{\rm Int}$, but we anticipate that more ambitious parameters should be attainable, e.g., by increasing $N$, the number of atoms in each cloud, and that there would be advantages in adding more sources along the vacuum tube. These would help control the impact of GGN, particularly in the presence of different rock strata. 

%%%%%%%%%%%%%%%%%%%%%%%%%%%%%
\begin{table}[h]
 \caption{Indicative experimental parameters for a 100~m Atom Interferometer.}
 \label{tab:AION100parameters}
  \centering
  \begin{tabular}{c|c|c|c|c|c|c|c}
   L [m] & T [s] & $n$ & $\Delta z$ [m] & $N$ & $\Delta t$ [s] & Phase noise [1/$\sqrt{\rm Hz}$] & $T_{\rm{Int}}$ [s] \\
   \hline
%   100 & 1.4 & 1000 & 85 & $10^8$ & 1.5 & $10^8$ \\
   100 & 1.4 & 5494 & 50 & $10^8$ &  1.5 & $10^{-5}$  & $10^8$
  \end{tabular} %\\
\end{table}
%%%%%
%\vspace{0.5cm}

\begin{figure}[h!]
\centering
%~~\\
%\vspace{-0.6cm}
\includegraphics[width=0.43\textwidth]{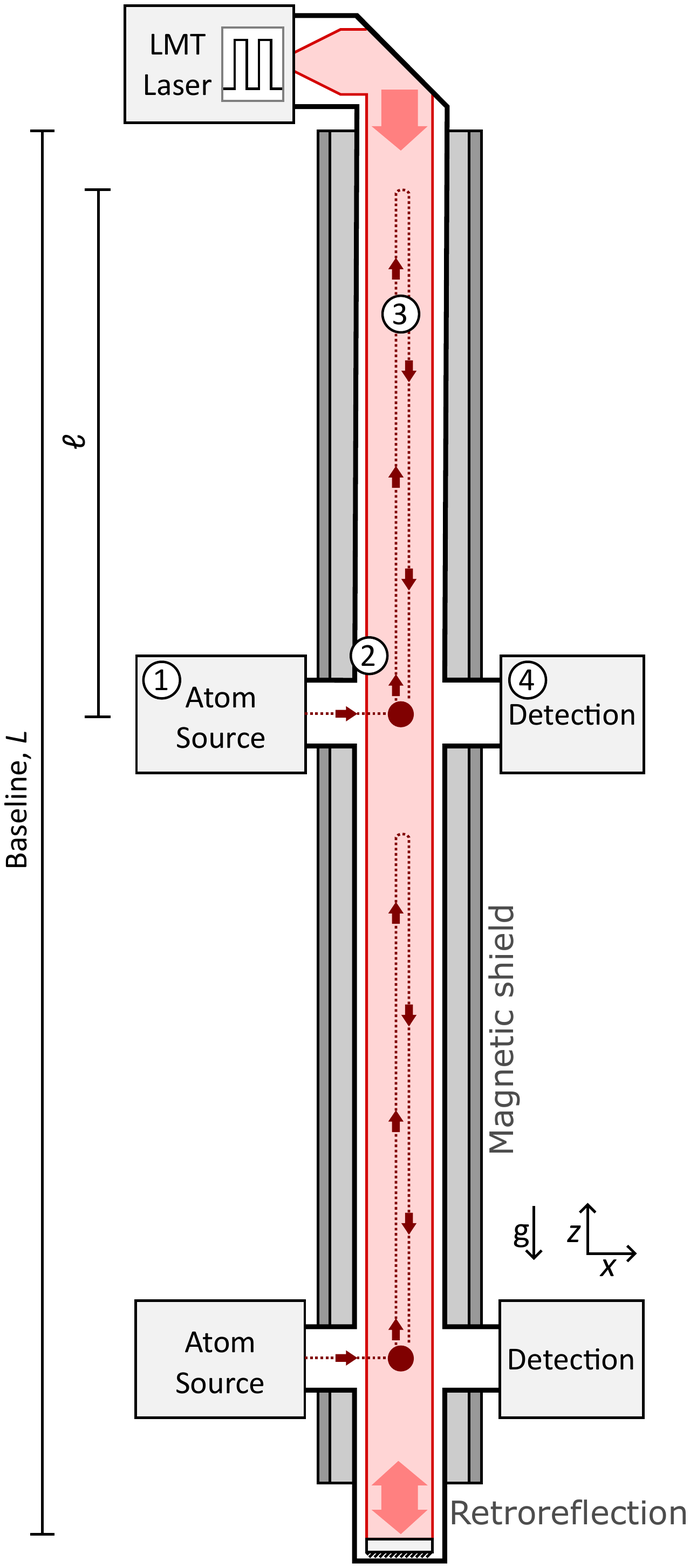}
%\vspace{-0.4cm}
\caption{Conceptual scheme of an Atom Interferometer (AI) experiment with two atom sources that project clouds vertically, addressed by a single laser source (this diagram is not to scale).}
\label{fig:Schematic}
\end{figure}

The conceptual scheme we consider for a vertical AI with two atom sources is shown in Figure~\ref{fig:Schematic}. 
%\begin{wrapfigure}[24]{t}{0.45\textwidth}
%\end{wrapfigure}
There is a single laser source, shown here as located at the top of the vertical vacuum tube, though a location at the bottom could also be considered. One of the atom sources is located at the bottom of the vacuum tube, whereas the location of the upper source is only indicative. This upper atom source is labelled by (1), the trajectories of the atoms it launches are labelled by (2,3), and the detector to measure interference patterns is labelled by (4). Additional sources and detectors may be added as desired. We note that the vacuum pipe is surrounded by a magnetic shield.

%%%%%%%%%%%%%%%%%%%%%%%%%%%%%
\begin{figure*}[h!]
 \centering
 \includegraphics[width=0.48\textwidth]{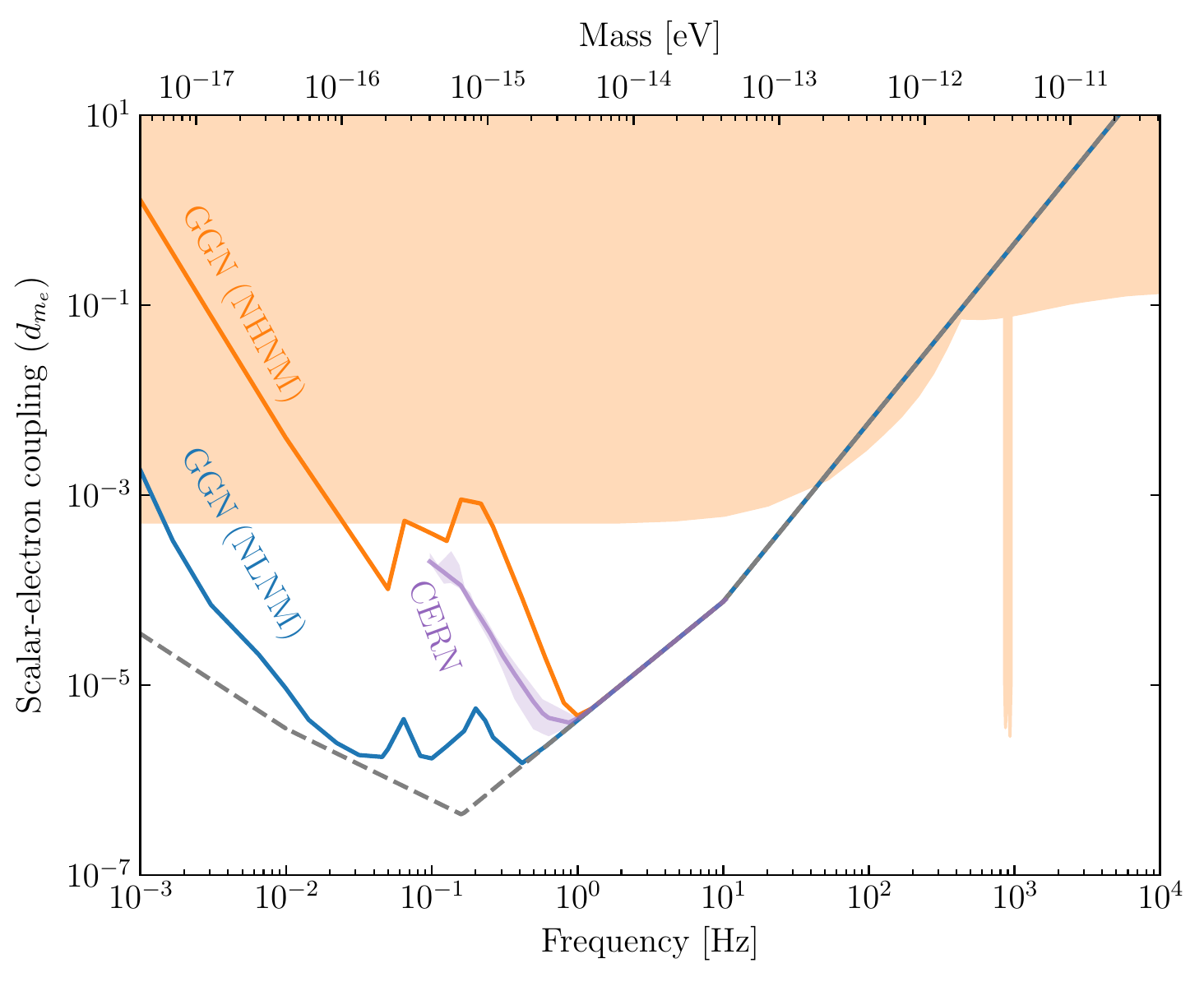}
 \includegraphics[width=0.50\textwidth]{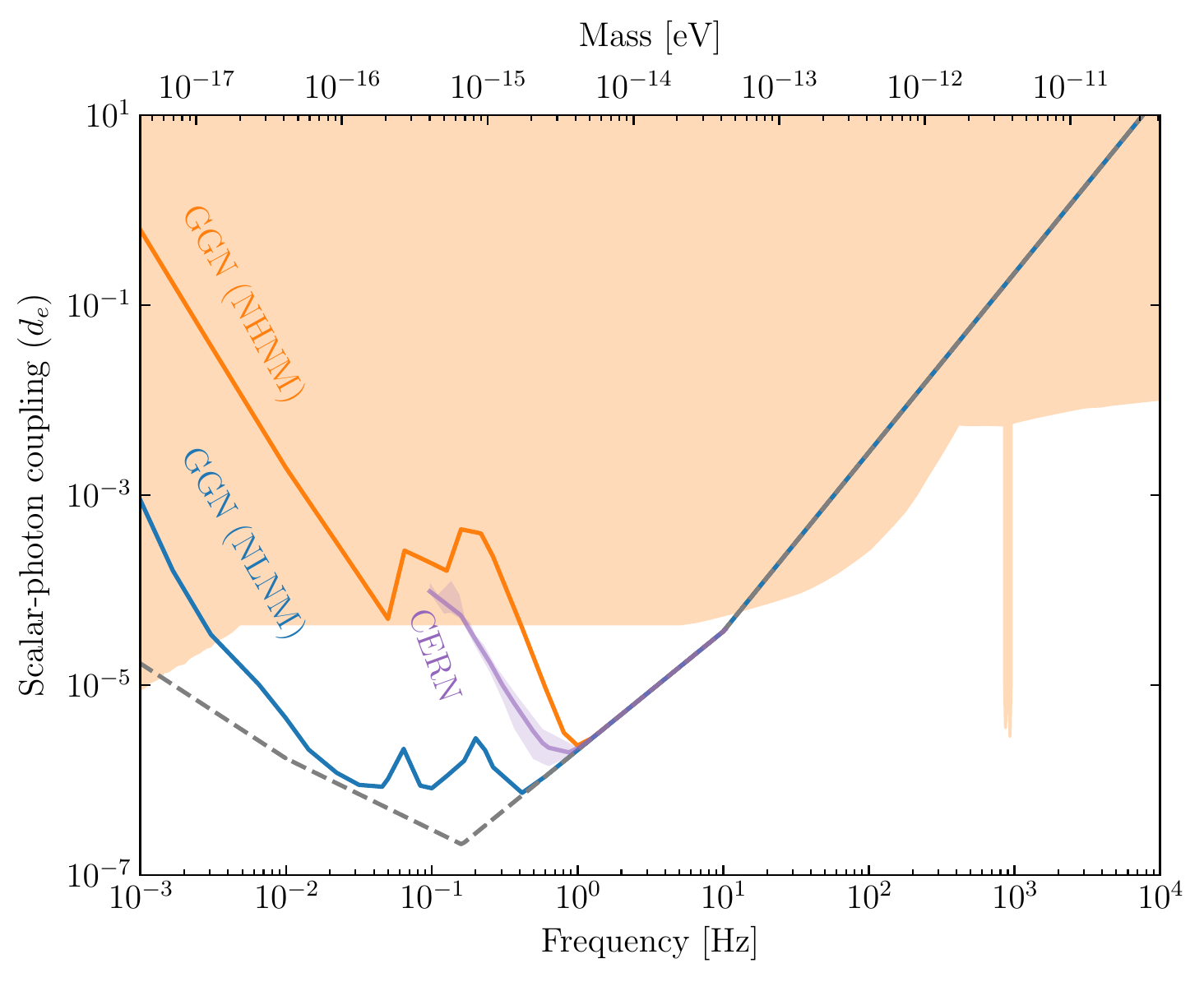}\\
 \includegraphics[width=0.48\textwidth]{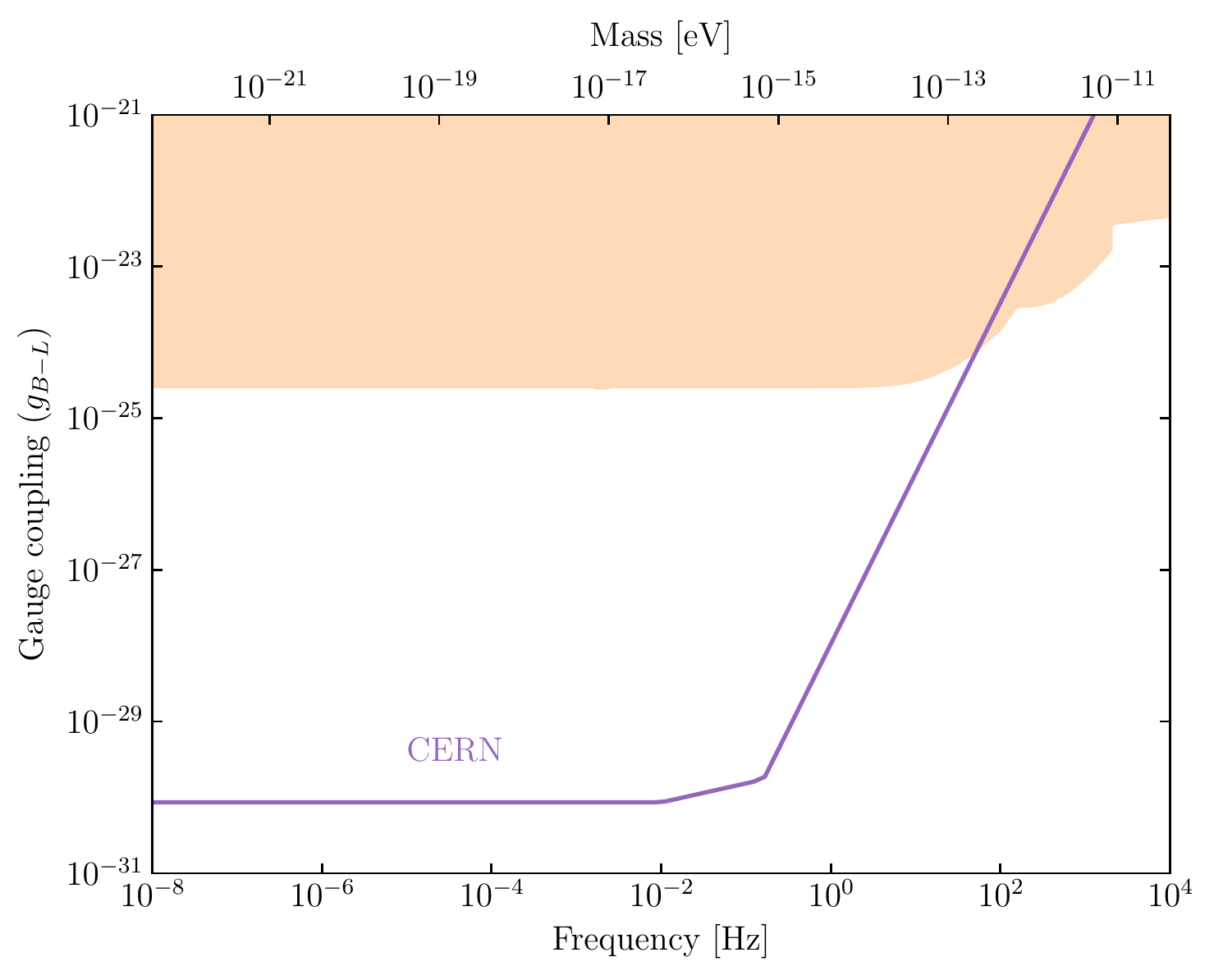}
 \includegraphics[width=0.50\textwidth]{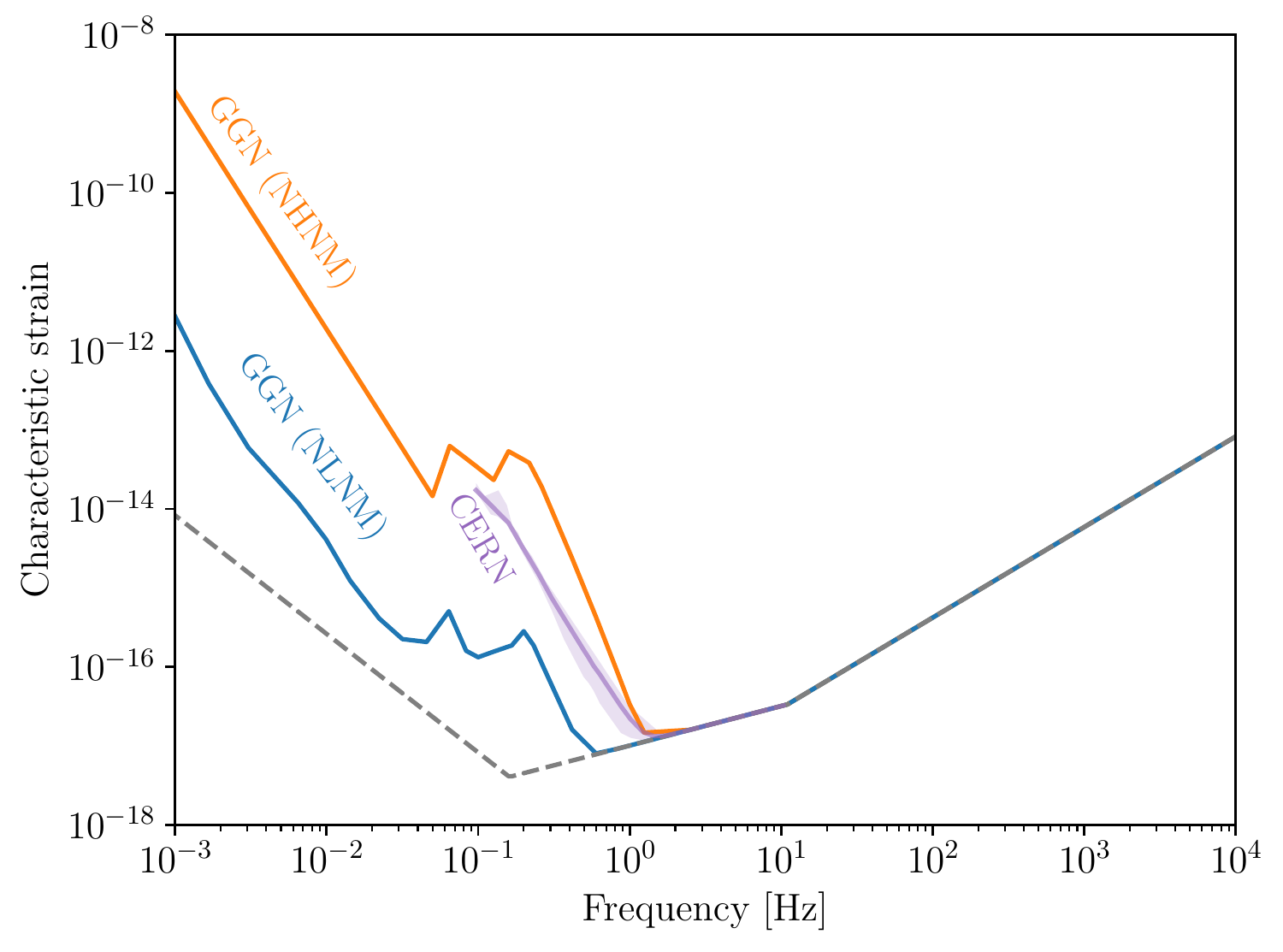}
 \caption{ \label{fig:AION100Sensitivities}Sensitivities of
 a 100-m Atom Interferometer to a coupling of scalar ULDM to the electron ({\it upper left panel}), to the photon ({\it upper right panel}), and to ultralight vector dark matter coupled to $B - L$ ({\it lower left panel}). The {\it lower right panel} shows the sensitivity to GW strain. In all panels, the atom shot noise calculated from the indicative experimental parameters in Table~\ref{tab:AION100parameters} is indicated by black dashed lines, and the new high-noise model (NHNM) for gravity gradient noise (GGN) is indicated by solid orange lines. In the upper panels and in the lower right panel the new low-noise model (NLNM) for GGN is indicated by solid blue lines and the GGN calculated on the basis of seismic measurements at the PX46 site is indicated by violet lines, and the shading corresponds to the diurnal fluctuations in vertical surface motion shown in Figs.~\ref{fig:RMS powerSpectral Density Min Max Case} and \ref{fig:noisemodels}.
 The surrounding rock is assumed to have properties similar to sandstone (molasse).}
\end{figure*}
%%%%%%%%%%%%%%%%%%%%%%%%%%%%%

Figure~\ref{fig:AION100Sensitivities} illustrates the capabilities of a 100-m vertical AI detector assuming the indicative experimental parameters listed in Table~\ref{tab:AION100parameters}, and broadband operation. The upper panels illustrate the sensitivities of such an AI experiment to scalar ULDM~\cite{Arvanitaki:2016fyj}, see also~\cite{Badurina:2019hst,Badurina:2021lwr} and~\cite{Badurina:2022ngn}.
The upper left panel shows the potential sensitivity to an ULDM-electron coupling $d_{m_e}$ of a 100~m AI such as AION, and the upper right panel shows its potential sensitivity to an ULDM-photon coupling $d_{e}$. The irreducible background due to Atom shot Noise (ASN) is shown as a black dashed line, and the solid orange (blue) line in this and other panels indicates the GGN background calculated using the New High-Noise Model (NHNM) and the New Low-Noise Model (NLNM)~\cite{peterson1993observations} that are based on extensive surveys of noise levels at different sites, assuming that the surrounding rock is an isotropic layer of sandstone (molasse)~\footnote{A core sample from the PX46 location found a thin surface soil layer above a layer of glacial till (moraine) down to a depth of 48~m, with sandstone (molasse) underneath.}. The violet line in this and the two right panels is the level of GGN estimated on the basis of seismic measurements at the top and bottom of PX46 shaft, as described later, again assuming that the surrounding rock is sandstone (molasse)~\footnote{The shaded bands correspond to the uncertainty in the GGN due to the daily variations in ground motion discussed in Section~\ref{subsubsec:Vibrations}, see Figs.~\ref{fig:RMS powerSpectral Density Min Max Case} and \ref{fig:noisemodels}.}. These GGN curves are not substantially altered if the rock surrounding PX46 is assumed to be glacial till (moraine). However, as also discussed later, PX46 is known to be surrounded by two strata, and the geological environment is anisotropic, so more complete calculations and seismic measurements will be needed to refine the illustrative GGN curves in Figure~\ref{fig:AION100Sensitivities}~\footnote{We note that the estimated GGN could in principle
be mitigated by
using an array of seismic sensors~\cite{Coughlin:2014yda} in the neighbourhood of PX46 to measure seismic perturbations, but this possibility has not yet been explored.}.

The lower left panel of Figure~\ref{fig:AION100Sensitivities} illustrates the potential sensitivity
to ultralight vector dark matter coupled to $B - L$, assuming two co-located interferometers using $^{88}$Sr and $^{87}$Sr isotopes, and 
the lower right panel of Figure~\ref{fig:AION100Sensitivities} illustrates the potential sensitivity to characteristic GW strain of a 100-m AI in PX46. The shaded regions in the upper parts of the upper panels and the lower left panel are excluded by present measurements, including by MICROSCOPE~\cite{MICROSCOPE:2022doy}, torsion balance experiments~\cite{Hees:2018fpg}, LIGO and Virgo~\cite{LIGOScientific:2021ffg}, AURIGA~\cite{Branca:2016rez} and atomic clock experiments~\cite{Filzinger:2023zrs}.

{\it We emphasise that no proposed experiment would be as sensitive as a 100~m AI for probing the `mid-mass gap' for scalar ULDM searches.}

%In this case, in addition to the NHNM model for the GGN (solid orange line) we also display the New Low-Noise Model (solid blue line) as well as an estimate of the GGN based on seismic measurements in PX46. We see that the NLNM and PX46 calculations yield noise levels that are very close to the ASN over the frequency range displayed.

In addition to these topics and searches for other forms of coupling between ULDM and SM particles~\cite{Badurina:2019hst}, additional possible science objectives for a 100~m vertical AI experiment are under 
active consideration. For example, there is the possibility of measuring the
gravitational Aharonov-Bohm effect, as pioneered in~\cite{Overstreet:2021hea},
which may provide an interesting window on quantum aspects of 
gravitation~\cite{Overstreet:2022zgq}.

These examples illustrate the potential scientific capabilities of a 100~m vertical AI experiment that could be located in an LHC access shaft. In the following Section of this report, we give an overview of such an experiment and the required technical infrastructure, and the following Section assesses the feasibility of installing such an experiment in the PX46 access shaft.

%~~\\
%~~\\
%~~\\

%\Section{Overview of the experiment and required infrastructure (O. Buchmuller, R. Hobson, J. Mitchell)}
\section{Overview of the experiment and of the required infrastructure}\label{sec:Overview}

% {\it 4-5 pages: description of what a long-baseline vertical atom interferometer could look like, based on the proposed AION-100 experiment, with emphasis on the technical and infrastructure requirements. Discuss preferred operational scenarios (24h/24 7d/7 access), maximum acceptable level of vibration and EM noise. Needed utilities.}

The design of a \SI{100}{m} AI is guided by balancing engineering requirements, technical limitations, and ambient sources of systematic noise. A full conceptual design would require careful discussion and cooperation between physicists and engineers to converge on an effective detector. However, one can make a reasonable first pass based on the current development of the similar MAGIS-100 detector~\cite{MAGIS-100:2021etm}. Many engineering questions are being actively pursued and answered with the development of MAGIS-100, and the final design of a detector like AION-100 should only diverge significantly if this were required by fundamentally different physics or technical demonstration goals.%not diverge too drastically.

At a high level the detector can be categorized into specific subsystems that are useful for focusing engineering efforts and understanding the technical requirements and the overall infrastructure needs. These subsystems are depicted in Figure~\ref{fig:aion-layout} and listed here:
\begin{itemize}
    \item Laser/control laboratory;
    \item Laser link;
    \item Interferometry region;
    \item Side-arms (atom sources);
    \item Mirror platform.
\end{itemize}
Each subsystem interfaces with the others and can also be further subdivided into smaller systems.

\subsection{Subsystem descriptions}
\label{sec:subsystems}

\begin{figure}[h!]
    \centering
    \includegraphics[width=0.95\textwidth]{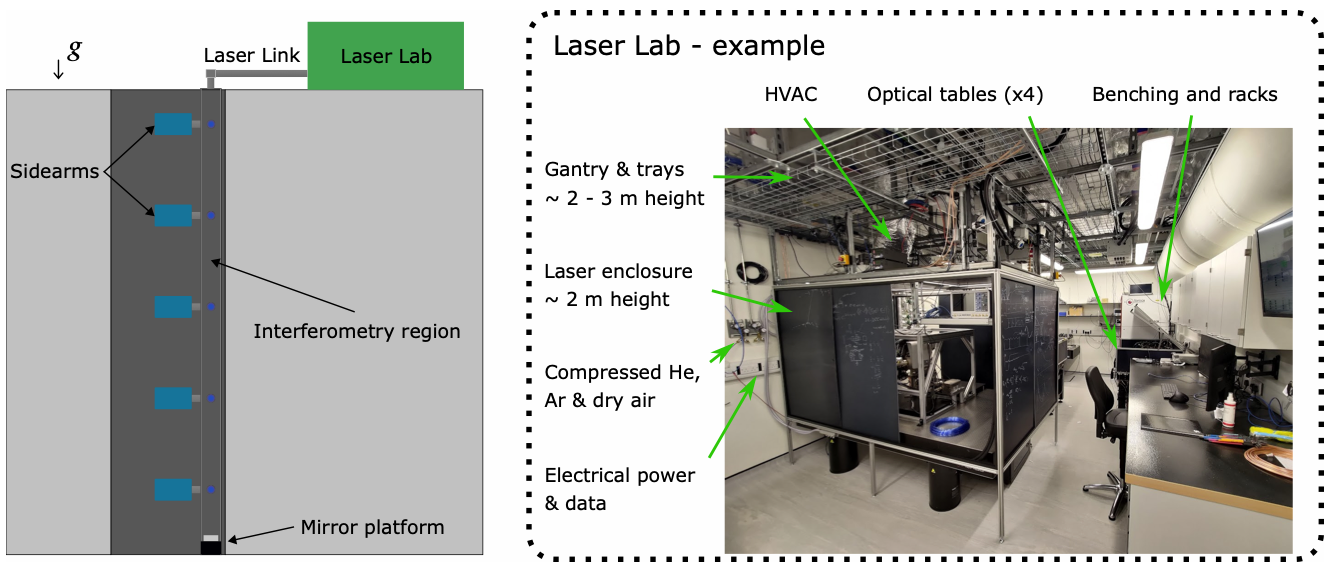}
    \caption{\textit{Left:} Diagram of the high-level subsystem layout. A laser laboratory in the surface building would connect to the shaft via a laser link which then guides the laser light for the atom interferometry down into the interferometry region (shown here in a cutaway drawing). The interferometry region would ideally be on the order of \SI{100}{m} with potentially 5 or more side-arms down the baseline. \textit{Right:} Reference example for a laser laboratory (the Imperial College AION laboratory), including the Heating, Ventilation, and Air Conditioning (HVAC) system.}
    \label{fig:aion-layout}
\end{figure}

Figure~\ref{fig:aion-layout} shows a diagram of where each subsystem would be located around the shaft.

{\bf Laser Laboratory:} The laser laboratory at ground level will house the specialised instruments needed for laser cooling and manipulation of atoms in the detector, e.g., lasers, optics, control electronics, computers, and coil drivers. It should have a floor area of at least \SI{50}{\square\meter}, allowing space for these instruments and for testing of each side-arm before installation into the 100-m detector. An example picture of a laser laboratory with suitable features is presented in Figure~\ref{fig:aion-layout}, and specifications for services are listed in Table~\ref{tab:tech-reqs}.

{\bf Laser Link:} Nearly all laser light, and all data and electrical signals, will be delivered from the laser laboratory to the detector through optical fibres and cables. The route for the optical fibres should be kept fairly short ($\lesssim \SI{50}{\meter}$), and isolated from acoustically noisy equipment, to limit optical absorption loss, phase noise and stimulated Brillouin scattering~(see Section~\ref{sec:env-reqs}).

Because of its high power and higher sensitivity to phase noise, the atom interferometry beam requires special treatment: For all but $\sim\SI{1}{\meter}$ of the distance between the laser lab and the interferometry, the link must propagate through free space, in a hermetically-shielded and laser-safe enclosure. A short length of single-mode fibre is required near the top of the interferometry region in order to reduce beam pointing fluctuations, but the fibre must be limited in length to approximately \SI{1}{\meter} in order to limit phase noise and stimulated Brillouin scattering. The short fibre will then launch directly into the interferometry region, propagating into the tube via a custom telescope and in-vacuum beam steering optics.

{\bf Interferometry region:} The atom interferometry region consists of an ultra-high-vacuum tube with internal length $\sim 100$~m and clear diameter of at least 150~mm. The vacuum tube is wrapped in field coils and surrounded by multi-layer magnetic shields, and has up to ten connection nodes to which side-arms will be attached. The specifications for services are relatively light because most of the interferometry region is passive, but some services are required, e.g., for environmental monitoring as listed in Table~\ref{tab:tech-reqs}.

{\bf Side-arms (atom sources):} The atom sources are attached to the interferometry region as ``side-arms''. In these side-arms, a hot atomic beam is slowed, captured, and laser-cooled, before being transported sideways into the interferometry region. Each side-arm consists of an ultra-high-vacuum chamber surrounded by field coils, optics, cameras, and local lasers (e.g., 461~nm laser diodes). These elements, combined with their support structure and enclosure, occupy a volume of approximately $1\times 1 \times 2$~\si{\cubic\meter} with a mass of order \SI{200}{\kilogram}. In addition, every side-arm requires local control electronics for sequence coordination, data acquisition, lasers, and active optics. The service specifications for each side-arm are listed in Table~\ref{tab:tech-reqs}. %Water cooling is required for the highest-powered field coils, and a heat exchange (either through water or ventilation) is required to carry away the tens of W generated by other active elements.

{\bf Mirror platform:} The interferometry beam is sent downwards from the top of the interferometry region, and must be retro-reflected by a mirror at the bottom of the tube in order to allow both upward and downward photon momentum kicks to be imparted to the atoms. The mirror must be in-vacuum, and installed on a piezo-tunable platform to allow for dynamic compensation of the rotation of the earth. For the purposes of this feasibility study, the service specifications for the mirror platform can be approximated as being the same as for a side-arm, though in practice the requirements for the mirror platform will be less extensive.

\subsection{Technical and infrastructure requirements}
\label{sec:infrastructure-reqs}

An early draft of the proposed technical requirements for these subsystems is shown in Table~\ref{tab:tech-reqs}. Estimates for temperature control are guided by the local noise requirements of the laser systems and the side-arms. There are also considerations for infrastructure requirements set by power estimates and safety factors.

Cold-atom technology is typically complex and delicate, requiring daily intervention such as optical alignment in order to maintain reliable operation. To support these interventions, access to the laser laboratory and side-arms is required for at least 12 hours per day. The access method must be practical for daily use (especially during initial commissioning), and must be safe from radiation, oxygen deficiency and fire hazards.

%%%%%%%%%%%%%%%%%%% Tech req table
\begin{table}[htbp]
\small
\centering
\caption{\label{tab:tech-reqs} Preliminary technical and infrastructure requirements. It is anticipated that 5 to 10 side-arms will be required.}
\smallskip
\bgroup
\def\arraystretch{1.25}
 \begin{tabular}{|>{\centering\arraybackslash}m{2.5cm}|>{\centering\arraybackslash}m{4cm}|>{\centering\arraybackslash}m{4cm}|>{\centering\arraybackslash}m{4cm}|}
 \hline
 Requirement & Laser Lab & Interferometry region & Side-arm (per side-arm)\\ \hline
 \hline
  Volume & Floor area $>$ \SI{50}{m^2} & \SI{1}{m^2} cross-sectional area &  1~m $\times$ 1~m $\times$ 2~m \\ \hline
  Mains power & $\sim$ \SI{35}{kW} (three- and single-phase outlets) & $\mathcal{O}(\SI{100}{W})$ diagnostic and monitoring electronics & $\mathcal{O}(\SI{10}{kW})$\\ \hline
  Control cables & Ethernet, fibre, coaxial & Magnetic coils, diagnostic and monitoring electronics & optical fibres, coaxial, high-power steel-clad fibers\\ \hline
  Temperature stability & \SI{22}{\degreeCelsius} w/ $\pm$ \SI{1}{\degreeCelsius} pk-pk & $<\SI{1}{\degreeCelsius\per\hour}$ & Temperature controlled, NEMA rated enclosure, $<\SI{0.5}{\degreeCelsius}$ pk-pk\\ \hline
  Water cooling & \SI{30}{kW} cooling capacity & n/a & \SI{5}{kW} cooling capacity, $<\pm \SI{1}{\degreeCelsius}$ stability\\ \hline
   Laser safety & Engineering (enclosures, interlocks); admin (training); PPE (glasses) & Already safe (enclosed) & Engineering (enclosures); admin (training); PPE (glasses) \\ \hline
  Gases & Helium, compressed air, Argon & n/a & Helium for commissioning\\ \hline
  Cryogenics & n/a & n/a & n/a \\ \hline
  Ventilation & Air-handling unit capable of temp. spec. & Air-flow to maintain temp. spec. & Air-flow to move \SI{5}{kW} of heat\\ \hline
  Access & Year-round ($>$ 12 hrs/day) & Access for maintenance (more access during calibration and commissioning) & Year-round $\sim$ 12 hrs/day (more R\& D for fully autonomous atom sources)\\ \hline
  Smoke detector & Yes & Yes & Yes\\ \hline
  Oxygen depletion monitor & Yes & During maintenance & n/a\\ \hline
  Hoisting \newline equipment & n/a & Modular sections $<$ \SI{907}{kg} & n/a \\
 \hline
 \end{tabular}
\egroup
\end{table}
%%%%%%%%%%%%%%%%%%%%%%%%%%%%%%%%%%%%%
\subsection{Environmental requirements}
\label{sec:env-reqs}

Many of the systematics affecting atom interferometers can be found in reference~\cite{MAGIS-100:2021etm} and references therein. In addition to the systematics that constrain design choices for the laser systems and the cold atom cloud production, other environmental systematics set limits on potential sites where the experiment can be located.

Environmental systematics include ambient seismic activity, infrasound and atmospheric temperature fluctuations. Seismic motion of the Earth couples directly to the detector through vibrations of the optics steering the interferometry laser beam and the local side-arm systems that generate the cold atom clouds. Based on estimates of the effect of vibration impacting the main interferometry laser beam, local vibrations of the main optics should be kept to acceleration fluctuations $\delta a \leq \SI{e-4}{(m/s^2)/\sqrt{Hz}}$. The full impact of vibrations on the side-arms requires further focused research efforts to be characterised fully. There exists a secondary coupling of the density perturbations of the Earth to the atoms (which act as test masses) through perturbations of the gravitational field. These effects are known as gravity gradient noise (GGN) and generate a low-frequency noise floor for long-baseline terrestrial AI. The effects of GGN on GW detectors have been studied in the high-frequency regime for LIGO and VIRGO~\cite{Harms:2019dqi} and have also been investigated for AIs~\cite{Canuel:2016pus,Mitchell:2022zbp,Badurina:2022ngn}, with further work ongoing. Just as seismic density perturbations lead to noise in the AI phase shift, infrasound waves have a similar impact in the very low-frequency range and warrant further study. These noise sources set requirements on the proposed installation site, which should be as passively quiet as possible. This can be established through long-duration seismic surveys. Active mitigation can be implemented using purpose-built instrumentation for seismic and weather monitoring. For the long-lived signals we are searching for, the predictability of the weather and ambient density perturbations are a key requirement for obtaining quality data for analysis. As a requirement on the site, one would search for a site where the target frequency band seismic amplitude spectral densities are as close to the NLNM as possible, throughout multiple seasons, and any spectra above the NHNM would be detrimental to the science reach of the detector. 

For all subsystems, the ambient magnetic fields and RF fields present will place requirements on the necessary levels of magnetic and RF shielding. Most static magnetic fields are not of great concern, as the experiment will use active bias fields and magnetic shielding to control the field uniformity. The concern then arises from time-varying magnetic fields around the laser laboratory and in the shaft around the interferometry region and side-arms. From previous analysis and estimates, the target sensitivity of the AI will require magnetic field fluctuations to be limited to $\delta B \leq \SI{100}{\nano\tesla/\sqrt{Hz}}$. This relatively relaxed constraint relies on operation of the detector in a magnetically insensitive mode, probing simultaneously the $M_F = \pm 9/2$ states of \textsuperscript{87}Sr. However, to allow for simpler modes of operation, magnetic field fluctuations below $\delta B \leq \SI{100}{\pico \tesla/\sqrt{Hz}}$ would be desirable. These constraints apply most strictly within the peak detector sensitivity band \SI{50}{\milli\hertz} to \SI{10}{\hertz}, though strong field noise peaks at Fourier frequencies outside this band could potentially alias into the detector band. Drift or steps in magnetic field of $\gtrsim\SI{50}{\nano\tesla}$ are also a potential concern for the side-arms and the detector, due to their effect on the Magneto-Optical Trap~(MOT) position and the interferometer transition, but slow changes of background magnetic field of up to several \si{\micro\tesla} could straightforwardly be mitigated by shielding and active field control if required.\\
% \textbf{\textcolor{red}{Are the variations shown in Fig. 15 manageable?}}

%\section{LHC Point 4 as preferred site}
%\section{Infrastructure and safety}
\section{Infrastructure and safety}
\label{sec:InfraSafe}

% {\it Short introduction explaining the scope of the following subsections: identify the present state of the site and its infrastructure, its compliance with the technical requirements set in section 4, any upgrade needed, with particular focus on safety aspects due to the LHC machine environment and the related risk analysis and mitigation measures. Bear in mind that this is a feasibility document, thus the analysis remains at a general level while still addressing all key elements}

The infrastructure at the LHC point 4, and in particular the PX46 shaft, has been selected as the best candidate site that CERN could offer as a location for a vertical AI. The following Section~\ref{subsec:Site} describes the selected site and the present status of the infrastructure at the LHC point 4, with particular reference to the environmental noise due to seismic vibrations and to electromagnetic interference. As the PX46 shaft is connected to the entire LHC machine infrastructure, it poses some constraints in terms of radiation protection and some hazards may arise in case of fire developing in the underground areas connected to the PX46 shaft or of accidental helium release from the LHC cryogenic infrastructure. These aspects are discussed in Sections~\ref{subsec:RP} and~\ref{subsec:Safety}. Finally, Section~\ref{subsec:NewInfra} describes the new infrastructure that has to be built, installed or adapted in order to make the PX46 shaft fully compliant with the technical requirements for an AI set out in Section~\ref{sec:Overview} and the safety requirements described in this Section.

%\subsection{Proposed site and characteristics (K. Balazs)}
\subsection{Proposed site and characteristics} %(K. Balazs)
\label{subsec:Site}

% 4-5 pages: identification of the site at LHC point 4 as best option compared to other LHC locations, and describe the general civil infrastructure above and below ground. Describe also present ventilation system and mention conditions of access and escape paths in case of fire/helium release, with reference to later sections for details. Consider impact of operational scenario 24h/24. Discuss incompatibility of staircase and elevator with EN-HE LHC handling requirements, to motivate choices discussed later}

\begin{figure}[htp]
    \centering
    \includegraphics[width=.5\textwidth]{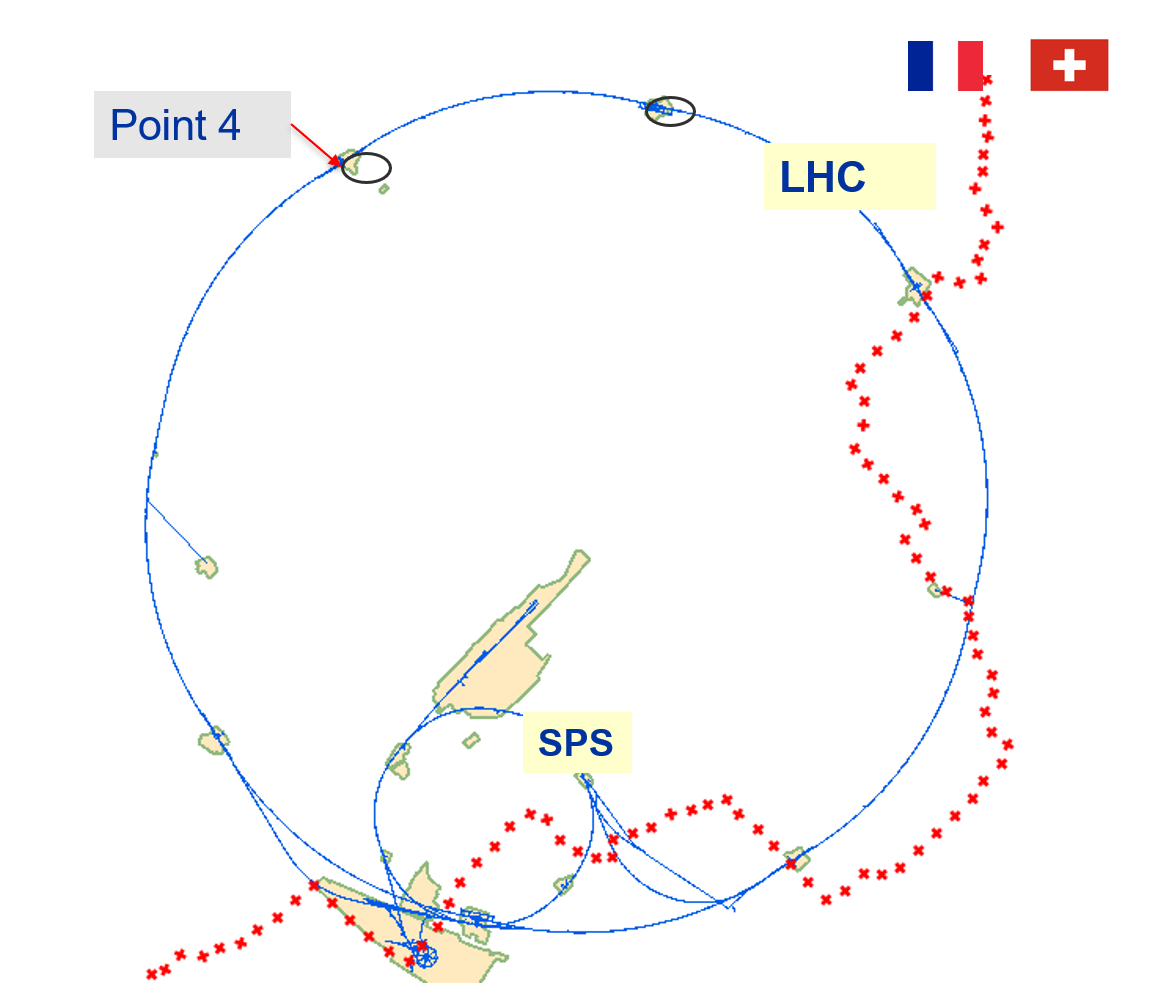}
    \caption{\label{fig:CE:CERN}
    The general layout of CERN, showing the locations of laboratories 1 and 2, the border between France and Switzerland, and the LHC access shafts. The location proposed for an AI experiment such as AION-100 is Point 4 (highlighted).}
\end{figure}

The general layout of the CERN site is shown in Figure~\ref{fig:CE:CERN}, including the locations of laboratories 1 and 2, the border between France and Switzerland, and the LHC access shafts. Following an initial study of the CERN infrastructures, the PX46 shaft located at Point 4 of the LHC tunnel was identified as the most suitable site to house an AI experiment such as AION-100 at CERN.

\begin{figure}[!h]
\centering % \begin{center}/\end{center} takes some additional vertical space
\includegraphics[width=.4\textwidth]{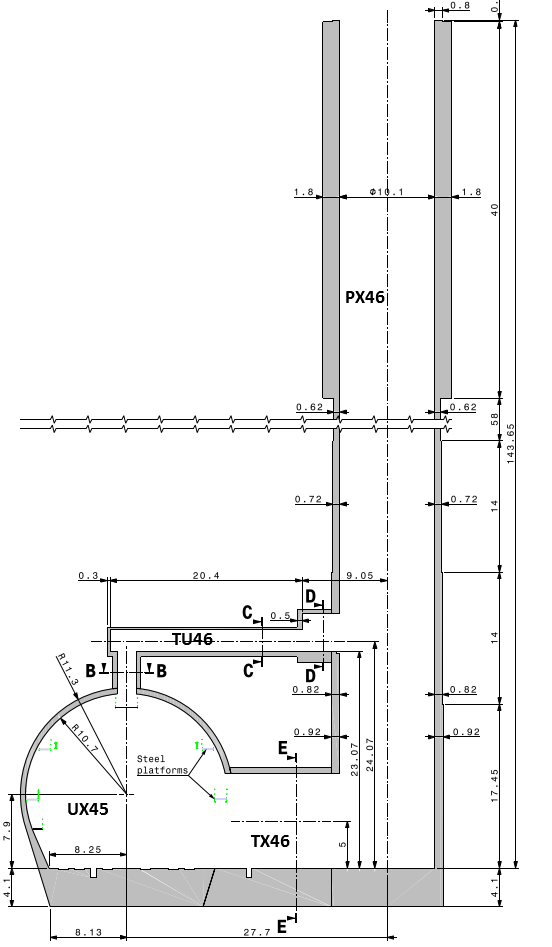}
\qquad
\includegraphics[width=.4\textwidth,origin=c,angle=0]{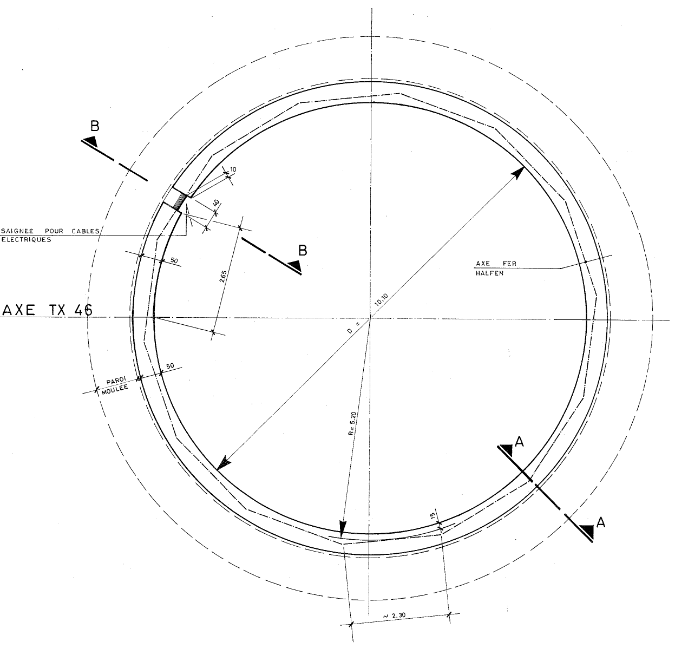}
% % "\includegraphics" from the "graphicx" permits to crop (trim+clip)
% % and rotate (angle) and image (and much more)
\caption{\label{fig:CE:PX46 existing} The general layout of the PX46 shaft, showing the TX46 tunnel that connects it to the UX45 cavern ({\it left}) and a section through the shaft ({\it right}).}
\end{figure}

PX46 is one of the deepest shafts at CERN with an internal height of 143~m. As shown in Figure~\ref{fig:CE:PX46 existing}, it has an internal diameter of 10.10~m, and is the only one 
with enough space to accommodate the installation of a 100~m long AI experiment while remaining available for its primary LHC-related functions.
The shaft is accessible on the surface via the existing SX4 building, which is equipped with a transport crane, and is connected directly to the UX45 cavern via the TX46 gallery. 

\begin{figure}[h]
    \centering
    \includegraphics[scale=0.4]{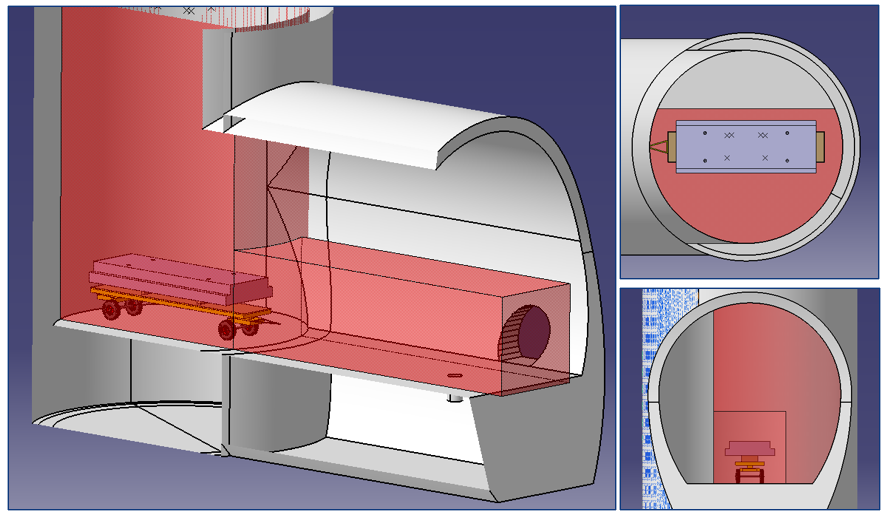}
    \qquad
    \includegraphics[width=.4\textwidth]{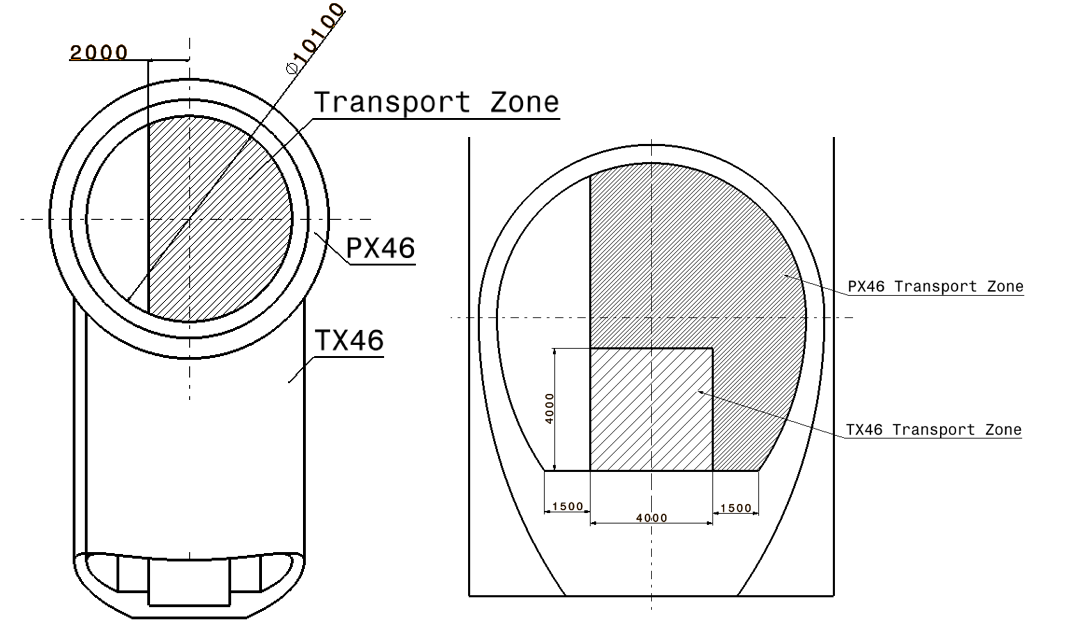}
    \caption{\label{fig:CE:transport}
    The results of a transport study showing the areas to be reserved in PX46 and TX46 for transporting LHC equipment.}
\end{figure}

\begin{figure}[h]
    \centering
    \includegraphics[width=.4\textwidth]{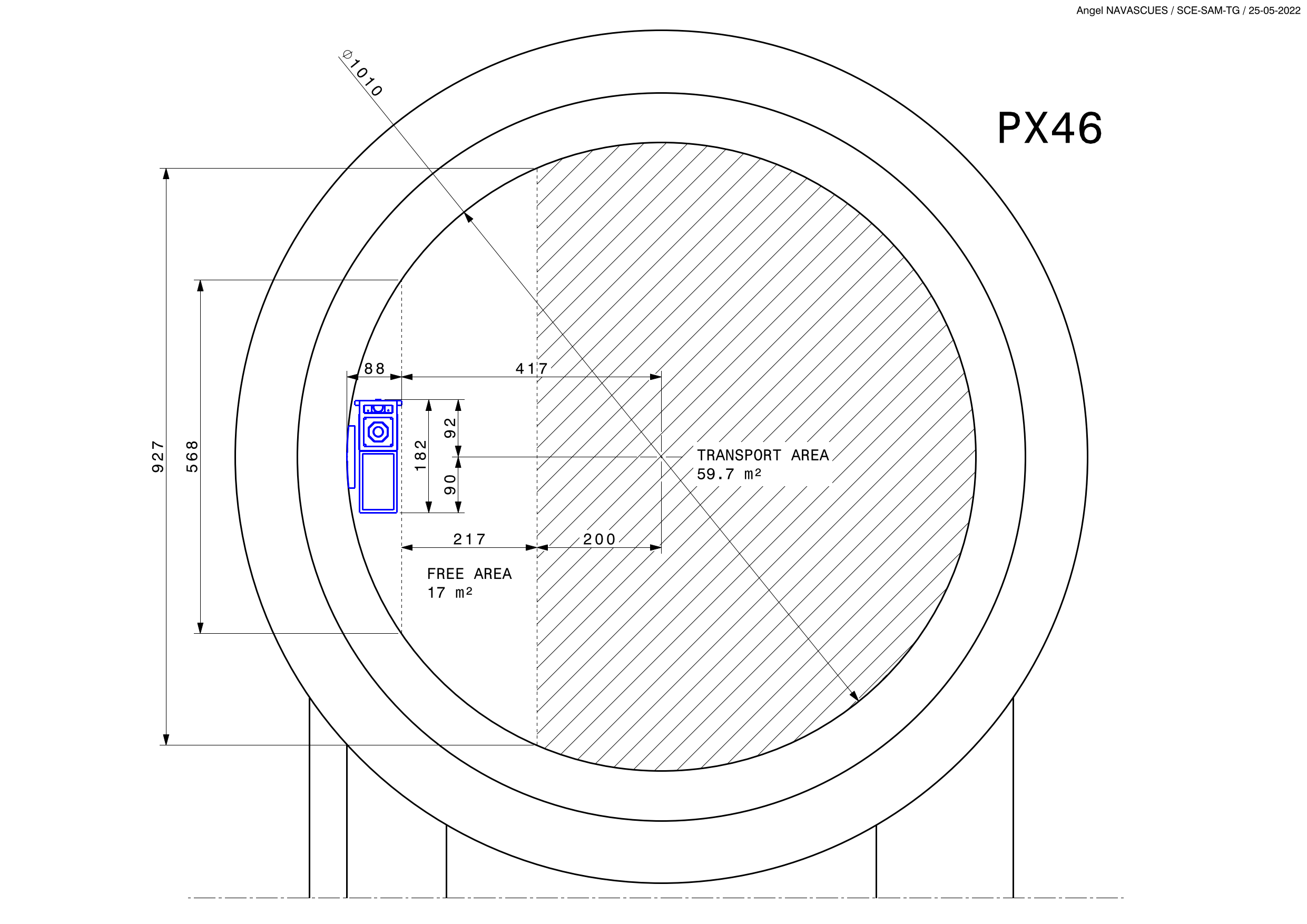}
    \caption{\label{fig:CE:Experiment}
    Cross section of the PX46 shaft showing the proposed location for the AI experiment and the space reserved for transporting LHC equipment.}
\end{figure}

The shaft is currently used for the transport of several types of LHC elements including superconducting RF cavities and the shielding roof of RUX45~(a platform, part of the LHC tunnel, traversing UX45 and connecting RB44 and RB46 --- see Figure~\ref{fig:PX46situation}), and it is planned that several new HL-LHC magnets and other equipment will pass through PX46. In order to assess the area required to be kept free continuously for the transport of such elements and to determine the free space left in the shaft, a preliminary transport study has been carried out by the EN-HE group at CERN, with the results shown in Figure~\ref{fig:CE:transport}. It was concluded that an area of approximately 17~m$^2$ should remain available in PX46, which would be large enough to accommodate an AI experiment. The proposed location of the experiment is shown in Figure~\ref{fig:CE:Experiment} together with the area reserved for transporting LHC equipment.

We conclude that an AI experiment such as AION-100 could be installed in PX46 without constraining or compromising LHC equipment transport in any way.

%\subsubsection{Vibration and seismic noise (M. Guinchard)}
\subsubsection{Vibration and seismic noise}% (M. Guinchard)
\label{subsubsec:Vibrations}

%{\it 2-3 pages: description of the measurements performed at LHC point 4 and compare with the requirements of the experiments set in Section 4}

Seismic activity from natural sources such as earthquakes, or cultural (human-made) sources including civil engineering
works, excite ground vibrations that can be transmitted to an atom interferometer experiment. Based on estimates of the effect of vibration impacting the main interferometry laser beam, local vibrations of the main optics should be kept to acceleration fluctuations below $\delta a \leq \SI{e-4}{(m/s^2)/\sqrt{Hz}}$. An additional effect is the Gravity Gradient Noise (GGN) imparted to the atoms by fluctuations in the local gravitational field, which should be below the New High-Noise Model (NHNM)~\cite{peterson1993observations} if the experimental measurements are to be competitive. This chapter describes the ground motion measurements performed to assess the site stability at LHC point 4.

\begin{figure}[!h]
\centering % \begin{center}/\end{center} takes some additional vertical space
\includegraphics[width=1\textwidth]{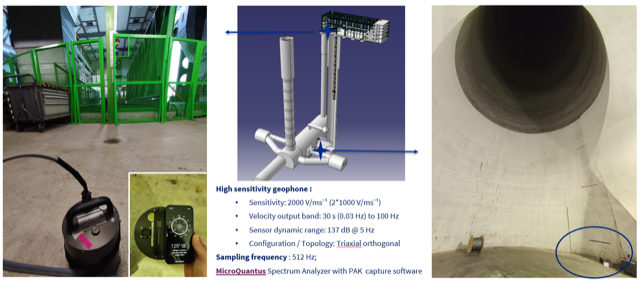}
\qquad
\caption{\label{fig:CE:PX46 SX46 Config} The general layout of the PX46 shaft ({\it centre}), showing %the TX46 tunnel that connects it to the UX45 cavern (left) and a section through the shaft.
the geophone locations in SX4 ({\it left}) and at the base of PX46 ({\it right}).}
\end{figure}

A series of ground measurements were performed at the surface in the SX4 building on top of PX46 and underground at the base of the PX46 shaft in January 2022, as shown in Figure~\ref{fig:CE:PX46 SX46 Config}. Ground motion was recorded in the three directions by geophones (Guralp CMG40T with a sensitivity of 2000~V/(ms$^{-1}$)). The data were recorded during six consecutive days from January 13$^{th}$ until January 18$^{th}$ 2022 with the LHC accelerator in operating conditions but without proton beams circulating.

\begin{figure}[!h]
\centering % \begin{center}/\end{center} takes some additional vertical space
\includegraphics[width=0.9\textwidth]{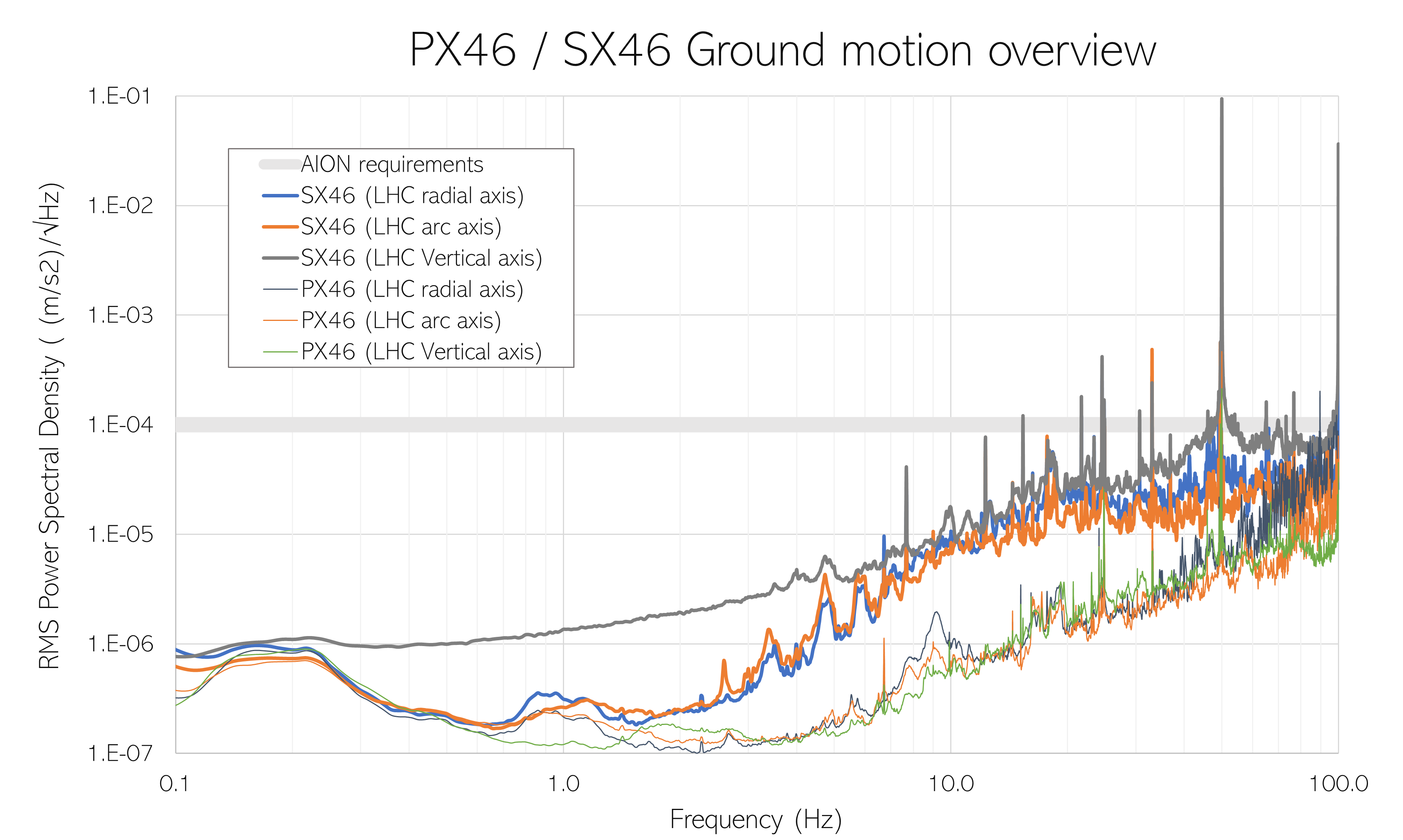}
\caption{\label{fig:RMS powerSpectral Density worst Case} RMS power spectral density (time block 64s, averaged over the worst day).}
\end{figure}

Figure~\ref{fig:RMS powerSpectral Density worst Case} shows the RMS power spectral density for both locations (in SX4 and the PX46 shaft) and for the three directions. The averages over one day of the spectra estimated from 64~s-long time intervals are shown in Figure~\ref{fig:RMS powerSpectral Density worst Case}. The worst day was selected. Below 1 Hz, and except for the SX4 Vertical directions, all the curves are very similar as expected (seismic activity response). The peak at 0.2 Hz corresponds to the microseism peak that is closely related to ocean wave energy coupling with the solid Earth.

\begin{figure}[!h]
\centering % \begin{center}/\end{center} takes some additional vertical space
\includegraphics[width=0.9\textwidth]{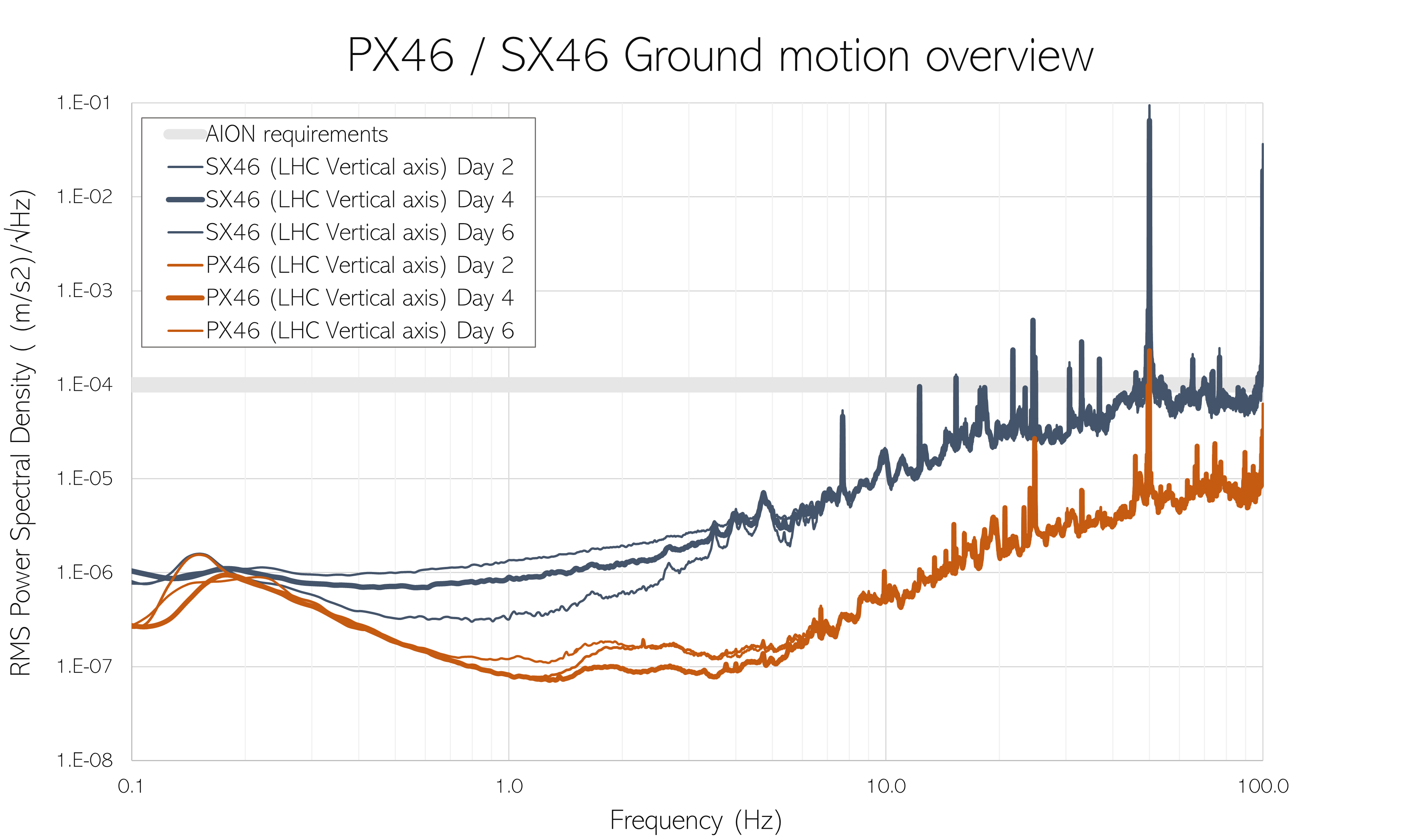}
\caption{\label{fig:RMS powerSpectral Density Min Max Case} The measured RMS power spectral density (time block 64s, single-day averages, minimum and maximum cases).}
\end{figure}

The vertical ground motion in SX4 varied over the 6 days of measurements, as seen in Figure~\ref{fig:RMS powerSpectral Density Min Max Case}, mainly in the frequency range between 0.2 and a few Hz, despite the fixed positions of the sensors. This variation is probably explained by the local seismicity and the effect of the microseismic peak. As also shown in Figure~\ref{fig:RMS powerSpectral Density Min Max Case}, similar behaviour was observed in PX46, but with less variability.

\begin{figure}[!h]
\centering % \begin{center}/\end{center} takes some additional vertical space
\includegraphics[width=0.7\textwidth]{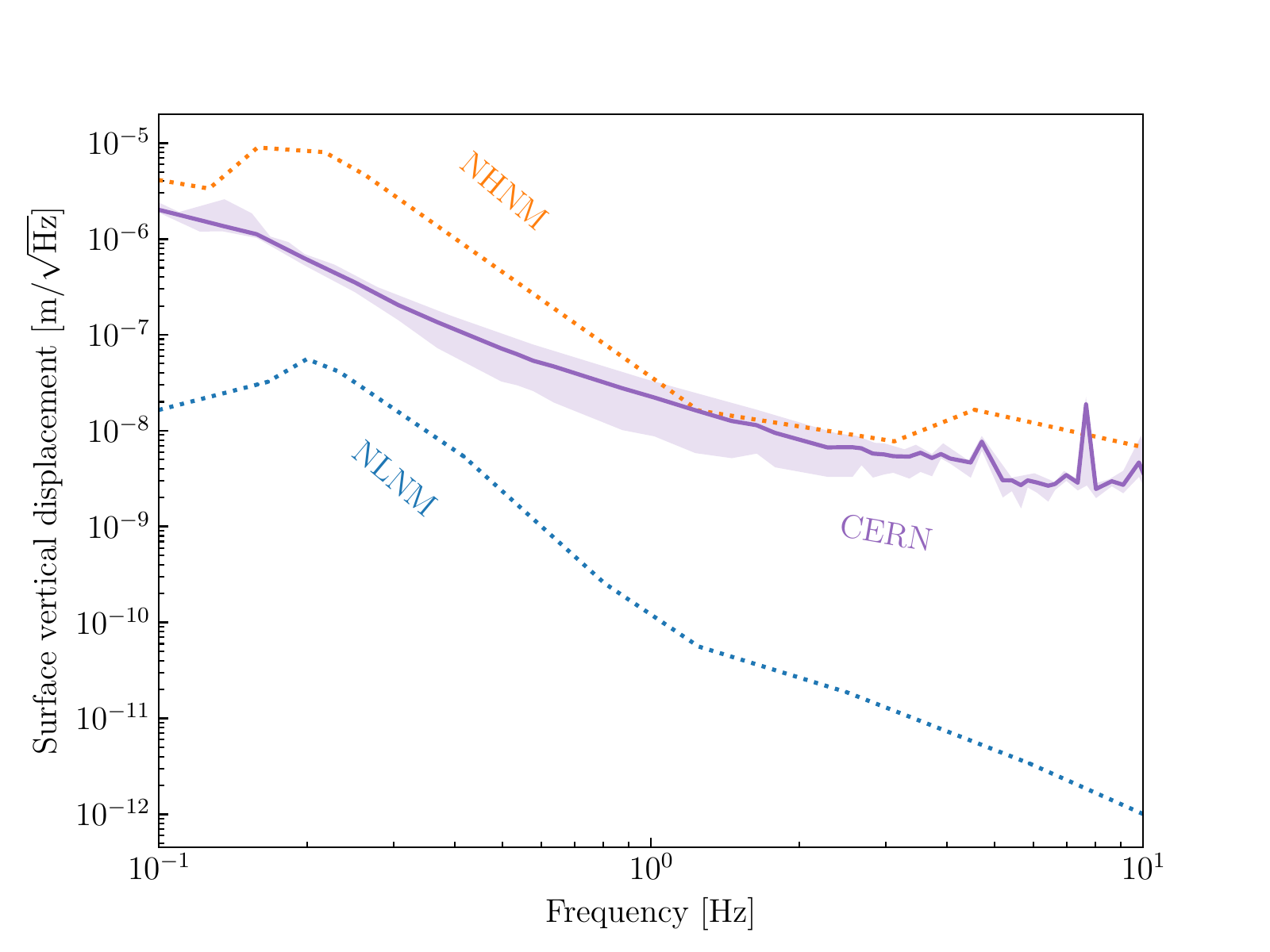}
\qquad
\caption{\label{fig:noisemodels} The RMS spectral density of surface  vertical displacement measurements, compared with the New High and Low Noise Models (NHNM and NLNM)~\cite{peterson1993observations}. The shaded band corresponds to the difference between the minimum and maximum daily measurements shown in Figure~\ref{fig:RMS powerSpectral Density Min Max Case}.}
\end{figure}

Above 1 Hz, the power spectral density varies by a factor up to 10, due to sources known as cultural noise. They can have two origins: local excitation sources (e.g., cooling systems or ventilation), or a variation of the amplitude at different times. The presence of noisy cooling and ventilation (CV) equipment was apparent in the measurements in SX4, leading to a strong peak in the spectrum at a frequency of 50 Hz.

Except for the 50 Hz disturbances coming from the CV equipment, which will require mitigation, all the ground motion measurements performed in SX4 and PX46 are below the AION requirements for the stability of the laser system defined in Section 4.3. 

As seen in Figure~\ref{fig:noisemodels}, the surface  vertical displacement measurements at frequencies $< 1$~Hz generally lie below the NHNM, albeit with fluctuations. The difference between the measured minimum and maximum power spectral density levels correspond to the shaded regions in Figure~\ref{fig:noisemodels} and the uncertainties in the GGN levels shown in Figure~\ref{fig:AION100Sensitivities}.

%\subsubsection{EM noise (D. Valuch, M. Buzio, M. Pentella)}
\subsubsection{Electromagnetic noise}% (D. Valuch, M. Buzio, M. Pentella)
\label{subsubsec:EMnoise}

The main sources of EM noise and background at LHC point 4 that need to be considered are  the RF accelerating system for the machine that is located in the UX45 cavern, the power converters powering the two adjacent LHC sectors 34 and 45 that are located in the UA43 and UA47 galleries parallel to the LHC long straight section, and a large cryoplant distributed between the surface and the underground cavern. Apart from these large systems, standard telecommunication equipment is installed and operated in the underground areas (Terrestrial Trunked Radio~--~TETRA, Long-Term Evolution~--~LTE, or 3G cellular networks) and the overhead crane in UX45 cavern is radio-controlled using Industrial, Scientific, and Medical~(ISM) frequencies in the 433~MHz band. 

The LHC RF power plant uses a total of 16 high-power RF klystrons, ferrite circulators, loads and a waveguide distribution system with a total combined power of 4.8~MW. The operating frequency is known and fixed (400.8~MHz). Such a large RF system is expected to emit EM energy at the operating frequency, as it is technically not feasible to design and manufacture ideal, RF-tight components or high-power transmission lines. Due to the very specific nature of the LHC equipment, there is no concrete technical standard defining the maximum permissible RF radiation level. The system was designed and the monitoring levels are set to limit the maximal radiated power to about 1~W. This value is about an order of magnitude lower than the health limit for human exposure to non-ionizing radiation, and it is similar to the level of power emitted by telecommunication devices. In industry, obtaining such a low level of RF emission for a power station of the size that we have opted for is quite challenging. 
The PX46 vertical shaft where the AI experiment would be installed is over 20~meters far from the RF power plant. The expected contribution to the EM field background at 400.8~MHz from the RF leakage would be less than 0.2~V/m.

The next most important contributors to the EM field background are the low-frequency fields driven by high currents within the cryoplant motor drives, stray magnetic fields from the power converter bus bars, return/ground currents from uninterruptible power supplies, and other return currents through the highly-interconnected metal structure (common bonding network) of the LHC machine, and ground currents in the volume of soil surrounding the LHC point 4~\cite{BRAVIN19989}.

The AI experiment is mostly concerned about low-frequency magnetic fields in the peak detector sensitivity band \SI{50}{\milli\hertz} to \SI{10}{\hertz} (see Section~\ref{sec:env-reqs}). A dedicated measurement campaign was launched in 2022 to characterize the low-frequency magnetic field background. Measurements were performed in spring 2022 before the LHC start-up after a year-end technical stop and later on multiple occasions in 2022 during a standard LHC run. The magnetic field was measured at the bottom of the PX46 shaft (2~m above the floor level) and at the top of the PX46 shaft (5~m below the steel lid of the shaft).

The environmental magnetic field was measured by combining the data from two probes: a three-axis fluxgate magnetometer to measure the magnetic field in the range from DC to 500~Hz, and an induction coil to measure higher frequencies. 
The fluxgate magnetometer was a Bartington Mag-13MS70 \cite{Bartington:2023}, capable of measuring magnetic fields up to 0.7~G (\SI{70}{\micro\tesla}), with a field resolution in the range of 1-2~nT, and a noise level of $\SI{10}{pT/\sqrt{Hz}}$. %The probe was operated with a full-scale range of 0.7~G ($\SI{70}{\mu T}$). 
Higher-frequency magnetic field components were measured by a screened coil, model Schwarzbeck ESP 5133-7/41 \cite{Schwarzbeck:2023}. 

Due to access constraints, all measurements needed to be performed remotely. A remotely-operated high-resolution LeCroy 8108HD oscilloscope was employed \cite{LeCroy:2023}. The digital oscilloscope enabled sampling the analogue signals from the probes at 1~kS/s and storing \SI{50e6}{points}, \textit{i.e.}, a data record length of \SI{13.8}{hours}. The very long record length is necessary to provide sufficient frequency resolution for the required measurements below 1~Hz. 

\begin{figure}
    \centering
    \includegraphics[width=0.9\textwidth]{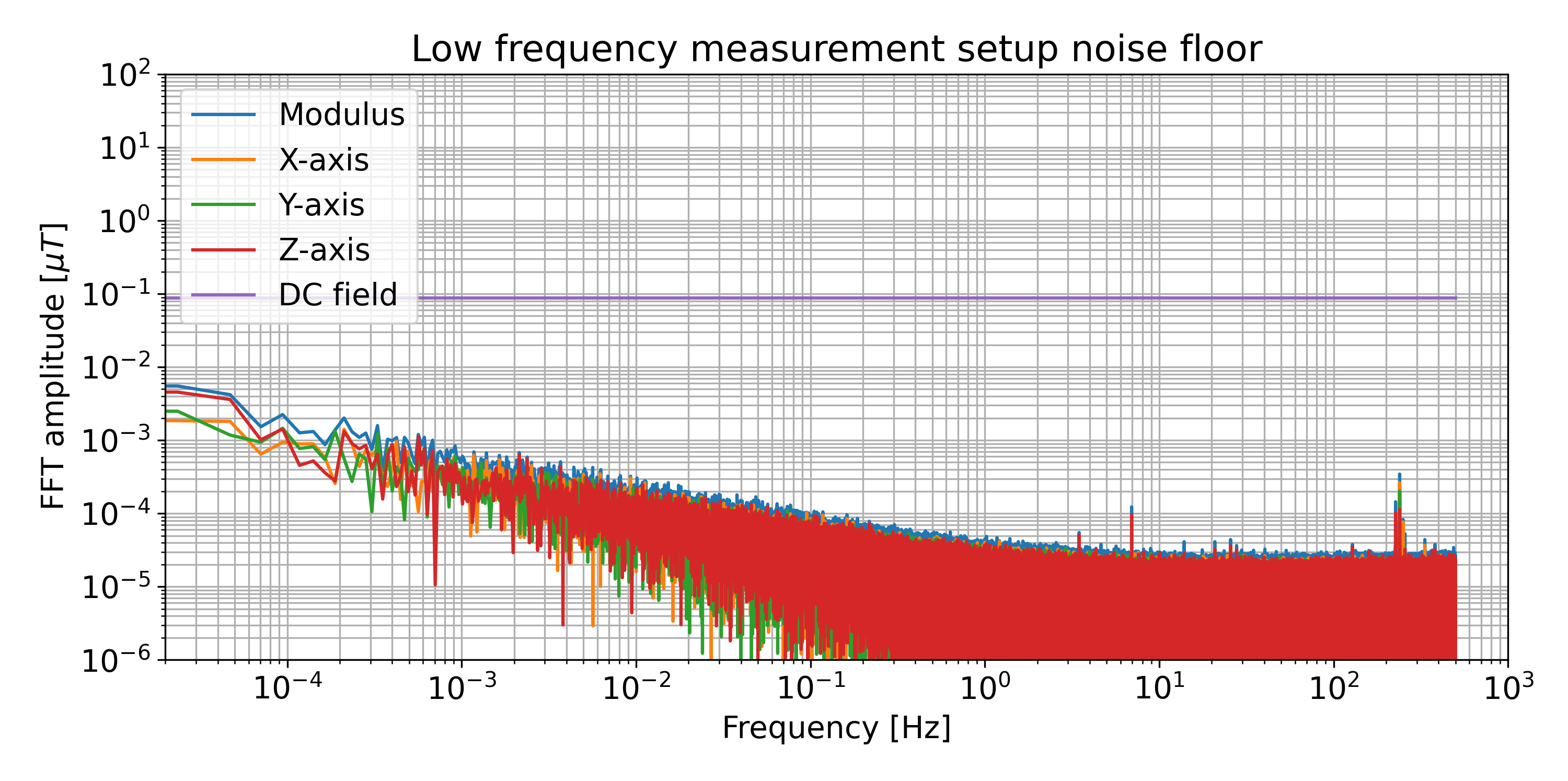}
    \caption{Noise floor of low-frequency measurement set-up (fluxgate magnetometer and digitizer).}
    \label{fig:setupEM-noisefloor}
\end{figure}

\begin{figure}[h!]
    \centering
    \includegraphics[width=0.9\textwidth]{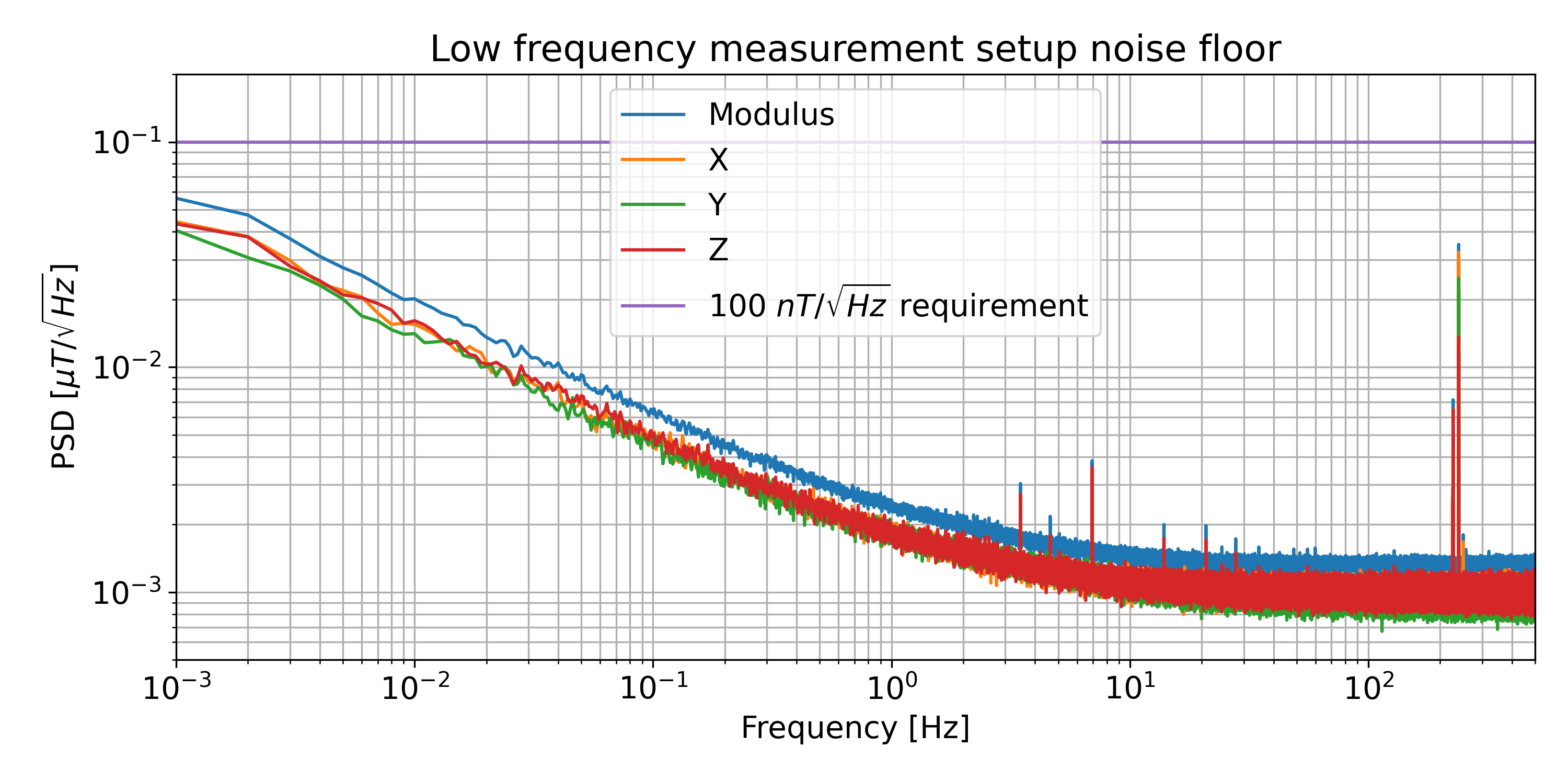}
    \caption{Noise floor of low-frequency measurement set-up (fluxgate magnetometer and digitizer).}
    \label{fig:setupEM-noisefloorPSD}
\end{figure}

The noise floor of the measurement set-up was first verified in the CERN ATS EMC laboratory. The whole set-up (oscilloscope, fluxgate magnetometer, and screened coil) was installed in the laboratory's Faraday cage to provide EM field shielding, and the fluxgate magnetometer sensor was, on top, inserted into a zero-gauss chamber. The measured noise floor is depicted in Figure~\ref{fig:setupEM-noisefloor}. The power spectral density of this measurement with the limit set in the Section~\ref{sec:env-reqs} is shown in Figure~\ref{fig:setupEM-noisefloorPSD}.

\begin{figure}[h!]
    \centering
    \includegraphics[width=0.53\textwidth]{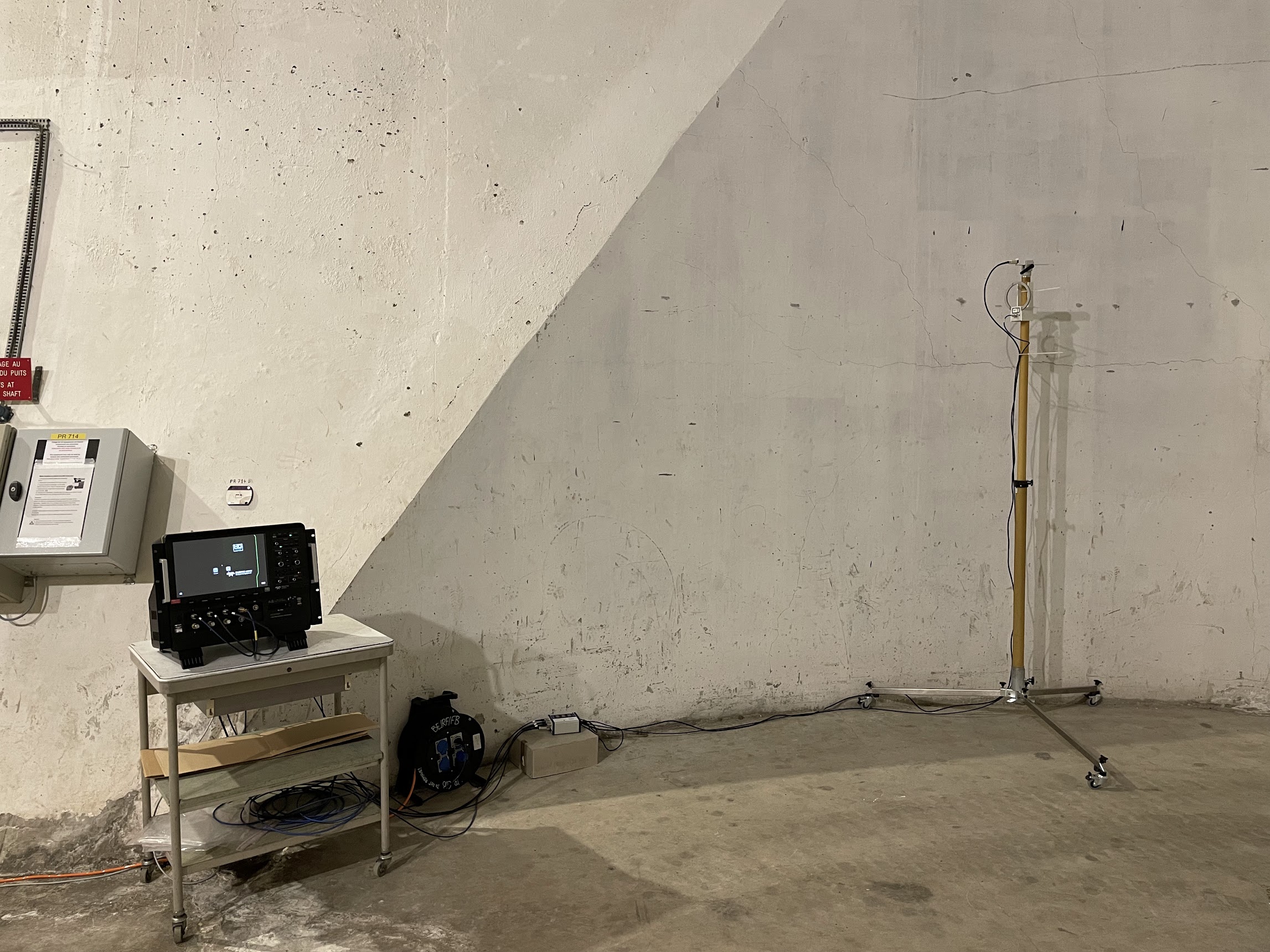}
    \includegraphics[width=0.45\textwidth]{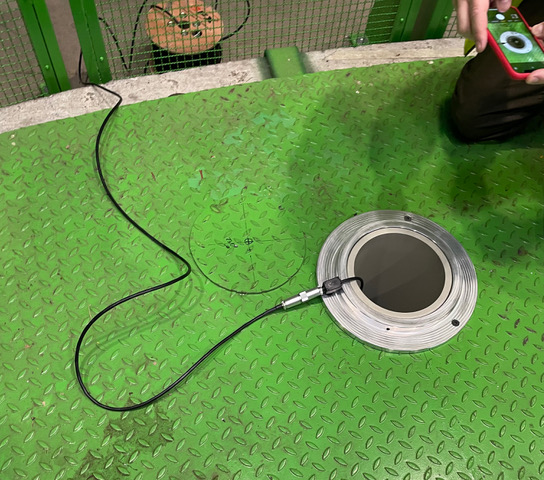}
    \caption{{\it Left}: measurement station for environmental magnetic field monitoring at the bottom of the PX46 shaft: the two probes are on the right side. {\it Right}: measurement station for environmental magnetic field monitoring at the top of the PX46 shaft: the two probes were suspended through a tube installed on the shaft lid. }
    \label{fig:setupEM-bottom}
\end{figure}

The measurement set-up was subsequently installed at the bottom of the PX46 shaft, see Figure~\ref{fig:setupEM-bottom}~(left), and towards the end of the 2022 physics run it was moved to the top of PX46, see Figure~\ref{fig:setupEM-bottom} (right).

% \begin{figure}
%     \centering
%     \includegraphics[width=0.7\textwidth]{Figures/5-1-2-EMnoise/setup_top_PX46.jpg}
%     \caption{A photograph of the measurement station for environmental magnetic field monitoring at the top of the PX46 shaft. The two probes were suspended through a tube installed on the shaft lid. }
%     \label{fig:setupEM-top}
% \end{figure}

\begin{figure}[h!]
    \centering
    \includegraphics[width=0.9\textwidth]{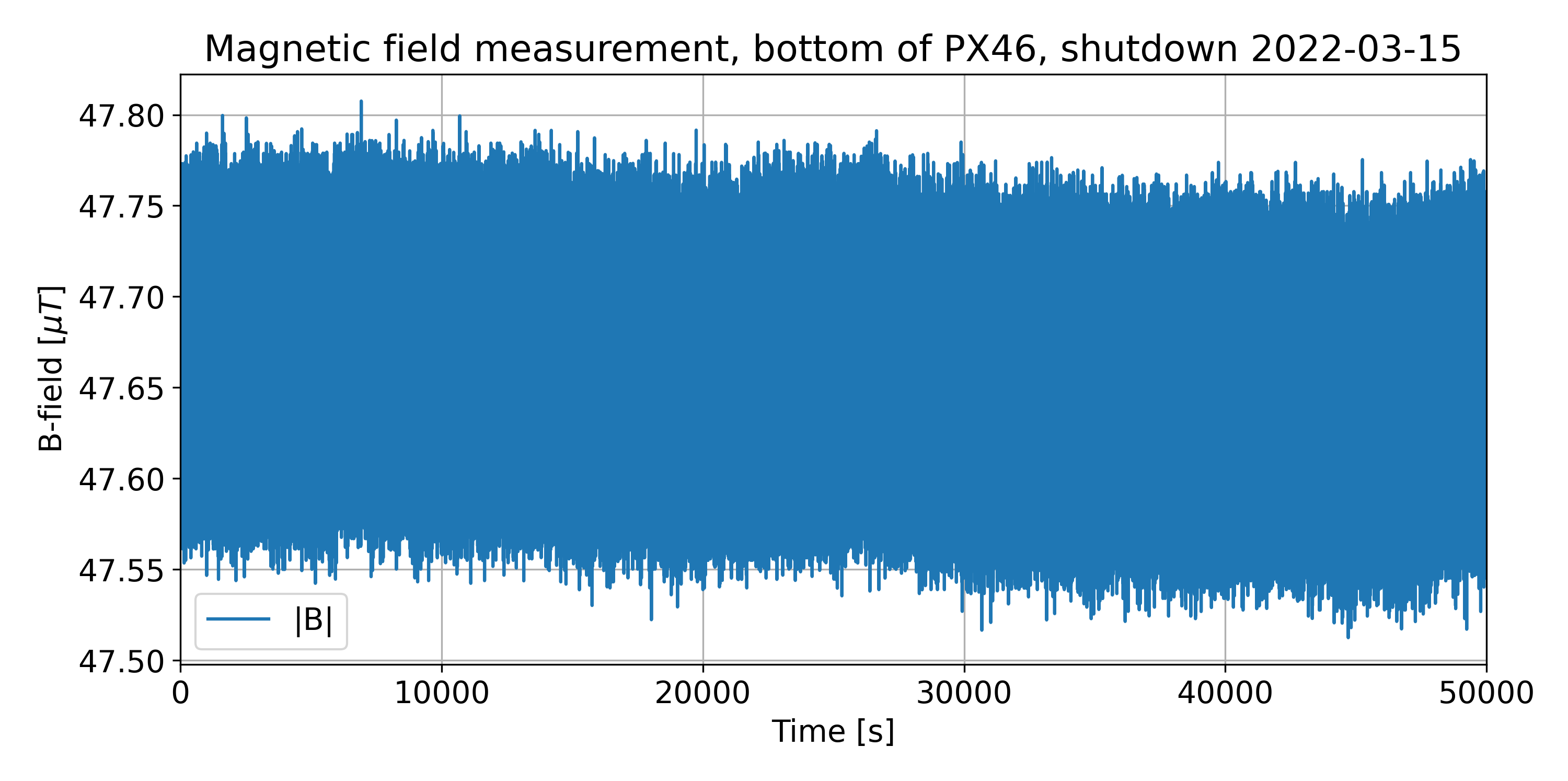}
    \caption{Low-frequency magnetic field measurement during the LHC shutdown.}
    \label{fig:MagField_shutdown_time}
\end{figure}

\begin{figure}[h!]
    \centering
    \includegraphics[width=0.9\textwidth]{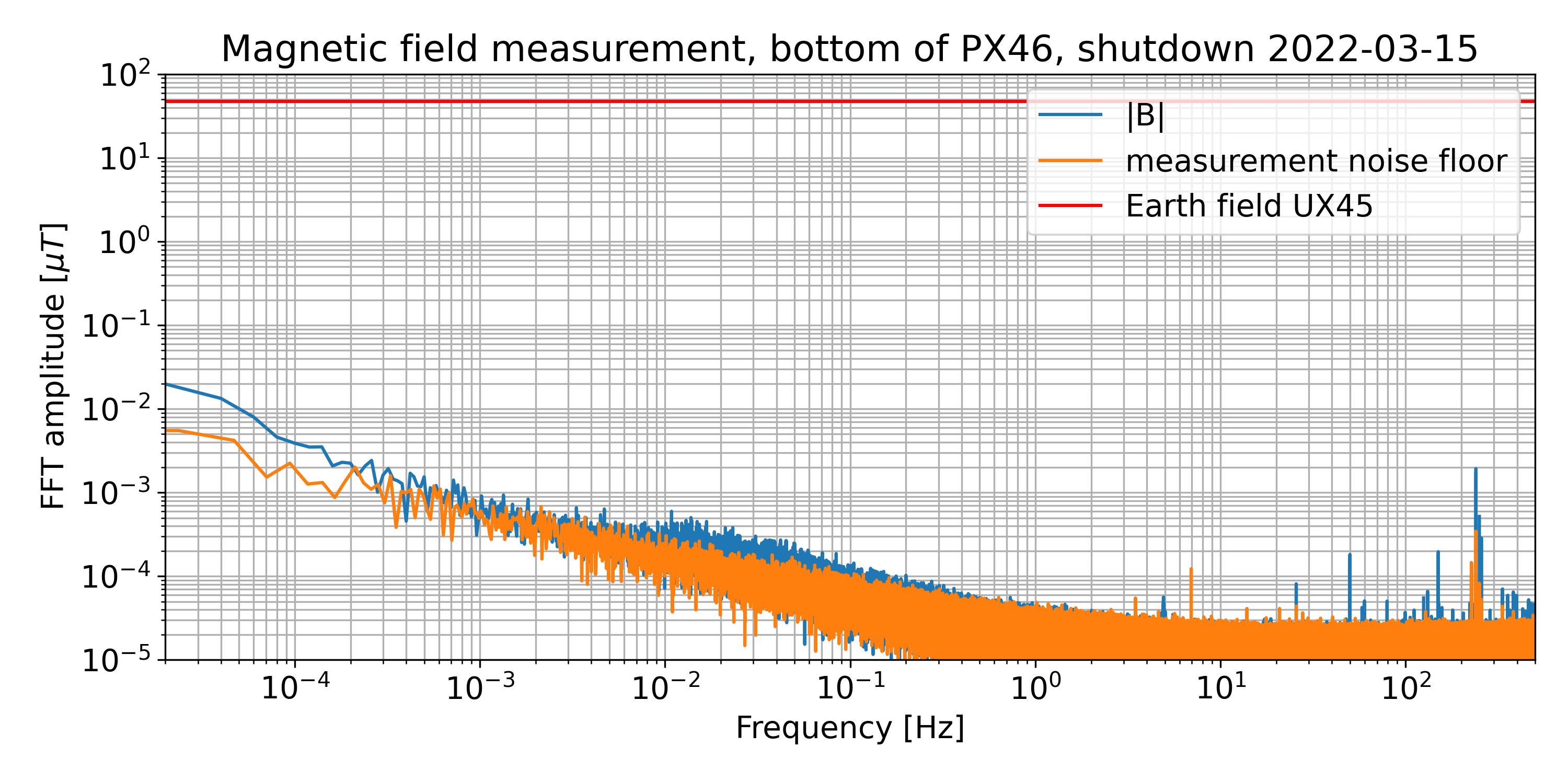}
    \caption{Low-frequency magnetic field measurement during the LHC shutdown, with the Earth's field shown for scale.}
    \label{fig:MagField_shutdown_freq}
\end{figure}

\begin{figure}[h!]
    \centering
    \includegraphics[width=0.9\textwidth]{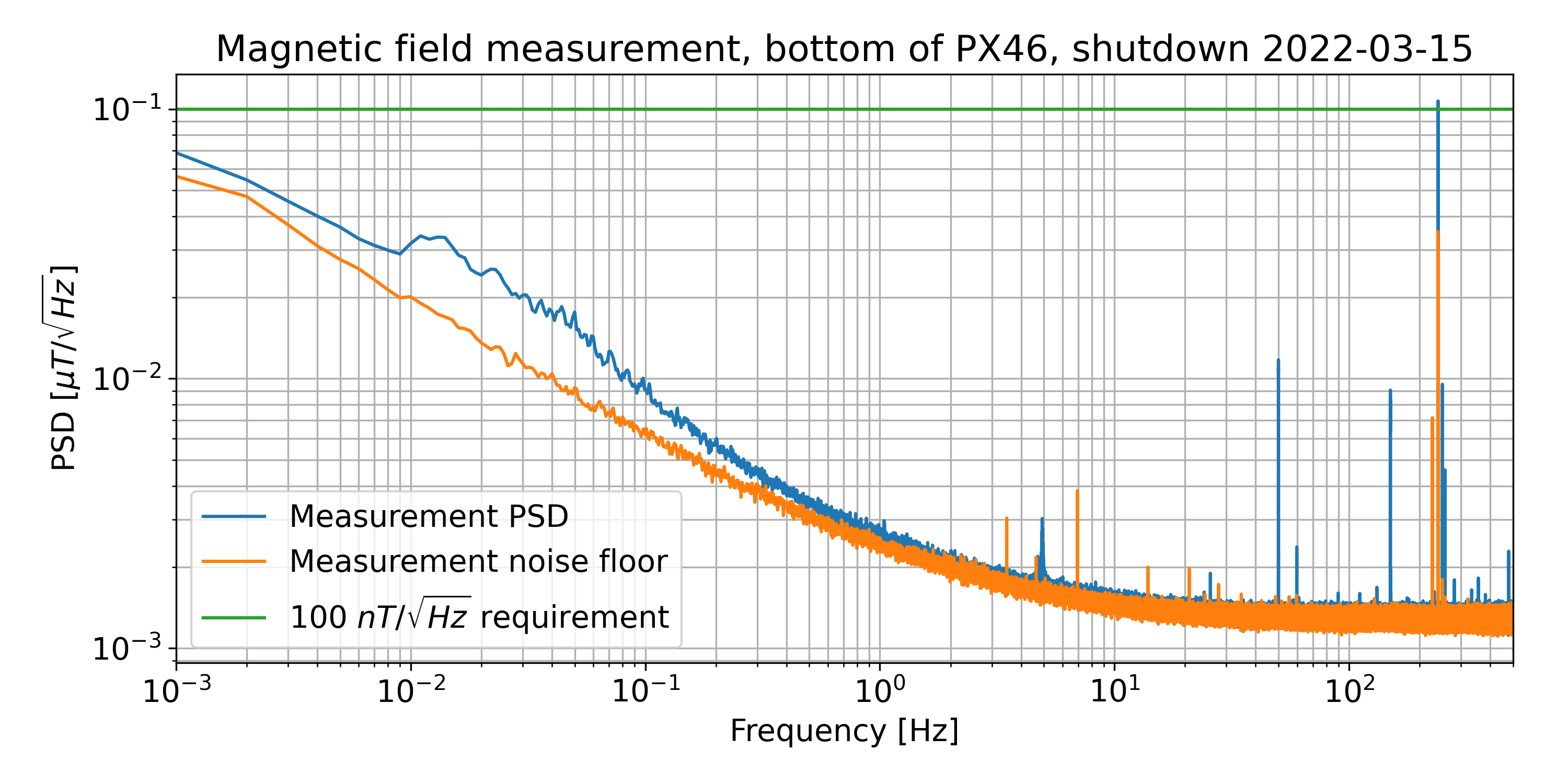}
    \caption{Low-frequency magnetic field measurement during the LHC shutdown.}
    \label{fig:MagField_shutdown_freq_PSD}
\end{figure}

A first set of measurements was done during the LHC shutdown, just before the machine recommissioning after the Year-End Technical Stop~(YETS) had started. Most of the services were already operational, but neither the RF systems, nor the power converters were running yet. Figure~\ref{fig:MagField_shutdown_time} shows the magnetic field measured in the time domain, and Figures~\ref{fig:MagField_shutdown_freq} and \ref{fig:MagField_shutdown_freq_PSD} in the frequency domain. The vertical scale is chosen to show the Earth's magnetic field for reference. 

Measurements taken in the machine show a slightly higher level than the noise floor measured in the Faraday cage and the zero-Gauss chamber. An excess of a few $10^{-2} {\mu T}$ is visible at very low frequencies (time scales of thousands of seconds and longer). The environment in the UX45 cavern is not tightly controlled in temperature or humidity. Due to access constraints, it was not possible to perform another noise floor measurement with the zero-Gauss chamber in the machine. Therefore, the origin of this excess could not be exactly determined. It might be due to the drift of the fluxgate magnetometer temperature, oscilloscope drift, or the diurnal variation in the Earth's magnetic field.

\begin{figure}[h!]
    \centering
    \includegraphics[width=0.8\textwidth]{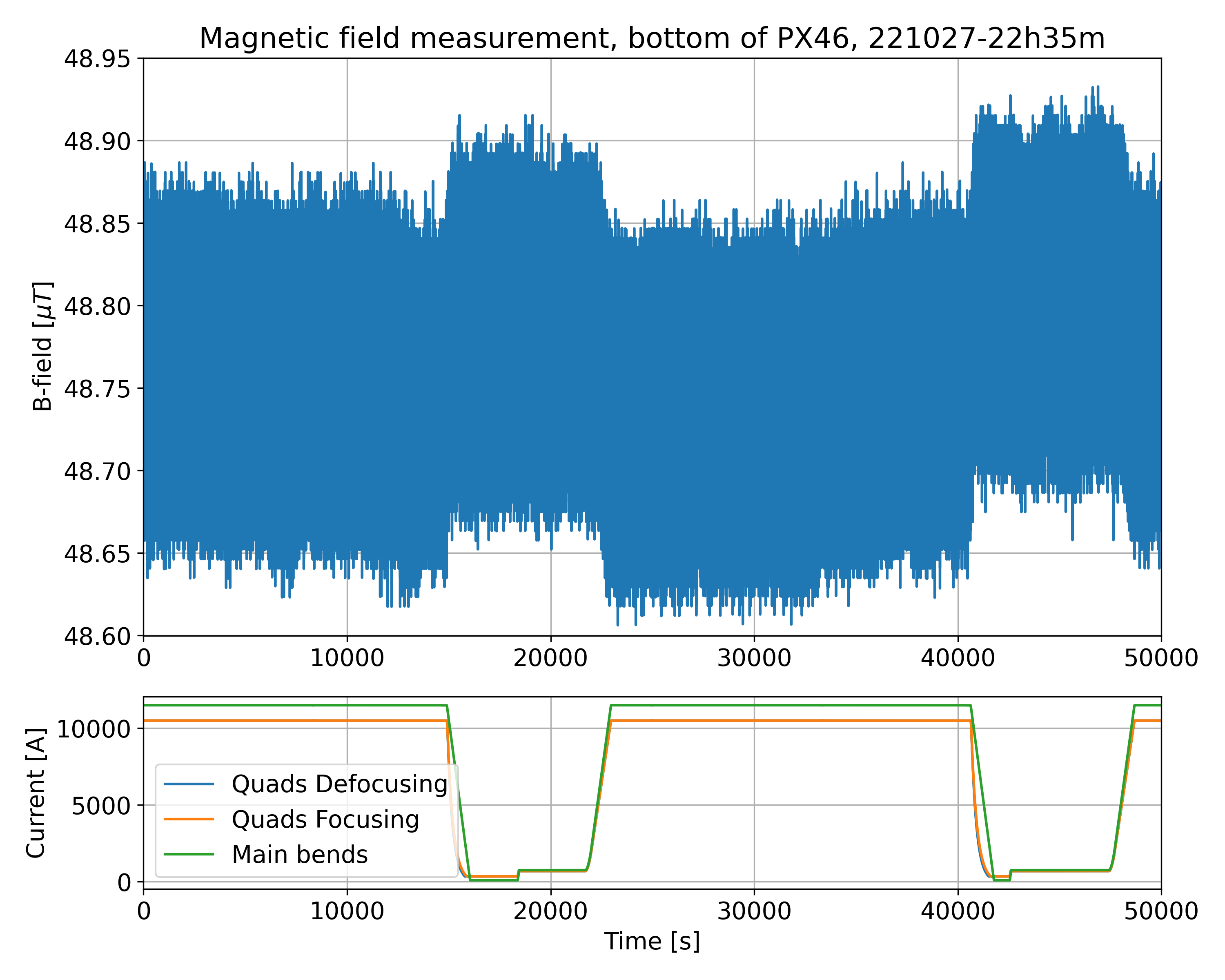}
    \caption{Low-frequency magnetic field measurement during the LHC machine cycle.}
    \label{fig:MagField_cycle_time}
\end{figure}

\begin{figure}[h!]
    \centering
    \includegraphics[width=0.9\textwidth]{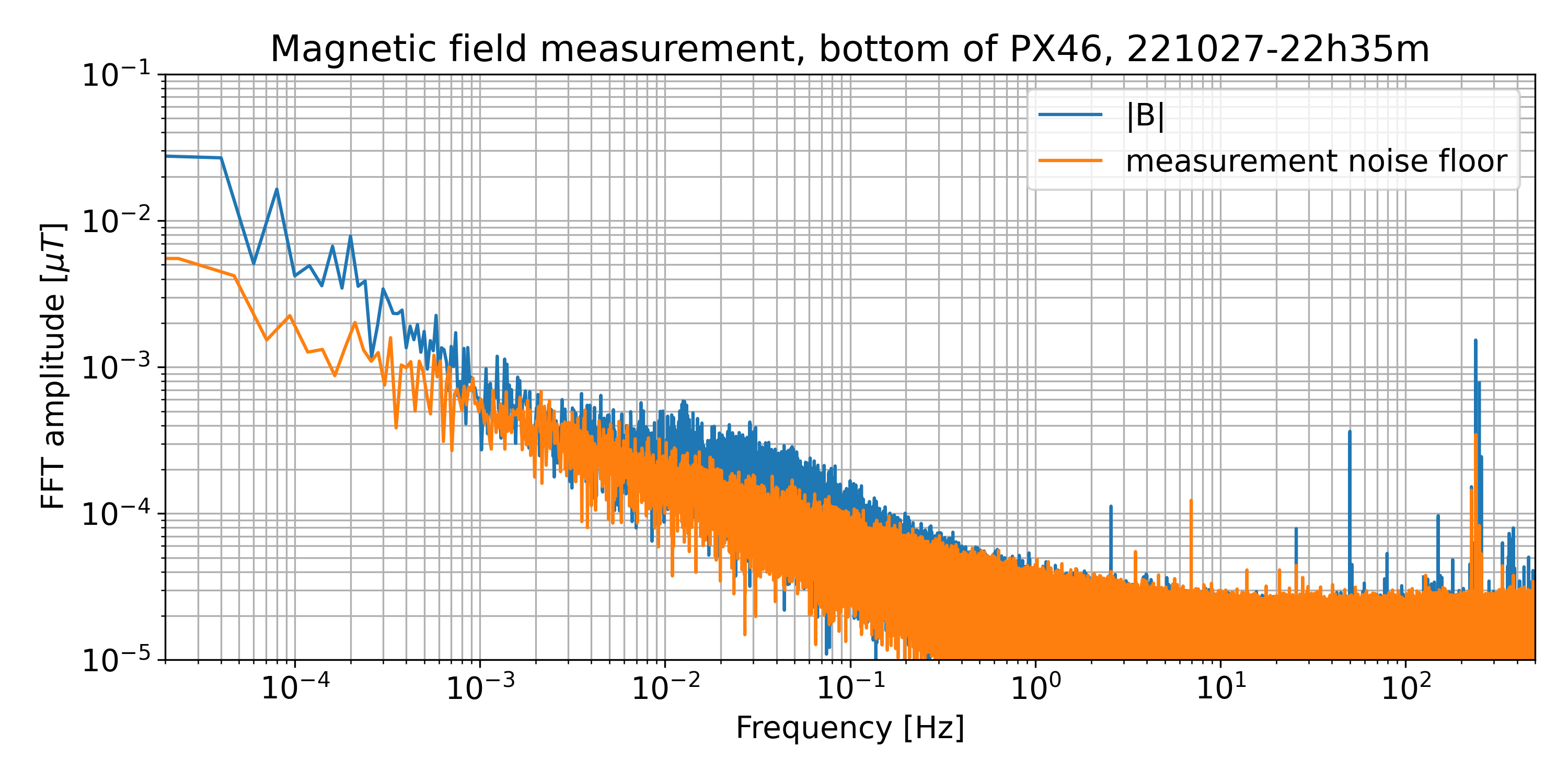}
    \caption{Low-frequency magnetic field measurement during the LHC machine cycle.}
    \label{fig:MagField_cycle_freq}
\end{figure}

\begin{figure}[h!]
    \centering
    \includegraphics[width=0.9\textwidth]{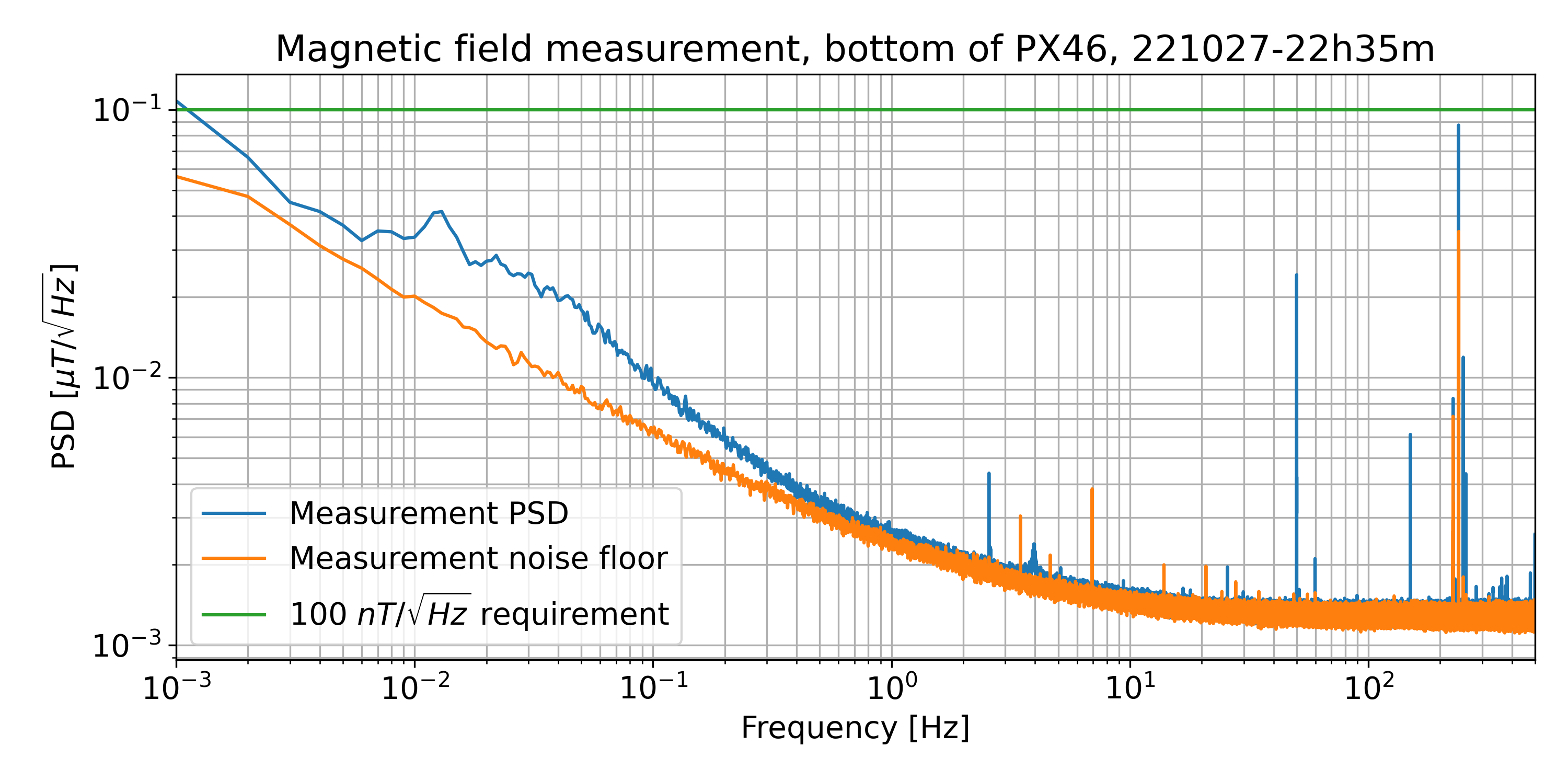}
    \caption{Low-frequency magnetic field measurement during the LHC machine cycle.}
    \label{fig:MagField_cycle_freq_PSD}
\end{figure}

More measurements were performed after the LHC restart, when all the LHC systems were operating. Figures~\ref{fig:MagField_cycle_time}, \ref{fig:MagField_cycle_freq} and \ref{fig:MagField_cycle_freq_PSD} show a typical machine cycle with two flat-top plateaux and two ramp-downs, including the pre-cycle.

The magnet system foreseen for the experiment can be synchronized with the LHC machine cycle and used to compensate the low-frequency magnetic field variations during the  cycle that are shown in Figure~\ref{fig:MagField_cycle_time}.

Measurements in the machine show a few discrete spectral lines at the mains frequency and its harmonics. No significant 'broadband' magnetic field perturbation was identified within the dynamic range and the noise floor of our measurement set-up.  

We conclude from these studies that the EM noise generated by the LHC systems does not pose any problem for operating an AI in the PX46 shaft.

\FloatBarrier

\vspace{0.5cm}

%\subsection{Radiation protection (A. Infantino, H. Vincke)}
\subsection{Radiation protection} % (A. Infantino, H. Vincke)
\label{subsec:RP}

% {\it 2-3 pages: description of the RP risks associated to beam losses when working in PX46, and mitigation measures (shielding wall). Mention assessment of risk for normal beam conditions.}

In the conception of new facilities, different Radiation Protection (RP) aspects need to be taken into account at the design level, such as shielding requirements, radiation levels during operation (stray or prompt radiation) and technical stops (residual radiation), area classification, radiation monitoring, activation of radioactive wastes in view of future disposal, and more~\cite{Forkel-Wirth:2002}. Among them, the  estimate of the stray radiation field is crucial for drafting the preliminary project proposal, identifying possible showstoppers and in general verifying that the preliminary plan is consistent with CERN RP rules. 

The CERN RP rules are set out in the so-called "Safety Code F"~\cite{EDMS-335729:2006}. The objective of Safety Code F is to define the rules for the protection of personnel, the population and the environment from ionising radiations produced at CERN. Safety Code F is based on and updated to meet the most advanced standards in the European and other relevant international legislation, including the legislation of the CERN host states France and Switzerland. 

The RP studies conducted in the context of an AI project aim to:

\begin{itemize}
    \item determine the prompt radiation levels in the PX46 shaft for different scenarios (normal and abnormal High Luminosity-LHC (HL-LHC) operation);
    \item determine the shielding requirements to accommodate the need to access the PX46 shaft during HL-LHC operation.
\end{itemize}

\noindent These aspects are particularly relevant for any requirement to access an AI experiment during HL-LHC operation. Areas inside CERN’s perimeter are classified~\cite{EDMS-810149:2007} as a function of the effective dose a person receives during their stay in the area under normal working conditions and routine operation. The potential external as well as internal exposures have to be taken into account when assessing the effective dose persons may receive when working in the area considered. The exposure limitation in terms of Effective Dose $E$ is ensured by limiting correspondingly the ambient dose equivalent rate, $\dot{H}^{*}(10)$, the operational quantity  for exposure to external radiation. Activation levels of specific airborne radioactive material (airborne radioactivity) and specific surface contamination at the corresponding workplaces for exposure from incorporated radionuclides must be considered. In addition, the exposure of people working on the CERN site, as well as the public, must remain below the dose limits under abnormal as well as normal conditions of operation. Table~\ref{tab:RP:area_classificaiton} shows the limits for area classification of Non-Designated and Supervised Radiation Areas at CERN: in order to deal with the request to access the underground experimental area during HL-LHC operation, PX46 should be classified in one of the levels of Table~\ref{tab:RP:area_classificaiton}. 

\begin{table}[htbp]
    \centering
    \caption{\label{tab:RP:area_classificaiton} Relevant effective dose limits for area classification at CERN~\cite{EDMS-810149:2007} for Non-Designated and Supervised Radiation Areas. Dose limits for Controlled Radiation Areas are not reported, since they are not relevant for this study. PX46 is considered a low-occupancy area, i.e., $<20\%$ working time.}
    \smallskip
    \begin{tabular}{|c|c|c|c|} 
    \hline
        \multirow{3}{*}{Area} & Annual dose limit & \multicolumn{2}{c|}{Ambient dose equivalent rate} \\
        & $E$ [mSv] & \multicolumn{2}{c|}{$\dot{H}^{*}(10)$ [$\mu$Sv/h]}  \\
        & year & permanent occupancy & low occupancy \\
        \hline
        Non-designated & 1 & 0.5 & 2.5 \\ 
        Supervised & 6 & 3 & 15 \\
        \hline
    \end{tabular}
\end{table}

\begin{figure}[ht]
\centering % \begin{center}/\end{center} takes some additional vertical space
\includegraphics[width=.4\textwidth, trim={0cm 0cm 0cm 1cm}, clip]{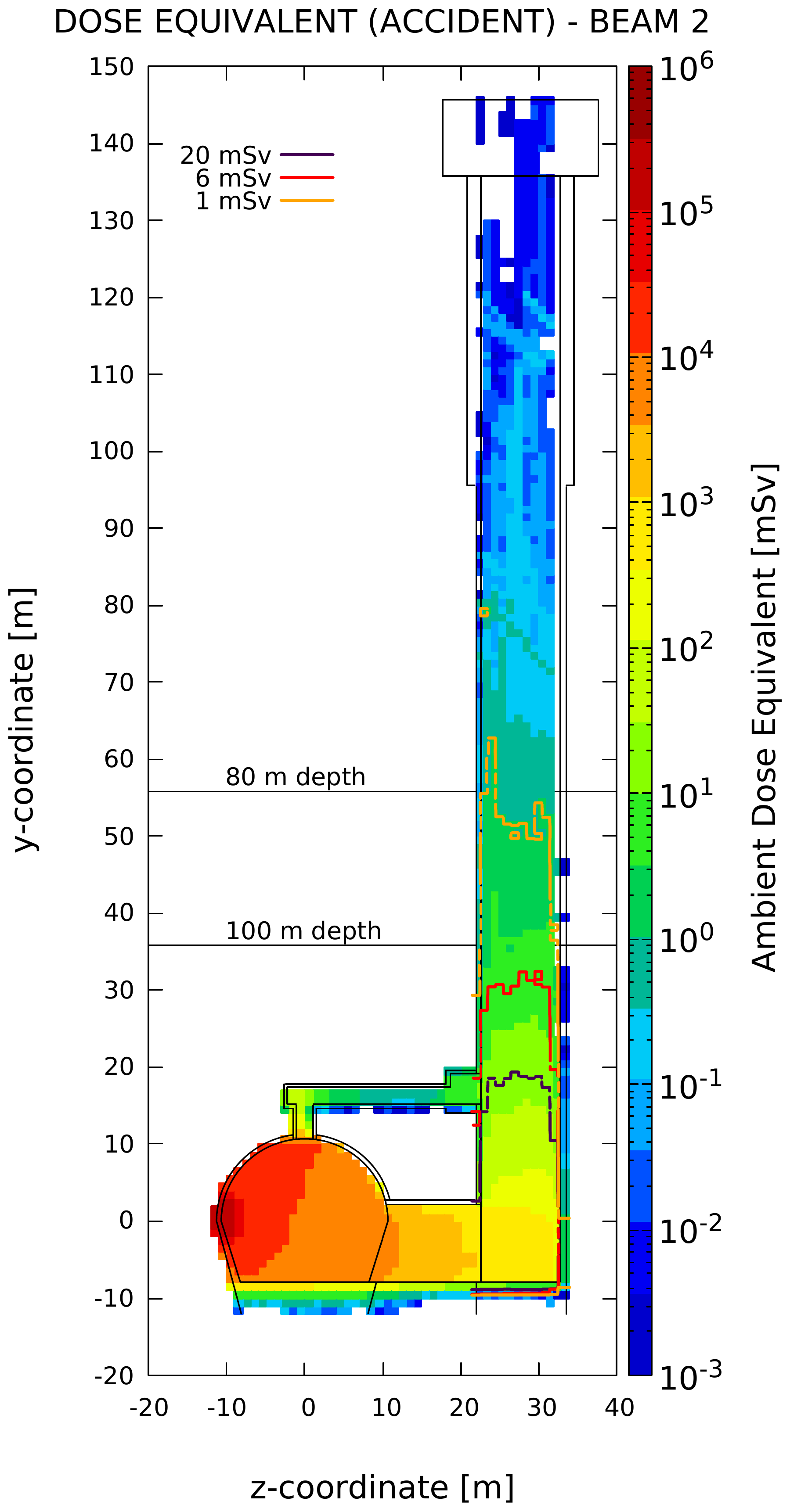}
\qquad
\includegraphics[width=.4\textwidth, trim={0cm 0cm 0cm 1cm}, clip]{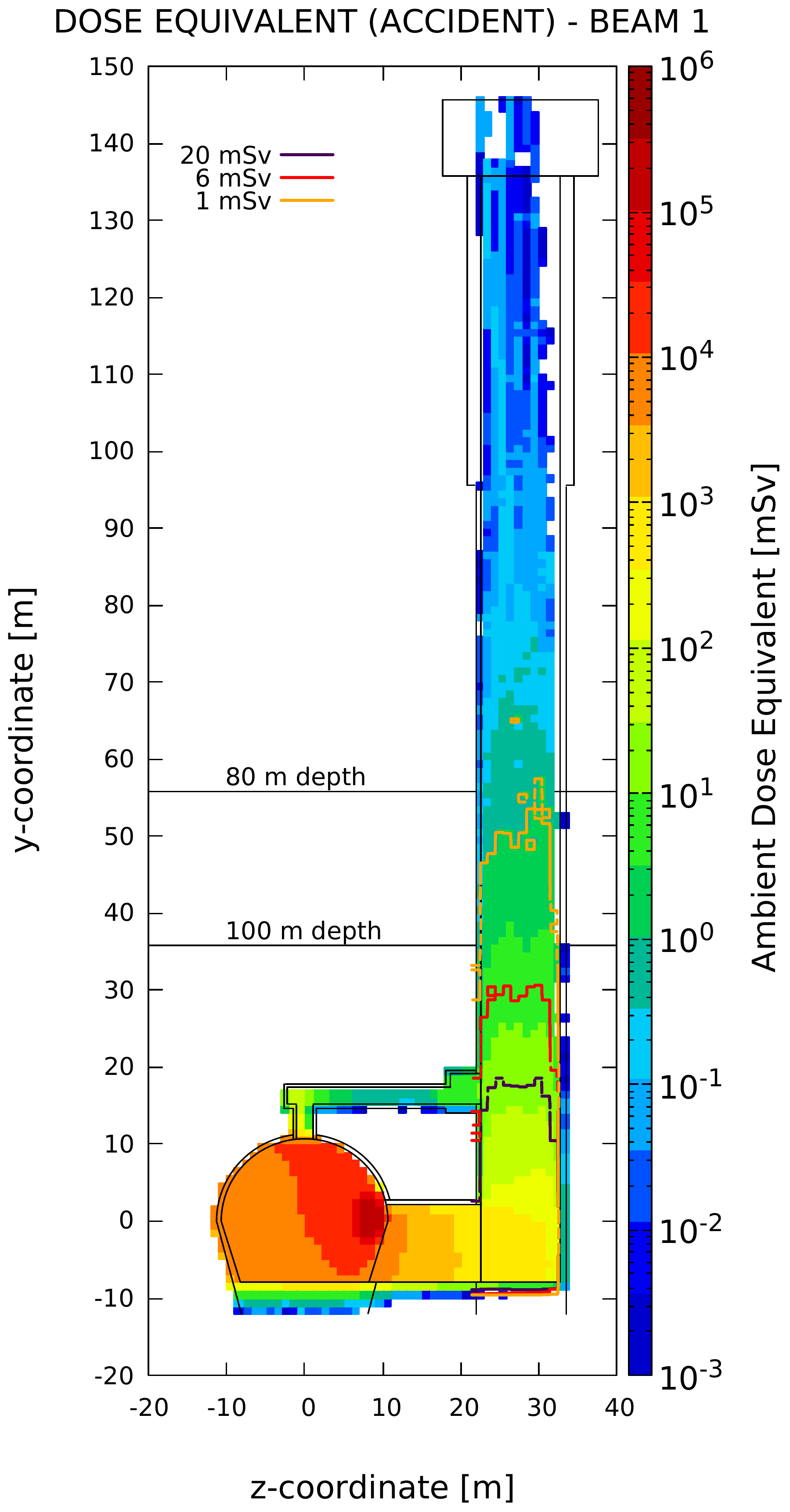}
% "\includegraphics" from the "graphicx" permits to crop (trim+clip)
% and rotate (angle) and image (and much more)
\caption{\label{fig:RP:dose_PX46_ZX} Ambient dose equivalent in the PX46 shaft when an LHC beam (beam 1 on the left, beam 2 on the right) is lost in one of the RF elements in LSS4 and no shielding is present in TX46 or in PX46. Full electromagnetic transport enabled.}
\end{figure}

In recent years, the CERN RP group has provided a series of studies related to the conceptual design of an AI experiment~\cite{Maietta:2020, Infantino:2021, Elie:2022, CERN-PBC-Notes-2022-003}, to which we refer the reader for more technical details. The preliminary RP study conducted for an AI experiment is summarized in~\cite{Maietta:2020}. The FLUKA~\cite{Ahdida:2022} geometry of the LHC Point 4 infrastructure, based on the dedicated drawings provided by the SMB department, includes: the UX45 cavern, the PX46 shaft, the TX46 and TU46 galleries, and air volume at the top of the PX46 shaft to mimic part of the overhead infrastructure (the SX4 building). FLUKA Monte Carlo simulations considering the full loss of the 7 TeV HL-LHC proton beam on an RF element in LSS4 were performed in order to evaluate the ambient dose equivalent field in the PX46 shaft in the absence of additional shielding (see Figure~\ref{fig:RP:dose_PX46_ZX}). For completeness, the simulations considered the loss of either beam 1 or beam 2 in order to assess possible differences of the resulting dose field due to the asymmetry of the infrastructure. The study concluded that a recommended maximum depth of \SI{80}{m} from the surface should be specified and a maximum depth of \SI{90}{m} depth should not be exceeded in the absence of additional shielding.

% \begin{figure}[htbp]
% \centering % \begin{center}/\end{center} takes some additional vertical space
% \includegraphics[width=0.25\textwidth,trim={19cm 2.0cm 25cm 2.0cm}, clip]{Figures/5-2-RP/FLUKA_3D_full.png}
% \qquad
% \includegraphics[width=.4\textwidth, trim={18cm 2cm 18cm 5cm}, clip]{Figures/5-2-RP/FLUKA_3D_cavern.png}
% % "\includegraphics" from the "graphicx" permits to crop (trim+clip)
% % and rotate (angle) and image (and much more)
% \caption{\label{fig:RP:FLUKA-geometry} FLUKA geometry used for the Radiation Protection assessments.}
% \end{figure}

\begin{figure}[ht]
\centering % \begin{center}/\end{center} takes some additional vertical space
\includegraphics[width=.4\textwidth, trim={0cm 0cm 0cm 1cm}, clip]{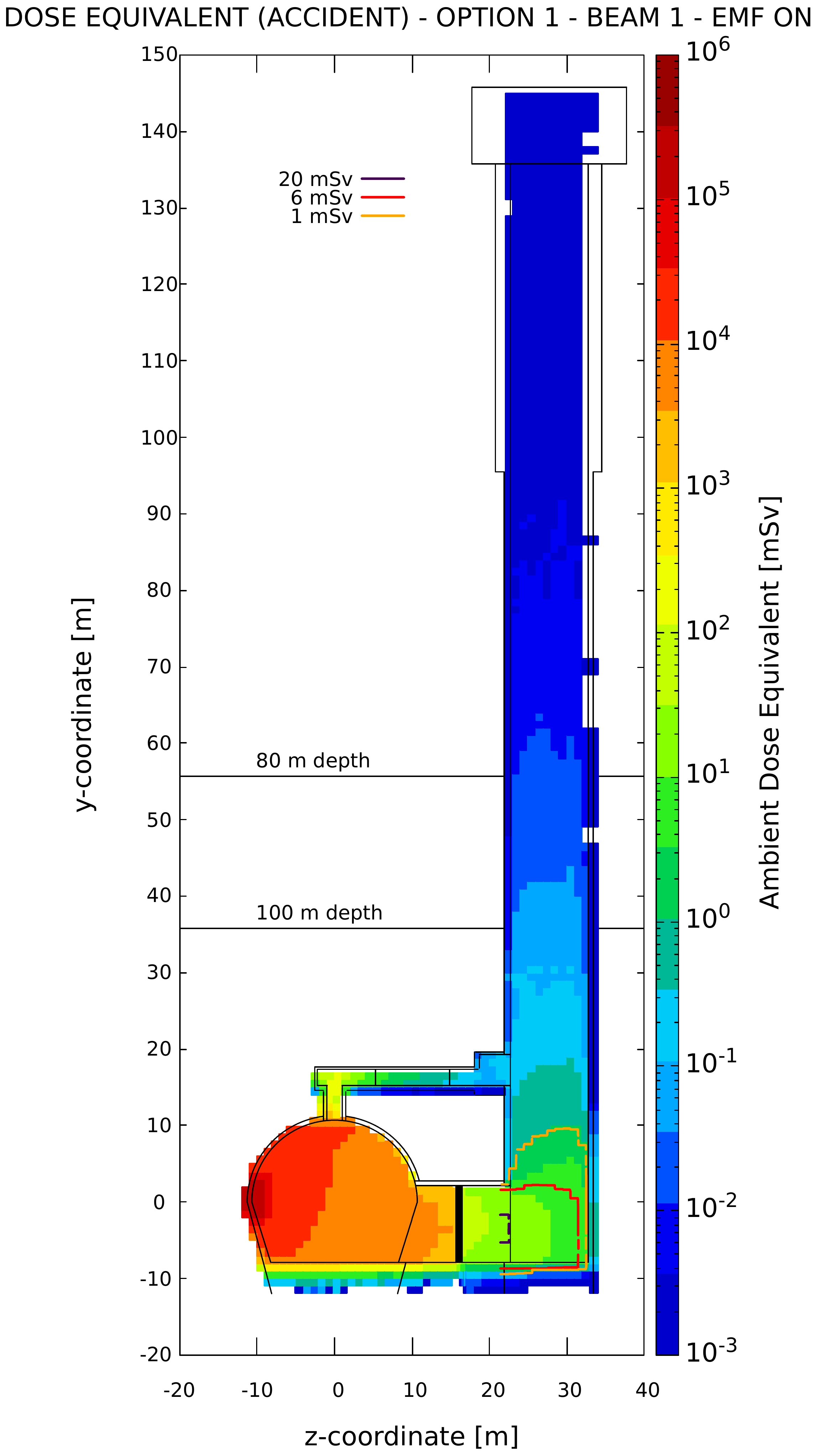}
\qquad
\includegraphics[width=.4\textwidth, trim={0cm 0cm 0cm 1cm}, clip]{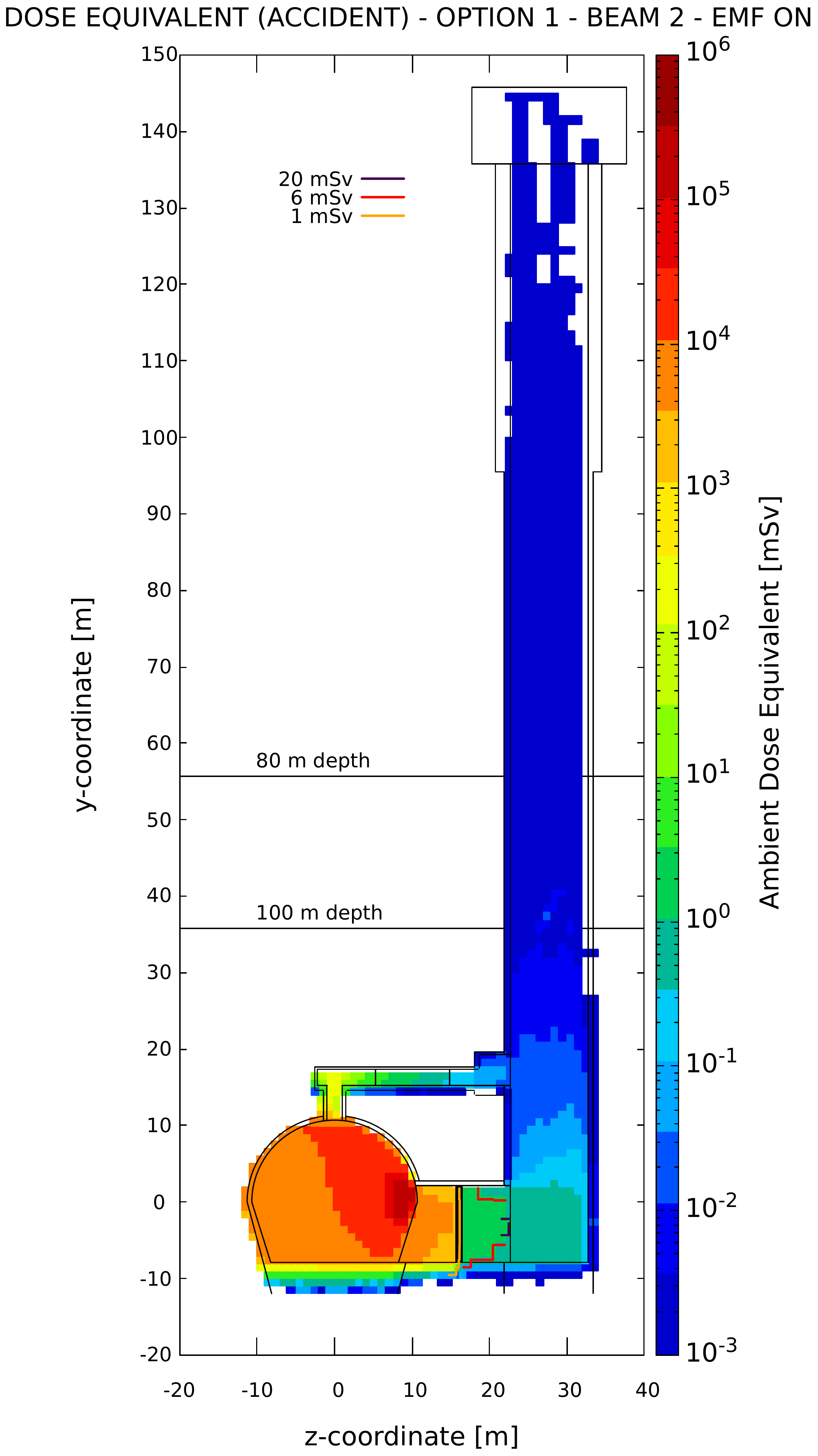}
% "\includegraphics" from the "graphicx" permits to crop (trim+clip)
% and rotate (angle) and image (and much more)
\caption{\label{fig:RP:dose_PX46_ZX_O1_ON} Ambient dose equivalent in the PX46 shaft when an LHC beam (beam 1 on the left, beam 2 on the right) is lost in one of the RF elements in LSS4 and shielding is present in TX46 (Option 1, 80 cm concrete shielding wall). Full electromagnetic transport enabled.}
\end{figure}

\begin{figure}[ht]
\centering % \begin{center}/\end{center} takes some additional vertical space
\includegraphics[width=.4\textwidth, trim={0cm 0cm 0cm 1cm}, clip]{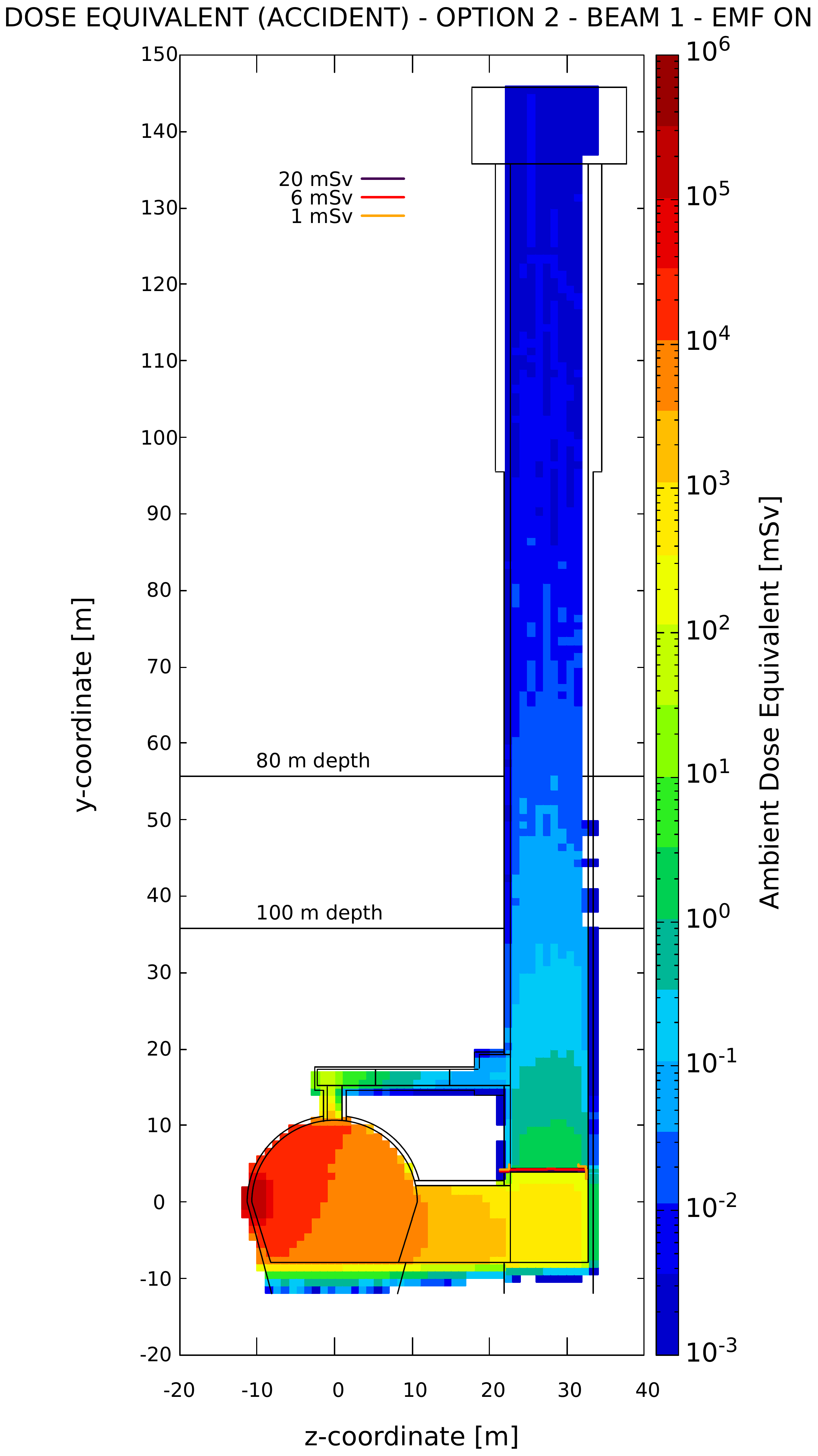}
\qquad
\includegraphics[width=.4\textwidth, trim={0cm 0cm 0cm 1cm}, clip]{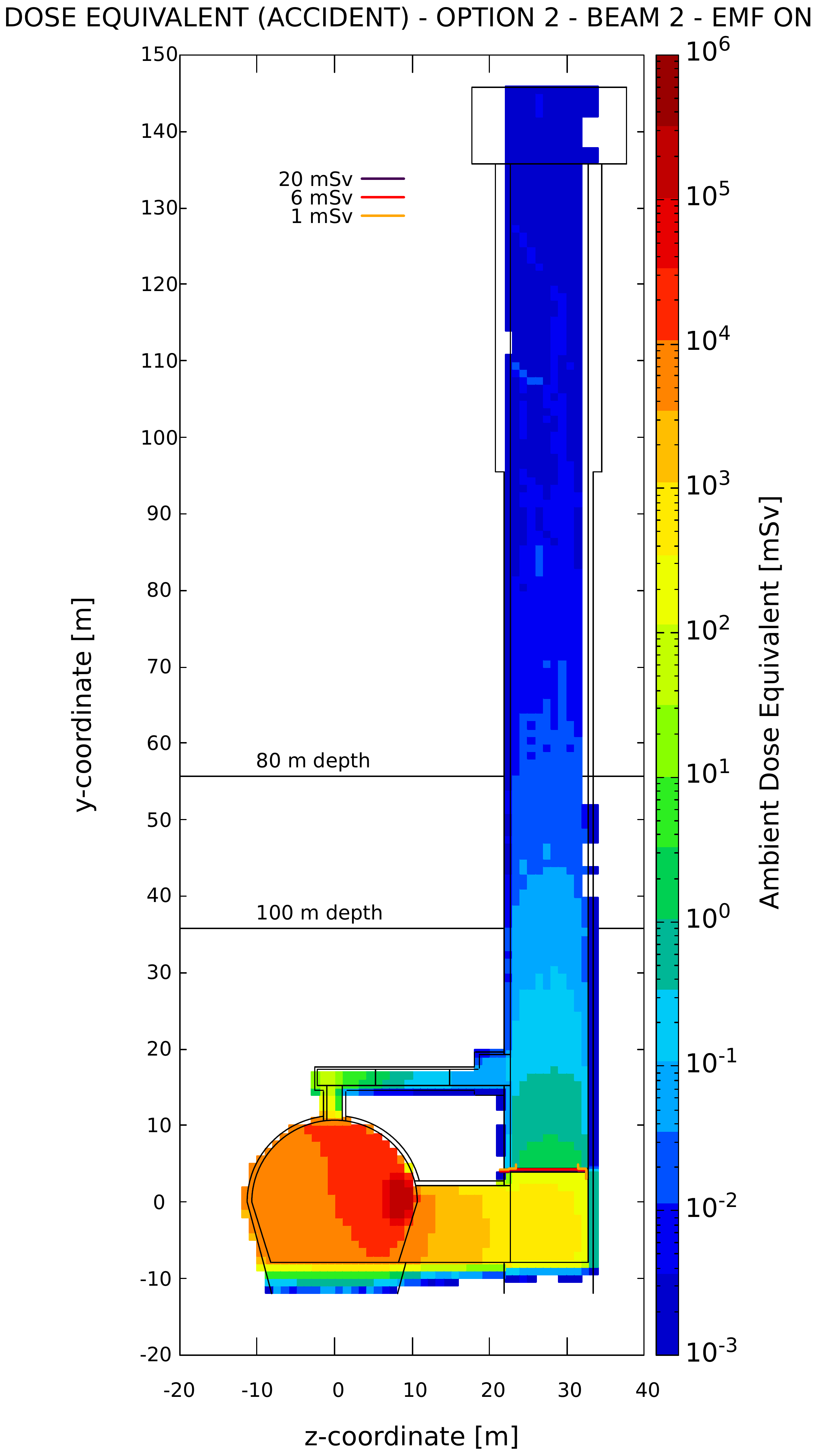}
% "\includegraphics" from the "graphicx" permits to crop (trim+clip)
% and rotate (angle) and image (and much more)
\caption{\label{fig:RP:dose_PX46_ZX_O2} Ambient dose equivalent in the PX46 shaft when an LHC beam (beam 1 on the left, beam 2 on the right) is lost in one of the RF elements in LSS4 and shielding is present in PX46 (Option 2, 40 cm concrete shielding wall in the PX46 shaft). Full electromagnetic transport enabled.}
\end{figure}

A new set of simulations have been performed and summarised in~\cite{Elie:2022}. These new simulations considered two shielding options: in TX46 (Option 1 - Figure~\ref{fig:RP:dose_PX46_ZX_O1_ON}) or in PX46 (Option 2 - Figure~\ref{fig:RP:dose_PX46_ZX_O2}),
respectively. In both options, the presence of the shielding significantly reduces the prompt ambient dose equivalent in PX46. However, it should be noted that the simulations so far consider perfect shielding, i.e., with no leakage. A more realistic representation of the shielding may be introduced in future simulations, once the integration of the new experiment into PX46 is better defined. As shown in Figs.~\ref{fig:RP:dose_PX46_ZX_O1_ON} and \ref{fig:RP:dose_PX46_ZX_O2}, the shielding allows the useful depth of PX46 to be extended, even as far as the floor in the case of Option 1. However, the results in Figs.~\ref{fig:RP:dose_PX46_ZX} and \ref{fig:RP:dose_PX46_ZX_O1_ON} represent two extreme cases, i.e., without and with perfect shielding, respectively. Any deviation from these two scenarios, e.g., the presence of a chicane in TX46, would lead to an intermediate situation in which PX46 might not be accessible down to the floor. More details are reported in~\cite{Elie:2022}.

% \textbf{\color{red} Any consideration about air activation? I believe that this is not an issue because the ventilation in UX45 and PX45 is separated from the ventilation in the tunnel. Correct? Might be worth clarifying}
Finally, the radiological risk due to air activation can be considered negligible, since only the air coming from UX45 is extracted from the PX46. Although the RUX45 platform and the UX45 cavern are not perfectly sealed due to the large aperture connecting the RF cryomodules and the klystrons, the air in UX45 normally has  a lower radiological risk compared to the air in the machine tunnel, which is injected in Point 4 and extracted via Points 3 and 5. Measurements from RP environmental monitors (e.g., VGM904) show releases during beam operation that are several orders of magnitude below the alarm threshold.

%\subsection{Fire safety and MCI compliance (All authors of the following subsections + tbd)}
\subsection{Fire and helium release safety risks and mitigation}
\label{subsec:Safety}

% {\it 1-2 pages: General description of the main causes of safety hazards at the proposed site: fire (in particular RF equipment in UX45), helium release in RF cryomodules, MCI incident in the LHC arcs. Present and discuss the criterion of 2 minutes for reaching safe exits in UP46 and UX45 and discuss the consequences on the people lifting equipment in the absence of a staircase. Requirements of PPE.}

In the previous Section~\ref{subsec:RP} it was demonstrated that the PX46 shaft could be made accessible when the LHC beam is running, with a limited radiation hazard even in case of beam loss, by construction of a suitable radiation shielding wall in the TX46 gallery (pending further detailed evaluation of the needed thickness of the wall and on the positioning of the chicane for the access doors). This complies with the requirements  for the operation of an AI set out in Section~\ref{sec:Overview}, where in particular access to the side-arms would in principle be needed  12~hrs/day, 365~days/year.
However, other safety hazards might be encountered in the PX46 shaft, in particular in the event of a fire or of a release of helium from the LHC machine.

This section addresses in some detail these two cases. From a general point of view, in the event of fire (smoke) or helium hazard, escape of the operators of the AI must be possible in the shortest time and safest way possible. Given the height of the PX46 shaft, it would be reasonable to assume the use of an elevator for normal access to the experiment and escape in case of emergency, with a parallel staircase for emergency evacuation in case of hazards and in case of failure of the elevator. However, the limited space available in the cross section of the PX46 shaft prevents the installation of both. This led to the choice discussed in the following sections and in Section~\ref{subsubsec:Platform} of having a single special lifting platform that can be used for normal access and for escape in all cases of emergencies, and which complies with all relevant requirements, regulations and rules.

%\subsubsection{Fire safety considerations (F. Corsanego)}
\subsubsection{Fire safety measures}% (S. Calatroni)
\label{subsubsec:Fire}

% {\it 2-3 pages: description of the requirement for fire safety prevention and escape measures, in particular fires originating in UX45 and escape routes via existing escape lanes in UP46 or via top of the shaft PX46}

\begin{figure}[h!]
	\centering
	\includegraphics[width=.5\textwidth]{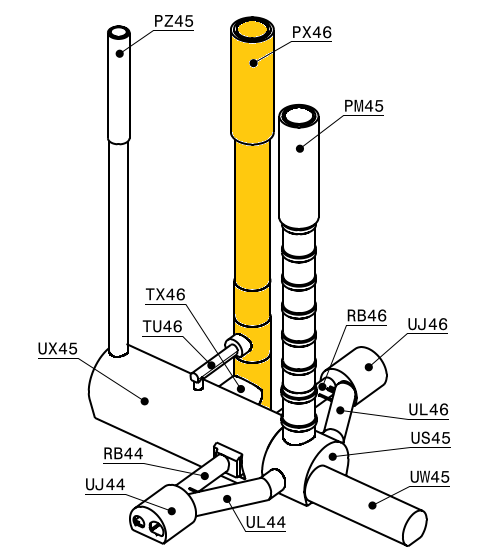}
	\caption{\label{fig:Fire1} Isometric view of the underground caverns, shafts and tunnels}
\end{figure}

The UX45 cavern contains high-power RF equipment such as the main klystrons and power converters for the LHC RF system. In the event of fire, the appropriate smoke detection and alarm systems are already in place, and evacuation routes are already prescribed, either through the PZ45 access shaft or through the PM45 access shaft via the UP46 connecting gallery (see Figures~\ref{fig:Fire1}~and~\ref{fig:PX46situation}). All relevant pre-existing information regarding fire safety prescriptions is documented in~\cite{LHC-0000006238:2013}.

An AI experiment in the PX46 shaft might also be impacted by any fire developing in the UX45 cavern and, after the construction of a radiation shielding wall in the TX46 gallery as discussed in Sections~\ref{subsec:RP} and~\ref{subsubsec:CE}, the smoke produced in the UX45 cavern will be evacuated primarily via the TU46 ventilation gallery and the PX46 shaft. To comply with the risk for the operators associated with the presence of smoke, it is anticipated that the access platform as described in Section~\ref{subsubsec:Platform} is used also as a means for escape, following the triggering of a fire alarm. Alarms from any emergency occurring in the UX45 cavern must be made visible and audible for the occupants of the platform and at the boarding stations on the surface, so that they can initiate the emergency evacuation or refrain from using the access platform if they are still on surface. Operators should put on appropriate Personal Protection Equipment (PPE --- self-rescue masks), which they should carry with them at all times when accessing the AI experiment, or which should be made available on the mobile platform in a suitable cabinet, and proceed with the evacuation. 

The mobile platform must be designed in such a way that in the event of an emergency it can reach, in full autonomy, both the surface and the lower level of the UX46 shaft within approximately 2 minutes.
A battery backup is foreseen, to allow this operation also in case of loss of the mains electrical supply.
In the case of a failure of both the mains electrical supply and of the battery backup, it is also foreseen that the platform can still descend, in a controlled, autonomous and safe way to the lower level of the PX46 shaft, within the same timeline of about 2 minutes mentioned before. This will leave enough time for the operators to safely escape through the access doors in the shielding wall in the TX46 gallery (see Section~\ref{subsubsec:Access}) and then via the pre-established routes mentioned above.

Specific auxiliary means of access for the fire brigade rescue teams will also be implemented, in order to allow the occupants to be reached and evacuated in the event of mechanical blockage of the platform and failure of the manoeuvre for reaching the ground level.
Different options, to be investigated, include access via a rescue cage, made either using the existing crane bridge, or a dedicated crane system to be installed. 
Bidirectional communication systems between the platform and the fire brigade control room are also foreseen, as is the case for all lifts at CERN.

Smoke from a fire in the LHC machine tunnel in sectors 34 and 45 or from the cryogenic plant in the US45 cavern should remain confined in the LHC tunnel region thanks to the physical separation from the UX45 cavern (due to the fact that the ventilation doors in the RB44, UL44, RB46 and UL46 galleries are normally closed), and either be evacuated from the PM45 access shaft or be blown away from LHC point 4 in the direction of the odd LHC points 3 and 5, following the standard air flow in the LHC tunnel and depending on the fire action matrix for such events. Alarms will anyhow trigger, and the AI operators will escape following the procedure described above.

The AI experiment itself is foreseen to consist of equipment with a relatively low combustible load; a complete assessment will have to be carried out at the stage of the detailed technical design. As for any CERN equipment, the materials used for fabrication will have to comply with the relevant Safety Instructions and Safety Guidelines, such as \cite{IS23:2005}, \cite{IS41:2005} and \cite{SG-FS-2-1-1:2021}, so as to minimize fire-induced hazards. As an example, the vacuum pumping stations should preferably rely upon dry pumps instead of oil-sealed pumps. No major differences to the escape procedures mentioned above are anticipated, but the installation of other smoke detectors and fire alarms will have to be assessed, and a detailed matrix of events and conditions that should trigger alarms and evacuation should be defined. Adequate operator training will be mandatory (and relevant courses to be defined) before granting access to the experiment, so that the final decision on the preferred directions and routes to take for evacuation can be left to the operators (thus avoiding automatic servo triggering of the movement to the mobile platform, for example).

Concerning the surface laser laboratory, standard fire and smoke detectors will have to be installed depending on the assessed risks, as for other laboratories in CERN technical buildings. Fire extinguishers and other similar fire mitigation devices will be provided, as defined by the final detailed technical study.

\subsubsection{Helium release incidents safety measures} %(S. Calatroni, tbd)
\label{subsubsec:Helium}

% {\it 2-3 pages: description of the consequences of a helium release incident (RF cryomodules, MCI from the LHC arc) and the needed safety measures}

Two major scenarios of uncontrolled helium release giving rise to a safety hazard are possible at LHC point 4: either a release from the superconducting RF cryomodules located in the RUX45 section of the tunnel~(see Section~\ref{subsec:Site}), or a release from the main LHC superconducting dipole strings in the arcs 34 or 45.

In the first scenario, the large apertures in the wall for the passage of waveguides between the RF cryomodules in the RUX45 platform and the klystrons in the UX45 cavern make it impossible to isolate these two volumes in case of helium release. Helium would then flow from the UX45 cavern via the ventilation gallery TU46 and finally escape via the PX46 shaft, as discussed also in Section~\ref{subsubsec:HVAC}. The total volume of liquid helium in the cavity cryomodules is $4 \times 320$~litres, corresponding to 972~m$^3$ of gaseous helium. Its release from the RUX45 platform to the UX45 cavern is slowed by the narrow waveguide passages, and the total volume of the UX45 cavern is about 18000~m$^3$: in first approximation the full release of helium from the cryomodules would not pose a significant hazard~\cite{Weisz:2009}. This is confirmed by recent findings following a helium release from the RF cryomodules in August 2022~\cite{Hakulinen:2022}, when measured oxygen levels in the TU46 ventilation gallery remained within acceptable limits. Moreover, it is anticipated that the AI operators would access the experiment with self-rescue masks, helping to minimize the risk. It should also be added that, at present, restricted access to the UX45 cavern (and thus to the PX46 shaft) is possible~\cite{Wenninger:2020} during LHC powering phase II, corresponding to full active hardware exploitation without beam~\cite{Solfaroli:2009}. This would remain unchanged upon installation of an AI experiment; the AI experiment itself being further separated by the foreseen radiation shielding wall in the TX46 gallery. However, a detailed assessment of any modifications of the air flows upon construction of the radiation shielding wall will have to be performed during the technical design phase.

 \begin{figure}[ht]
	\centering
	\includegraphics[width=.4\textwidth]{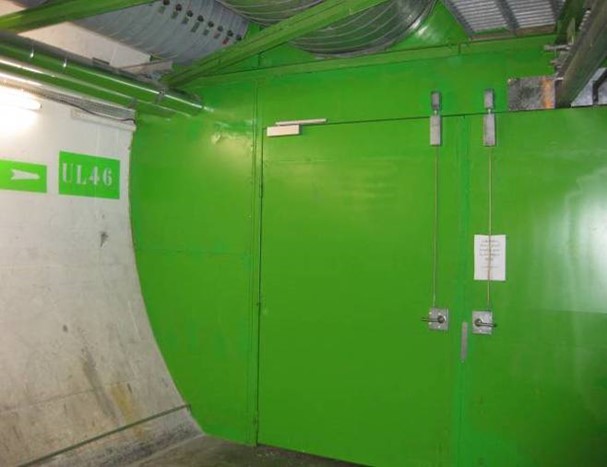}
	\qquad
	\includegraphics[width=.4\textwidth]{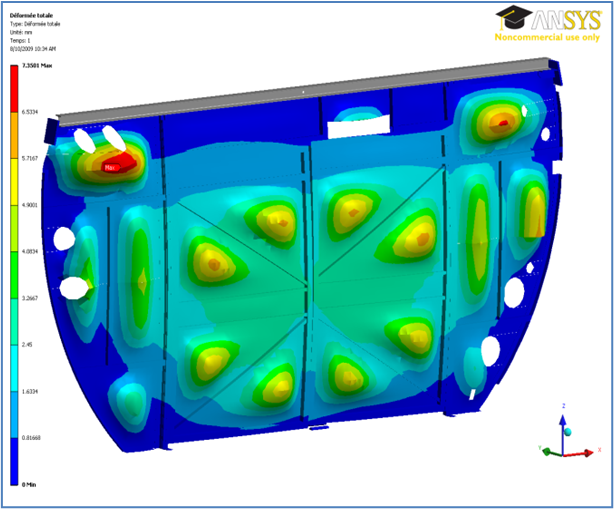}
	\caption{\label{fig:Helium1} Door in UL46 that acts as pressure relief valve. The right part is a simulation of the deformation under a 30~mbar overpressure (deformation = 7.4~mm max.)~\cite{Gaignant:2021}.}
\end{figure}

 The scenario of helium release from the LHC dipole string corresponds to an accident similar to the one that occurred in September 2008 and analyzed in the final report of an \textit{ad hoc} Task Force~\cite{Task34}. In such an event it is considered that an amount of 40~kg/s of liquid helium can be released in the LHC tunnel in a Maximum Credible Incident~(MCI) event. Following the recommendations of the Task Force, and in order to minimize safety hazards to personnel in case a similar event should repeat, confinement doors in the LHC tunnel have been installed or upgraded in order to properly route the helium flow, that are able to sustain the expected overpressure~\cite{Risk34} or alternatively open according to the different foreseen scenarios, as extensively discussed in~\cite{Weisz:2009,Gaignant:2021}. If such an event were to occur in either of the LHC sectors 34 or 45, the ventilation doors located in the UL44 and UL46 galleries would act as pressure valves that open under a 10~mbar overpressure from the tunnel side. They are reinforced to stand an overpressure of 30~mbar from the other side, to ensure that the helium flow would go up the PM45 shaft and not leak into the adjacent sector~(see Figures~\ref{fig:Fire1}~and~\ref{fig:Helium1}). It should also be noted that the general LHC ventilation pattern foresees that air is pushed into the tunnel from even points and extracted from odd points.
 %: in the event of an MCI happening at a large distance from Point 4 helium influx or oxygen deficiency is expected only over a few 100~m upstream of the ventilation flow. This has been confirmed during a specific helium spill test~\cite{spill}.

 Finally, a pressure-resistant partition and ventilation door has been installed in the RB44 gallery to separate sector 34 from the RUX45 platform, where the superconducting RF cavities are located. It ensures that, in the event of an MCI in sector 23 or 34, the helium flow goes through the UL44 gallery to the PM45 shaft and that it does not reach the RUX45 platform and furthermore the UX45 cavern. In much the same way, a pressure resistant partition and ventilation door in the RB46 gallery isolates the RUX45 platform in case of a MCI in sector 45 or 56.
 In conclusion, an MCI event in the LHC sectors adjacent to Point 4 will not create a significant hazard in the UX45 cavern and consequently in the PX46 shaft.

%{\bf I would create a new section (Section 6) here and include some of the following subsections (5.4/5.5/5.6.1/5.6.2 and 5.2.3) and the subsubsection on the lifting platform: the title could be "Proposed design of the Infrastructure" coordinated by Kincso with the following subsections:
%\begin{itemize}
%    \item Shielding and civil engineering (this should include the description of the supporting structure for the interferometer - Kincso)
%    \item Lifting platform (D. Lafarge)
%    \item Access control and safety systems (T. Hakulinen)
%    \item HVAC (R. Langlois)
%    \item Cooling (no need?)
%    \item Electricity (M. Parodi)
%\end{itemize}
%}

%\subsection{Proposed design on the new infrastructure (K. Balazs)}
\subsection{Proposed design of the new infrastructure}
\label{subsec:NewInfra}

% {\it 1 page: brief introductory text to next subsubsections.}
The technical requirements for an AI experiment have been discussed in Section~\ref{sec:Overview}. The proposed site at the LHC Point 4 satisfies these requirements in terms of the background seismic and electromagnetic noise, see Sections~\ref{subsubsec:EMnoise} and~\ref{subsubsec:Vibrations}. However, the added constraints due to the LHC accelerator environment, namely in terms of radiation protection (Section \ref{subsec:RP}) and of safety hazards due to fire  (Section~\ref{subsubsec:Fire}) and to helium release (Section~\ref{subsubsec:Helium}) impose some modifications, improvements and additions to the infrastructure in the PX46 shaft and in the TX46 gallery connecting to the UX45 cavern. These interventions are discussed in detail in the following Sections. In addition, in order to comply with the other requirements of the experiment in terms of technical infrastructure, some technical services have to be adapted and improved, as discussed in subsequent Sections.

%\subsection{Shielding and civil Engineering (K. Balasz)}
\subsubsection{Civil Engineering}% (K. Balazs)
\label{subsubsec:CE}

% 2-3 pages: description of the needed CE infrastructure to be built or modified as a result of the needs described in the previous sections (shielding wall, support structures, etc.).

\begin{figure}[h!]
    \centering
    \includegraphics[width=0.8\textwidth]{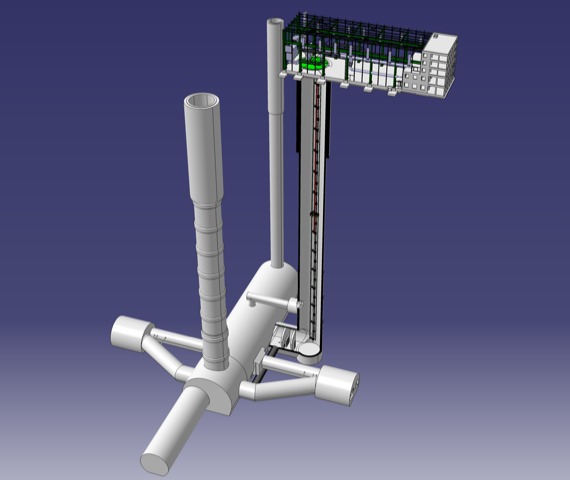}
    \caption{\label{fig:CE:proposed} 3D model of the existing infrastructure at Point 4 of LHC with the proposed AI located in the PX46 shaft and the shielding required in TX46.}
\end{figure}

Minor civil engineering works will be required to house an AI experiment such as AION-100 in the proposed location. Figure~\ref{fig:CE:proposed} shows the existing infrastructure at Point 4 of the LHC tunnel with the proposed AI located in the PX46 shaft.

 \begin{figure}[h!]
     \centering
     \includegraphics[width=.4\textwidth]{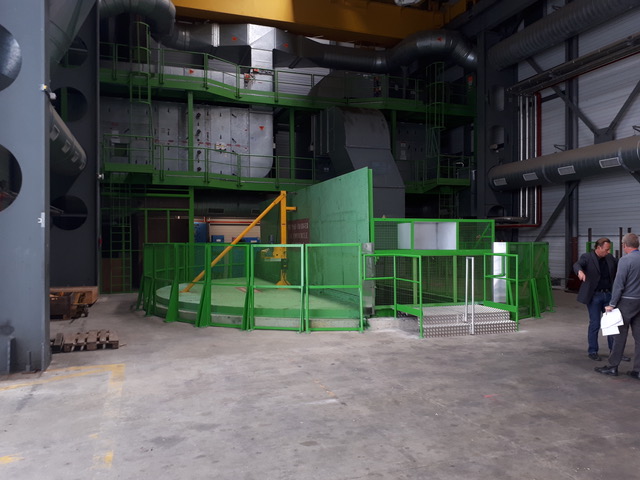}
     \quad
     \includegraphics[width=.4\textwidth]{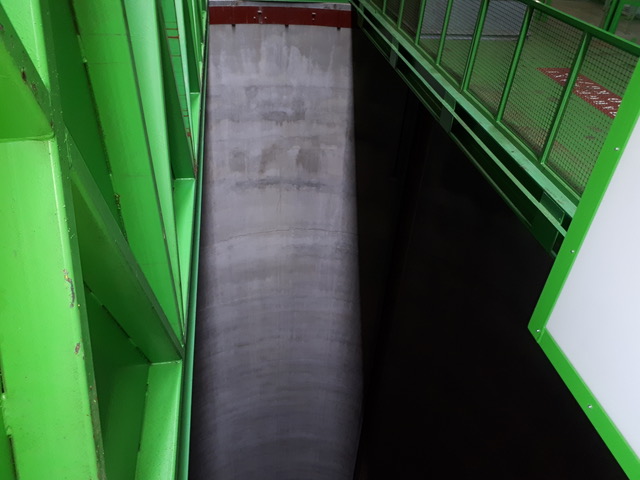}
     \caption{\label{fig:CE:steel cover} View from the building SX4 showing the existing steel cover on the top of the shaft}
 \end{figure}

Currently the top of the shaft is closed with a steel cover that can be opened in the middle with a hoist to allow transport into and out of the shaft, see Figure~\ref{fig:CE:steel cover}. This will require some modification of the design to allow the installation of the experiment with the proposed lifting platform and regular access from the top. A new enclosed access area will be created for this purpose with a door controlled by a badge reader connected to the LHC Access Control System~(see Section~\ref{subsubsec:Access}).

\begin{figure}[h!]
    \centering
    \includegraphics[width=.8\textwidth]{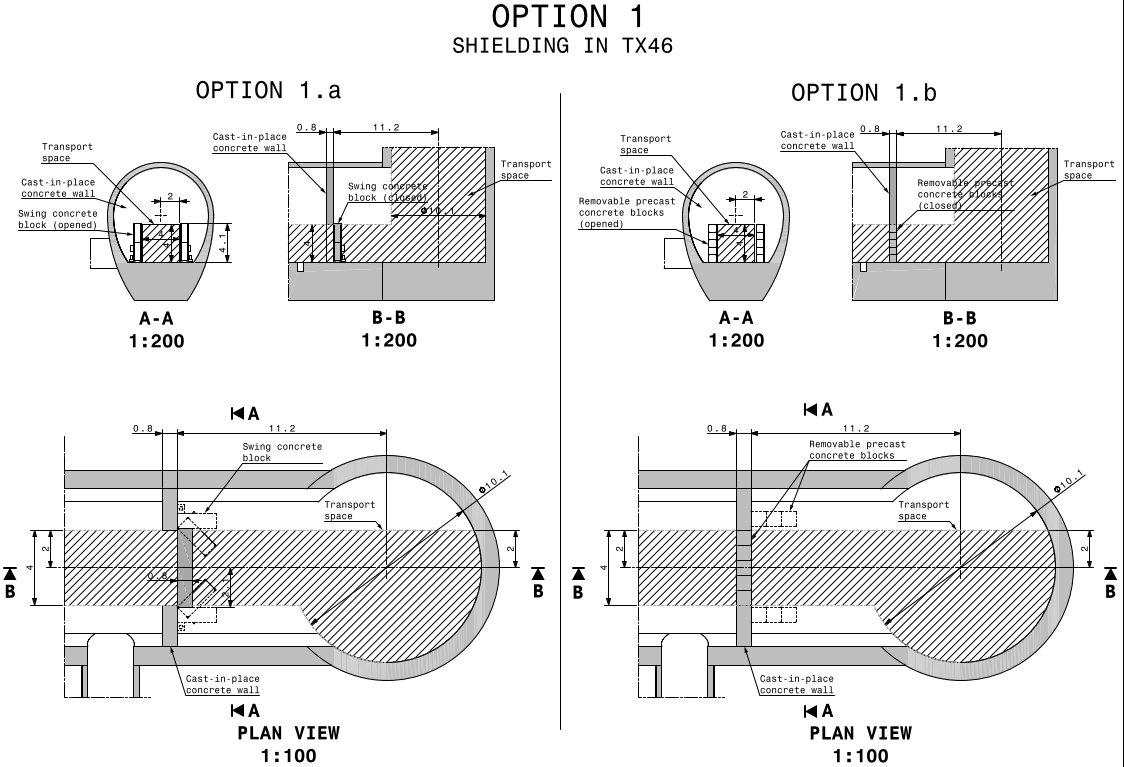}\\
    %\quad
    \includegraphics[width=.8\textwidth]{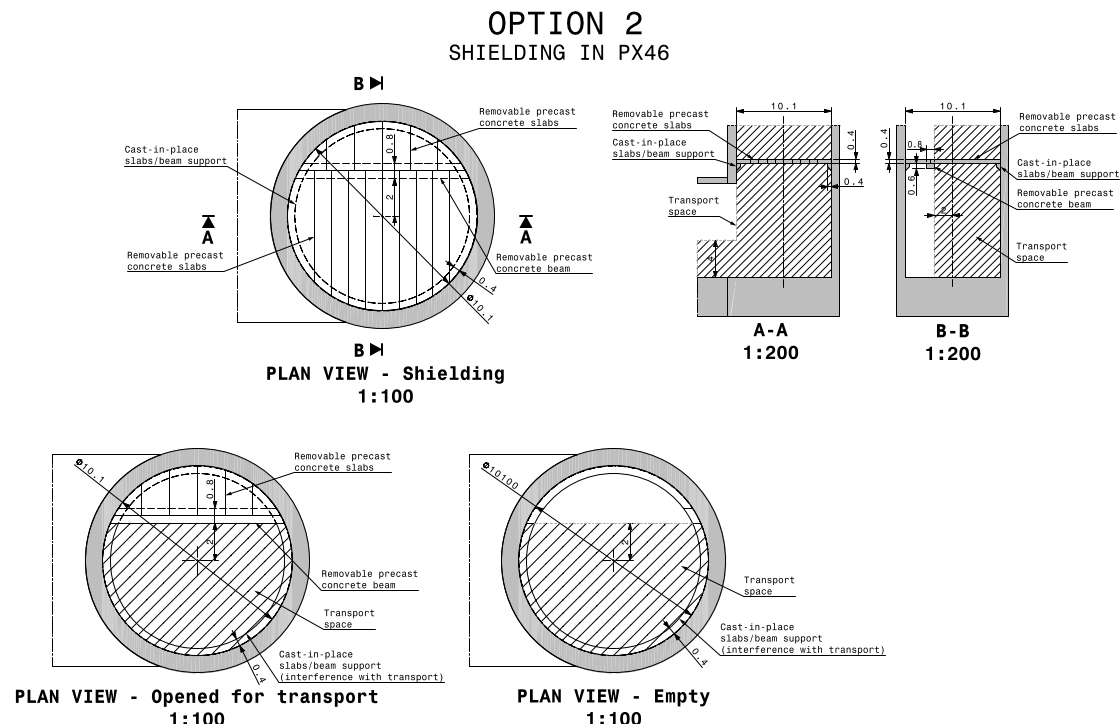}
    \caption{\label{fig:CE:Shielding} Options for shielding with removable blocks.}
\end{figure}

As a result of the radioprotection studies described in Subsection~\ref{subsec:RP}, installation of a shielding wall will be required for regular access including during LHC beam operations and to be able to use the full depth of the shaft. Initially two different solutions were proposed, one having a shielding wall in the TX46 tunnel and the other having a shielding slab at the bottom of the shaft as shown in Figure~\ref{fig:CE:Shielding}. 

Both options include removable blocks in their design to avoid blocking the area for transportation while fulfilling the RP requirements. 
In the TX46 tunnel a 0.8~m thick cast-in-place concrete wall was proposed, with either a shielding door or removable blocks covering approximately 16~m$^2$ that can be dismantled during long shutdowns, freeing the area for transportation and handling. 
The shielding in the PX46 shaft was designed as a 0.8~m thick concrete slab with removable elements covering an area of approximately 60~m$^2$. In order to allow free passage of the transported elements in and out of the shaft, the shielding would need to be dismounted by the surface crane existing in the SX4 building. However, this is impossible due to the limitation of the area served by the hook at the bottom of the shaft, therefore having the shielding in PX46 is not feasible. As a result, Option 1 has been retained. 

\begin{figure}[h!]
    \centering
    \includegraphics[width=1\textwidth]{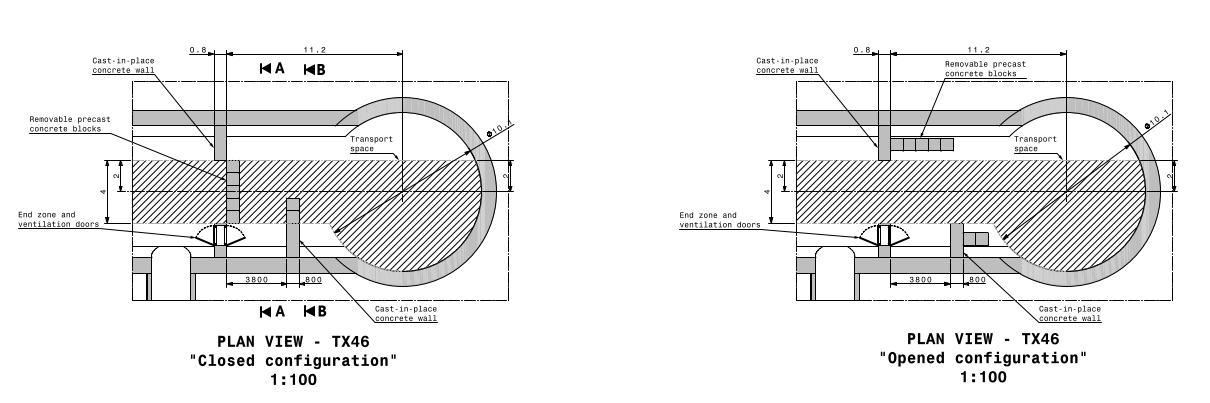}
    \includegraphics[width=0.6\textwidth]{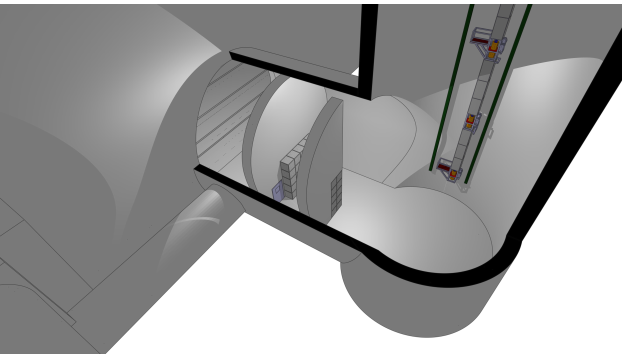}
    \caption{\label{fig:CE:chicane}Shielding wall with the proposed access doors and added chicane. {\it Top left panel}: closed configuration. {\it Top right panel}: open configuration for transport. {\it Bottom panel}: 3D view.}
\end{figure}

An escape route from the bottom of the shaft is envisaged via the existing UP46 side tunnel. The initially proposed concrete shielding would create a dead-end tunnel with the only access being from the top, a design which is not in compliance with CERN's Health and Safety requirements. In order to provide a safe evacuation path through the shielding wall, it is proposed to install two new access doors in the shielding wall in TX46: an end-of-zone door and a second ventilation door with an additional 0.8~m thick chicane wall, as shown in Figure~\ref{fig:CE:chicane}. In this configuration both the proposed doors and the added chicane could present a possible streaming path for radiation providing an intermediate situation with respect to the  scenarios studied in Section~\ref{subsec:RP}, Therefore the possibility of accessing the full depth of the PX46 shaft during machine operation remains to be validated. 

On the surface, an approximately 50~m$^2$ big laser room will be built in the existing building SX4 in proximity of the top of the shaft. The space is available and no technical issue is expected.

%\subsubsection{Lifting platform technical requirements (D. Lafarge)}
\subsubsection{Lifting platform technical requirements}
\label{subsubsec:Platform}

% {\it 2-3 pages: description of the technical requirement for the lifting platform and the proposed technical solution. Discuss compliance of this solution with fire safety requirements}

Following the preliminary study discussed in Section \ref{subsec:Site}, it was assessed that the free area of approximately 17~m$^2$ that is available in the cross-section of  PX46 does not leave enough space to accommodate an AI experiment together with both a lifting platform and emergency evacuation stairs. This posed the problem of finding an adequate compromise for accessing the experiment that satisfies the safety requirements (see Sections~\ref{subsubsec:Fire} and \ref{subsubsec:Helium}) while allowing reasonable easy access for normal operation. On the one hand, accessing the experiment only via a staircase 140~m high would be unmanageable in the long term, and pose the risk of difficult and lengthy evacuation. On the other hand, standardized lifting platforms following EN1495 norms would allow a maximum speed of 0.2~m/s in operation and of around 0.1~m/s in case of braked descent, thus not complying with the time requirements for safe evacuation in case of fire described in Section~\ref{subsubsec:Fire}.
A consultancy firm was commissioned to study the feasibility of designing and constructing an {\it ad hoc} platform that complies with the following requirements:

\begin{itemize}
    \item total maximum useful load: 500~kg;
    \item room for two operators in an adequate protective booth, to avoid risk of personal injuries during fast ascent or descent;
	\item electrically powered from a secure network with backup batteries in case of power failure;
	\item controlled descent in case of failure of both electrical network and batteries at a speed of 70~m/min to comply with the requirement of a 2-minute maximum descent time for evacuation set out in Section~\ref{subsubsec:Fire}.
\end{itemize}

\begin{figure}[h!]
	\centering
\includegraphics[width=.45\textwidth]{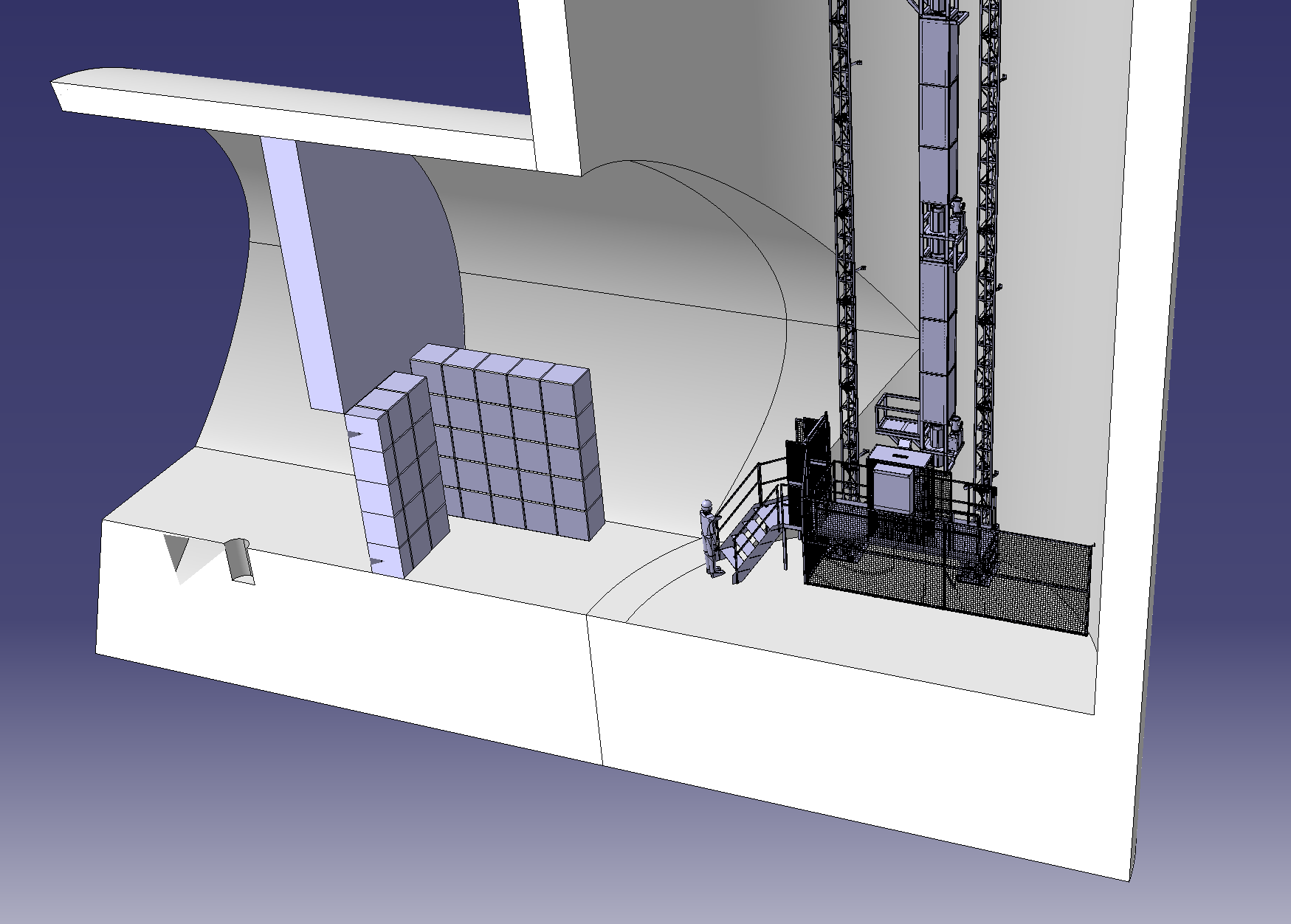}
    \quad
\includegraphics[width=.45\textwidth]{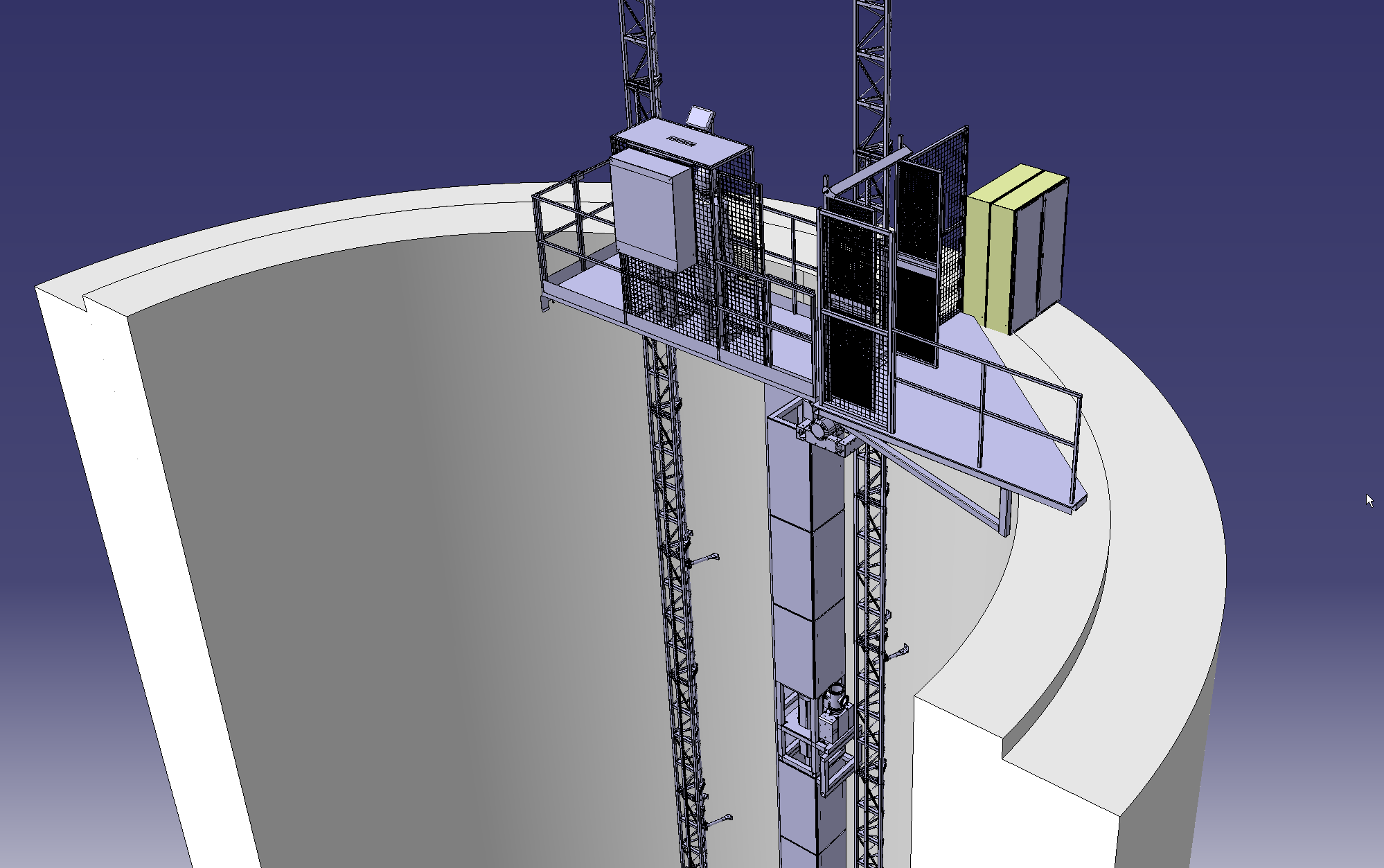}
\includegraphics[width=.5\textwidth]{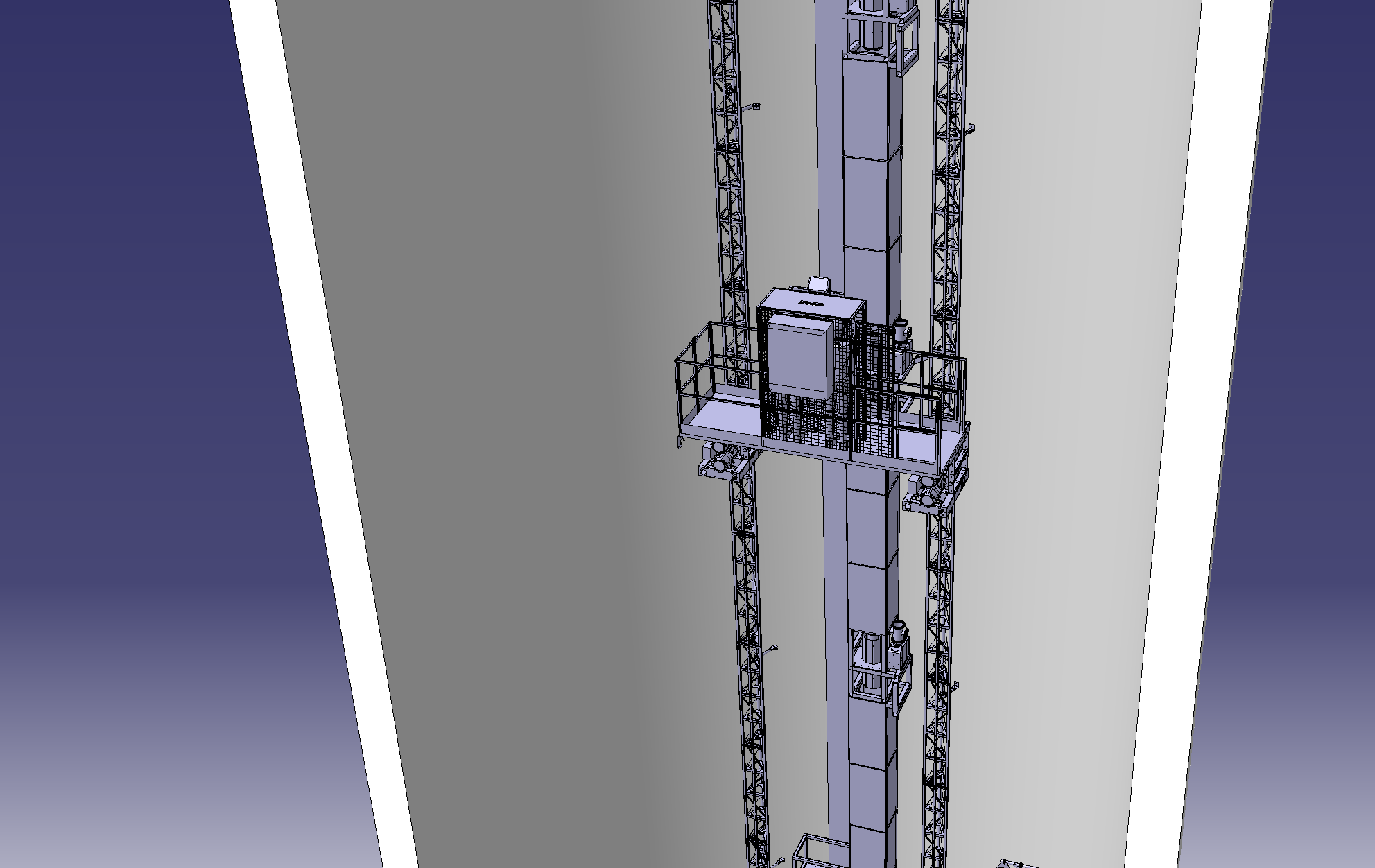}
	\caption{\label{fig:Platform1} {\it Top panels}: 3D view of the bottom and top landings of the dedicated lifting platform in PX46. {\it Bottom panel}: 3D view of the platform during operation~\cite{XLIndustries:2023}}
\end{figure}

It has been demonstrated that these requirements can indeed be satisfied~\cite{XLIndustries:2023}. The proposed purpose-designed lifting platform (see Figure~\ref{fig:Platform1}) then be used for routine access to the experiment, stopping at the location of the side arms of the experiment via an automatic operation mode like a normal elevator. In the case of an emergency evacuation, it could reach the surface with sufficient speed for escaping through the SX4 building and, in the case of a major power failure, it could descend to the bottom of PX46 with the required speed as specified.

The platform could in fact be used also during installation of the AI experiment itself. In such a case, a manual operation mode could be envisaged, where the operators could start and stop the platform manually (possibly moving at reduced speed) and thus be able to work at any point along the height of PX46. The heaviest module of the AI could be lowered in the PX46 shaft with the existing overhead crane in the surface building SX4 (possible load up to 5~tons, clearing under the hook of 10~m), and then assembled by the operators working from the platform. It would then be logical that the platform is among the first items of equipment to be built and assembled inside the shaft.

% \begin{figure}[h]
% 	\centering
% 	\includegraphics[width=.6\textwidth]{Figures/5-4-2-LiftingPlatform/Top_landing.png}
% 	\includegraphics[width=.6\textwidth]{Figures/5-4-2-LiftingPlatform/Bottom_landing.png}	
% 	\caption{\label{fig:Platform2} Schematic views of the top and bottom landings of the lifting platform in PX46 (\copyright XL Industries~\cite{XLIndustries:2023}).
 % \textcolor{red}{John: Do we need this second view of the bottom landing?}\textcolor{green}{ Sergio: I think we can remove the entire figure and keep only Fig.~\ref{fig:Platform1}}}
% \end{figure}

A further aspect to be considered is the possibility of rescuing the operators in the event of mechanical failure of the platform itself, the probability of this happening concurrently with a fire or helium release event being considered negligible. In such an event a rescue nacelle could be suspended and lowered from a hoist attached to the overhead crane in the SX4 surface building, or preferably from a dedicated small crane that could be operated independently by a Fire Brigade rescue team. The design of the platform should allow enough room and capacity to support the two operators and two rescuers in case of an intervention.

% \textbf{\color{red} Should we say anything about the installation of the modular sections of the interferometer. Do we have the hoisting equipment for 907 kg as requested in Table 2. Can we use the crane?}

%\subsection{Access control and safety systems (T. Hakulinen)}
\subsubsection{Access control and safety systems} % (T. Hakulinen)
\label{subsubsec:Access}

%{\it 2-3 pages: description of the access control system, doors and interlocks needed for safe experiment and machine operation, including egress in emergency situations. Mention alarms, ODH etc detection, if possible.}

Modifications to the access control system consist of two distinct parts: access control to the shaft PX46, which will normally be classified as a supervised radiation area, and access control to the surface building SX4 and the laser room by the shaft. 

\begin{figure}[htbp]
     \centering
     \includegraphics[width=.40\textwidth]{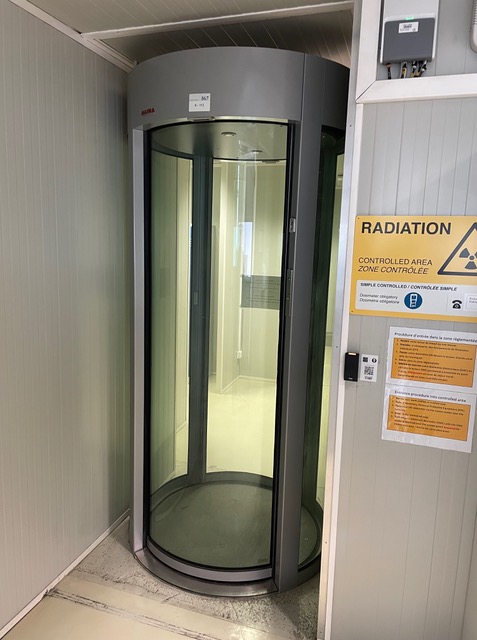}
     \quad
     \includegraphics[width=.50\textwidth]{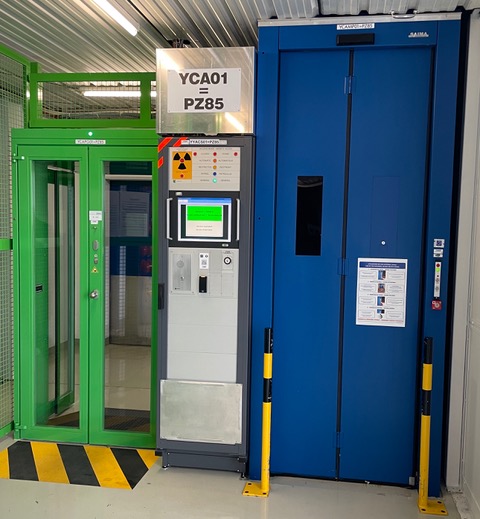}
     \caption{\label{fig:AC:badgepadmad} A simple airlock with a badge reader ({\it left}), PAD and MAD ({\it right}).}
\end{figure}

From the top of the shaft it will be possible to descend to the bottom of the shaft via the planned elevator, to which access will need to be appropriately controlled. Technical options for access control range from a simple door with a badge reader to a full Personnel Access Device (PAD) and Material Access Device (MAD).
The currently favoured access control solution is via a simple airlock controlled by a badge reader attached to the LHC Access Control System (LACS). This is a light-weight solution when compared to an installation of a full PAD and MAD, which are used at other access points of the LHC (see Figure~\ref{fig:AC:badgepadmad}), but it still allows accurate counting of personnel accessing the pit. If safety studies indicate that a full PAD/MAD solution is necessary, it can be installed albeit with a considerably larger footprint and expense.

The required access badge would be the standard CERN dosimeter badge used to access radiological areas and a specific access model would be created for the AI area in LACS with an assigned person responsible for granting this access. Also, specific on-line safety training will need to be prepared for this area to inform users on the specific safety and access particularities in all situations, and a set of required PPE will need to be defined (e.g., helmet, safety shoes, lamp, dosimeter, etc.).
LACS-specific video surveillance of the access device would normally also be installed, as is the case at all other access points.

Access control to the SX4 building and the laser room would be managed by the CERN standard site surveillance system (SUSI). It would need to be decided if the access control to the main building, SX4, needs to be modified. In any case, a specific access model and control to the laser room is likely to be needed. Even though the access control to the laser room would likely be connected to SUSI rather than LACS, the required badge can be defined to be the same dosimeter badge as for the LACS, if necessary.

\begin{figure}[htbp]
     \centering
     \includegraphics[width=.4\textwidth]{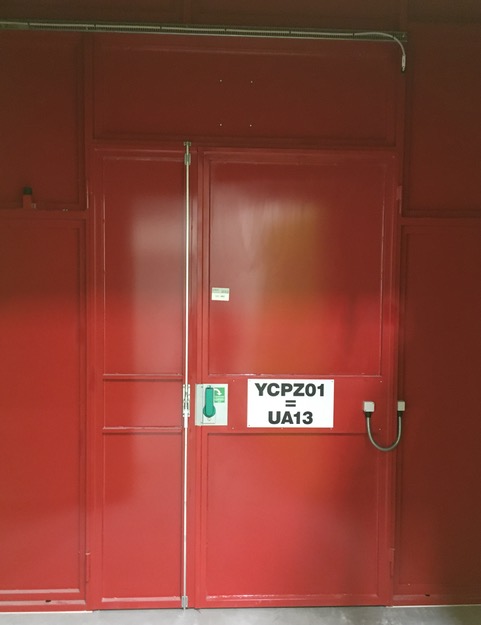}
     \quad
     \includegraphics[width=.4\textwidth]{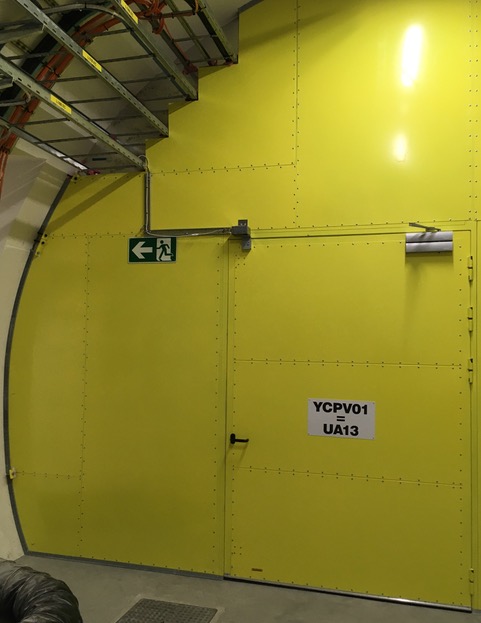}
     \caption{\label{fig:AC:eozventdoors} LASS end-of-zone door ({\it left}), ventilation door ({\it right}).}
\end{figure}

The LHC Access Safety System (LASS) will also need to be modified. These modifications mainly consist of changes to the sectorization of the UX45 area.
Currently the entire PX46 shaft belongs to the interlocked LASS zone PZ45 together with the UX45 cavern. During an LHC Run the top of shaft is covered to prevent access and to contain any prompt radiation in case of an accidental beam loss close to UX45.
The new end of the PZ45 zone will be at the TX46 connecting gallery, where sufficient radiological shielding will be installed as per an analysis by RP (see Section~\ref{subsec:RP}).
Two new access doors are to be installed successively in the passage through the shielding in TX46, an end-of-zone door and a ventilation door (see Section~\ref{subsubsec:CE}).
This arrangement follows the general principle of the LASS that any access from a non-interlocked area to an area where a radiological risk is present must pass through a minimum of two interlocked doors with separate safety contacts and cabling.
The end-of-zone door is a standard LASS door: red in color and either gridded or solid with double position contacts (connected to both LASS safety chains, PLC and relay-based), and including an emergency opening handle on both sides.
The ventilation door is a standard solid door with regular opening handles and LASS double position contacts (see Figure~\ref{fig:AC:eozventdoors}).
If either of these doors is opened, or even if the end-of-zone door emergency handle is turned, the LHC beam will be dumped and the system put into a safe state.

Installation of an end-of-zone door in TX46 means that PX46 shaft can be classified as a non-interlocked area, no longer supervised by the LASS and allowing access from the top via a simplified procedure by the LACS, as described above. 
This status is similar to other non-interlocked experiment areas at PX15 (ATLAS), PM54 (CMS), PZ85 (LHCb), as well as HL-LHC PM17 and PM57.
As described in Section~\ref{subsubsec:CE}, TX46 will act as an evacuation route via the UX45 cavern and PM45 or PZ45 if the access to the top of the shaft is unavailable or dangerous.

Alarm systems for fire, oxygen deficiency hazards and evacuation as well as emergency communication will probably also need to be installed. However, the exact detection and alarm needs are to be defined by a risk analysis by the competent safety officers. Fire detection is likely to be needed in the PX46 shaft with a probable effect on the LHC fire protection strategy in P4. An analysis by the CV group will be needed to understand the exact ventilation conditions affecting fire safety in the PX46. ODH detection will also likely be necessary in the new laser room in the SX4 building, where a certain gas inventory is foreseen. Whether a full ODH detection will be necessary also in the PX46 shaft remains to be seen. At this time it would seem that the probability of a large helium release in the event of an MCI in this area is small, as demonstrated by earlier real helium releases in the RF area, where the helium gas has had the tendency to be largely contained in the LHC tunnel and driven away from the UX45 towards other LHC sites due to the effect of the regular ventilation~\cite{Hakulinen:2022}. All this is again subject to a more detailed risk analysis.

%MCI:
%The helium release incident on Aug 23 in the RF area was indeed very useful for understanding the He-risk in the UX cavern. The ventilation scheme in the LHC is such that the air is taken in at the even points and extracted from the odd ones. Therefore, what was seen in terms of oxygen level variation was that the effect was almost entirely confined in the LHC tunnel. Two ODH detectors at the top of the UX45 cavern (one right above the beam line, another close to the wall of the US service cavern) registered barely visible dents in oxygen levels while the tunnel detectors saw sizable drops all the way to point 5. We don’t have ODH detection in the PX46 shaft, so we cannot tell what might have been seen there, but so far it looks encouraging. Analysis by CV is in any case necessary.

%\subsubsection{HVAC (R. Langlois)}
\subsubsection{Heating, ventilation and air conditioning (HVAC)}% (R. Langlois)
\begin{figure}[h!]
	\centering
	\includegraphics[width=1.0\textwidth]{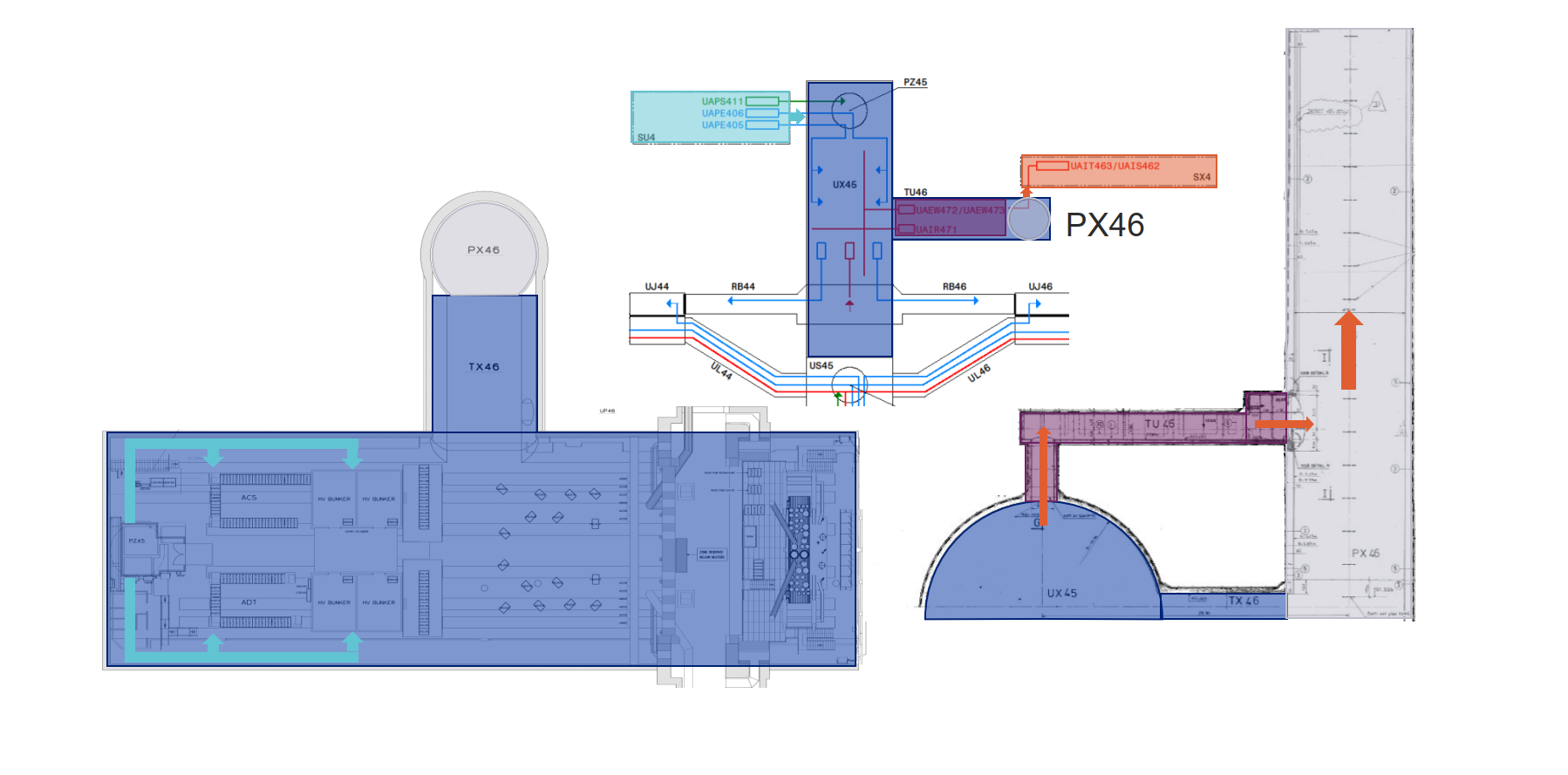}
	\caption{\label{fig:HVAC1} UX45 HVAC infrastructure.}
\end{figure}

\label{subsubsec:HVAC}
PX46 is part of the UX45 cavern ventilation infrastructure. Its main functions are air supply, extraction and smoke extraction. A schematic view of the HVAC installation for UX45 is shown in Figure~\ref{fig:HVAC1}.

The air supply is provided by two redundant units located in the SU4 surface building. They provide 45 000 m$^3$/h air flow with the temperature controlled at 17.0 °C (+/-1.0 °C) and the dew point controlled to be < 10.0 °C. Air is collected at the top of the UX45 cavern by ducts to ventilation units located in the TU46 technical gallery. These units blow air to the surface via the PX46 shaft without a duct. At the top of the PX46 shaft, a duct is connected to the steel cover of the SX4 building to feed the main extraction unit. The extracted air is filtered before being released to the atmosphere at a flow rate of 45 000 m$^3$/h.
In view of the characteristics of PX46, the air renewal rate is typically 3 times per hour and the air velocity is 0.16 m/s. The temperature in the PX46 shaft is stable during LHC machine operation (<1 °C/h), but the present measurement sampling does not provide a full-year study of temperature stability on a 1h scale. Once steady state is reached the temperature in the PX46 shaft presents a negative gradient  (typically -2.0°C over the height of the shaft). The shielding in the TX46 gallery would not affect the UX45 cavern main ventilation. However, the bottom of the PX46 shaft would become a stagnant air area; a dedicated system would need to be added to guarantee air renewal.

%If a fire is detected in the cavern, all ventilation units are stopped. 
If a fire is detected in the cavern, the fire brigade will intervene and change the ventilation mode from the fire brigade cubicle. 
Smoke extraction is manually activated, and all units put in operation for a total air flow of 90 000 m$^3$/h (for both supply and extraction).

%\textbf{\color{red} It would be good to clarify whether we expect any issue from the presence of the shielding in TX46. I seem to remember that from the discussions we had the shielding should not affect significantly the ventilation. Correct?} 
The PX46 shaft does not have cooling infrastructure. The main chilled water production and distribution station is located in the neighbouring surface building SU4, with chilled water distribution piping available in the SX4 building. However, due to the thermal load of the atomic sources (up to $\approx$50~kW), there is no margin in the existing system to add a new user, even without considering the conflicting requirements between the different connected systems during operation and maintenance.
%The configuration of the AI experiment requires cooling and ventilation capabilities on the surface and along the PX46 shaft.
A primary source of cooling is therefore necessary for the experimental systems. There are  two options: either a dedicated chiller for the whole experiment, or upgrading the existing SX4 distribution or creating a new distribution system. In both cases, new pumps in SU4 and new piping from SU4 to SX4 have to be added, and power distribution and control systems in SU4 have to be modified.

%Several solutions could be considered: connection to existing systems (primary water or chilled water) or installation of a new one. Apart from cost considerations, the available capacity and the operation schedule will have to be taken into account.
The laser laboratory located on the surface will have dedicated cooling and ventilation systems. Their design will be strongly influenced by the temperature stability goals and the presence of gases and any oxygen deficiency hazard in the volume.
A special feature of the experiment is the distribution along the shaft of side-arms requiring cooling and ventilation.
The cooling system will be located on the surface and water will be distributed to each side-arm. Vertical distribution piping will be installed along the full height of the experiment and each side arm will be fed by a dedicated manifold.
The design of the ventilation system will have to take into account many requirements: layout, temperature stability, heat load to be extracted, air quality (humidity control and filtration), low vibrations and maintainability. 

This situation is different from existing EN/CV installations. Further studies will be necessary to converge on a conceptual design for the cooling and ventilation of the side arms. Installation works, operation and maintenance will be constrained by the shaft location. Other issues to be considered include X-ray testing of the welding of the pipes, access for maintenance (both persons and material), etc.

%Can we say something about the temperature stability in PX46? Consistent with the requirement of temperature drift < 1C/h - See Table 2. Any comment on the compressed air distribution required for the Laser lab?

%\subsubsection{Electricity (M. Parodi)}
\subsubsection{Electricity} %(M. Parodi, C. Marcel)
\label{subsubsec:EL}

%{\it 1-3 pages: description of any extra infrastructure works needed for powering the experiment and the laser lab in SX46. Include any safety considerations, if relevant}

 The total electrical power required by this project is $\sim 135$ kW: 35 kW needed for the laser laboratory (for lights and sockets), and 100 kW needed for the shaft (for up to 10 atom sources that need 10 kW each, connected by socket boxes). There is already enough power available at Point 4. Indeed, there is a 1.25 MVA transformer in SX4 (EMT302/4X) that only feeds the main low voltage switchboard EBD1/4X, and a measurement performed during YETS 2022-2023 showed a consumption of 85 kVA - that needs to be confirmed during a run period - but leads us to think that the available power is sufficient to supply the needs of an AI experiment without major interventions on the upstream network. So it will be sufficient to install a feeder on the switchboard EBD1/4X to supply the needs of the experiment.

\begin{figure}[h!]
     \centering
     \includegraphics[width=.8\textwidth]{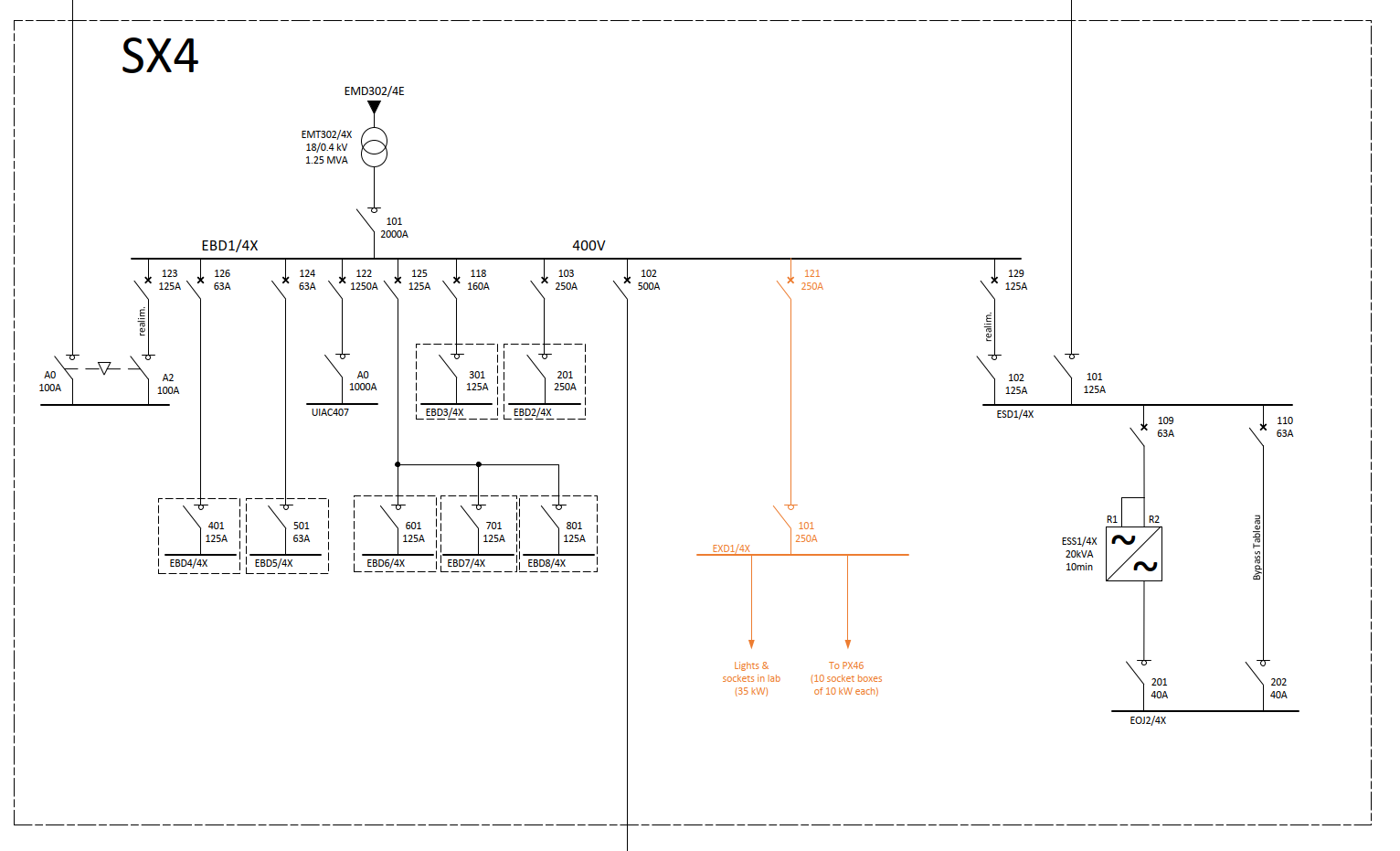}
     \caption{\label{fig:ElecSLD} New preliminary single line diagram of SX4, with the new feeder of EBD1/4X, the new switchboard EXD1/4X and its feeders in orange.}
\end{figure}

\begin{figure}[h!]
     \centering
     \includegraphics[width=.5\textwidth]{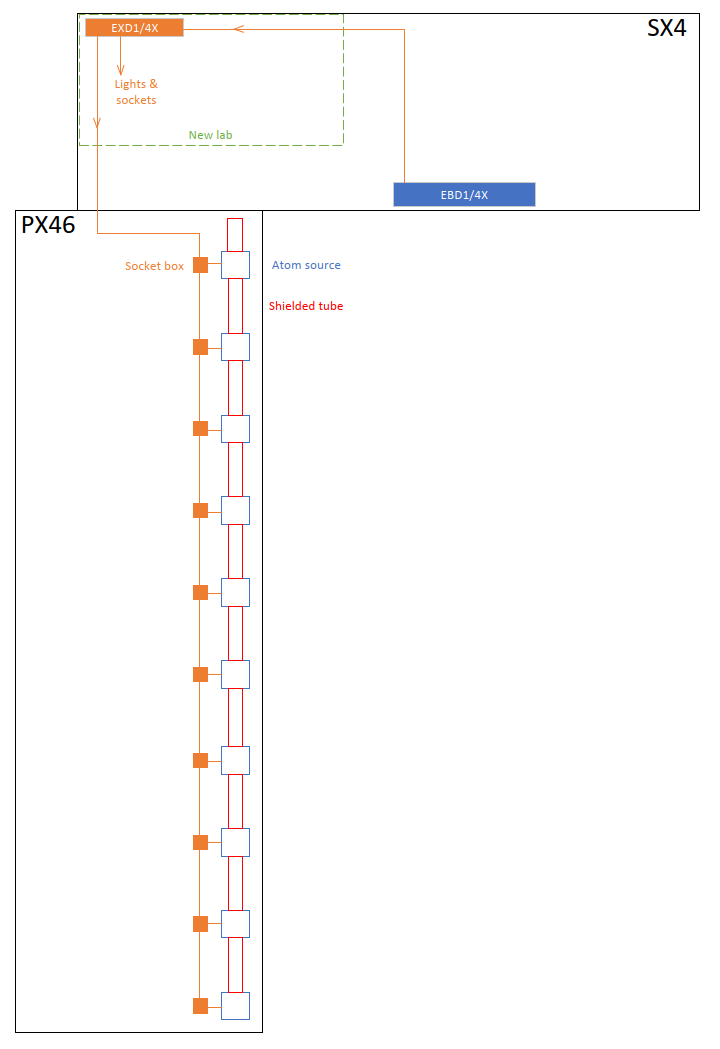}
     \caption{\label{fig:ElecLayout} Preliminary layout of SX4 and PX46 with the switchboard in the laser laboratory and the two feeders for the supplies to the laser laboratory and the atom sources.}
\end{figure}

The new feeder on the existing switchboard EBD1/4X will power a new switchboard in the lab EXD1/4X, that will itself have two main loads: one for the laser laboratory (lights and sockets, 35 kW) and one for the shaft PX46 (10 socket boxes for the atom sources, 100 kW) as shown in Figures~\ref{fig:ElecSLD}~and~\ref{fig:ElecLayout}.

\begin{figure}[h!]
     \centering
     \includegraphics[width=.5\textwidth]{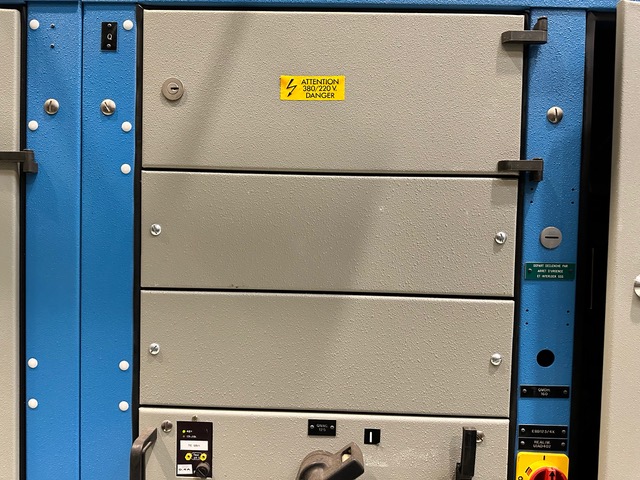}
     \caption{\label{fig:ElecEBD1} Picture of the spare feeders available in the existing EBD1/4X.}
\end{figure}

There is already available space in EBD1/4X that is large enough to hold the feeder going to the laboratory (minimum \SI{250}{\A}), as shown in Figure~\ref{fig:ElecEBD1}. Depending on the time scale for executing the project, it is possible that meanwhile the existing switchboard will be replaced in the framework of EN-EL consolidation. In that case, a dedicated feeder for the AI experiment should be considered in the new switchboard.

The first connection, between EBD1/4X and EXD1/4X, will be made by a new \SI{250}{\A} feeder with a circuit breaker and a copper cable around \SI{30}{\m} long, 4$\times$1$\times$120~mm$^2$ (Ph+N) + 1$\times$70~mm$^2$ (PE).

For connecting the new switchboard EXD1/4X to the different socket boxes supplying the atom sources, the best solution would be to have as few cables as possible. However, a single cable would not be possible, as its section would be too large to connect it to the socket boxes (ten 16 A socket boxes are needed). Therefore, for a preliminary design, 4 cables can be considered, each of them supplying between 2 and 3 socket boxes. They will all be copper cables, 5$\times$16 mm$^2$, which is the maximum acceptable section for a cable to be connected to a \SI{16}{\A} socket box. After a preliminary check, no cable trays are available to go along PX46, and the integration of new cable paths will be studied in accordance with project requirements.

For control cables and optical fibres, the requirements are not yet sufficiently detailed to allow a technical proposal and to establish a preliminary component list. Once the requirements are better developed, a preliminary study can be performed to include them in the technical solution.

For the laser laboratory itself and its electrical needs, there will be lights and sockets.
For the lights, the surface of the lab is expected to be around 50~m$^2$. Without further information, we estimate that about 12 ceiling lights (31~W each) will be needed, together with an emergency light for evacuation (7~W) above the door (379~W in total), so less than 500~W.  

% \begin{figure}[h!]
%      \centering
%      \includegraphics[width=.3\textwidth]{Figures/5-4-5-Electricity/Lights.PNG}
%      \caption{\label{fig:ElecSLD} Preliminary layout of the light distribution}
% \end{figure}

For the sockets, all we know for now is that the lab will need about two three-phase outlets and about 100 single-phase outlets. In the absence of more detailed information, we can anticipate:
\begin{itemize}
    \item 2 boxes of 3 single-phase outlets of 16A and 1 three-phase outlet of 16~A;
    \item 9 strips of 16~A of 10 single-phase outlets each;
    \item 1 strip of 16~A of 5 single-phase outlets.
\end{itemize}
We therefore arrive at the following preliminary list of required materials:
\begin{itemize}
    \item 1 new EXD1/4X switchboard;
    \item 1 new electric drawer in EBD1/4X;
    \item 1 new cable (30~m long, 4$\times$1$\times$120~mm$^2$~(Ph+N) + 1$\times$70~mm$^2$~(PE)) between EBD1/4X and EXD1/4X;
    \item 4 cables (all 5$\times$16 mm$^2$, approx. 500m in total) going from EXD1/4X to the shaft;
    \item 10 socket boxes in the shaft;
    \item Cables powering lights and outlets in the laser laboratory;
    \item 12 lights (Philips Coreline Slim Batten of 1135mm, 31W) + 1 emergency light (LUMATEC Straitbox 500 w/ pictogram);
    \item 2 boxes of 3 single-phase outlets of 16~A and 1 three-phase outlet of 16~A;
    \item 9 strips of 10 single-phase 16~A  outlets each;
    \item 1 strip of 5 single-phase outlets of 16~A;
    \item Cable trays to hold the new cables.
\end{itemize}

Another requirement for the project is to supply an elevator. For this, 22~kW are needed from the secured network, with a 63~A feeder. The current switchboard is not fitted for this. It would need to be replaced and consolidated to host a new 63~A feeder with a copper cable \SI{60}{\m} long, 5$\times$25 mm$^2$.

Concerning the planning, the recommendation is to consider the workload of the activities already scheduled for LS3 for other projects and avoid conflicts with them. The best option if feasible seems to perform the works for the electrical distribution and the cabling during the first years of run 4.

%\section{Cost and Schedule (All)}
%\section{Cost and Schedule (All)}
\section{Infrastructure Cost Drivers and Schedule Constraints} %(G. Arduini, S. Calatroni, O. Buchmuller, J. Ellis)
\label{sec:Cost}

When assessing the infrastructure costs associated with
installing a $\sim 100$~m AI experiment at CERN, one must
distinguish between the generic costs associated with
such an experiment - such as the establishment of a laser
laboratory, the support framework for the vacuum tube and
electricity distribution - from the specific infrastructure costs
associated with installing such an experiment at CERN.
The latter include the measures necessary for protection
against radiation and fire hazards, and to control access
to the AI experimental area.

As discussed earlier in this report, a key requirement
is to provide radiation protection for people working on
the experiment in the event of a catastrophic LHC beam 
loss close to the RF system at the base of the PX46 shaft.
This requires the installation of a shielding wall that
is configured to permit the transportation of LHC
machine components while offering robust radiation
protection. As described above, the preferred technical
solution is to locate the shielding wall in the TX46
access gallery.

Another key requirement for safe operation at CERN is
provision for evacuation from PX46 within $\sim 2$~minutes
in the event of a fire. This time restriction is 
incompatible with evacuation via stairs or a conventional
(relatively slow-moving) lift. Consulting engineers have
proposed a suitable technical solution.

Access to PX46, like other CERN areas and facilities,
will be subject to restrictions enforced by the LHC Access and Safety Control systems and monitored by adequate safety systems (smoke detection, ODH alarms, etc.).

\begin{table}[htbp]
%\small
\centering
\caption{\label{tab:Costs} Preliminary list of main cost items for the infrastructure required for an AI in PX46.}
\begin{tabular}{|c|c|}
 \hline
 \textbf{Item} & \textbf{Cost [kCHF]} \\ 
 \hline
 \hline
Shielding & 400 \\ \hline
Lifting platform & 400 \\ \hline
Access, safety systems and monitoring & 200 \\ \hline
% Safety Monitoring System (smoke, ODH, etc.) & xxx \\ \hline
General services and utilities & 500 \\ \hline
% Cooling and Ventilation & xxx \\ \hline
% Electrical distribution & xxx \\ \hline
% Compressed air & xxx \\ \hline
% Gas distribution & xxx \\ \hline
% Communication network backbone & xxx \\ \hline
\textbf{Total} & \textbf{1500} \\
 \hline
 
 \end{tabular}
\end{table}

The main cost items for the infrastructure required for the installation of an AI in PX46 are listed in Table~\ref{tab:Costs}.
In the above we have considered only the main civil engineering, access and safety systems, and the major services and utilities such as electrical supply, cooling and ventilation, communication network, gas infrastructures. It is anticipated that the laser laboratory (with the associated access interlock system) and the AI with its ancillaries will be entirely provided by the experimental collaboration. At this pre-conceptual stage of the design, the uncertainty range for the cost estimate is to be considered Class 5\footnote{A Class 5 estimate uncertainty has a lower range between -20\% and -50\% and an upper range between +30\% and +100\%.}~\cite{bib:DOEGuide}.
% These are the three principal drivers for extra costs
% associated with locating a 100~m AI experiment in PX46.
% Preliminary order-of-magnitude estimates indicate that
% the shielding wall might cost $\sim 500$~kCHF, the
% lifting platform $\sim 400$~kCHF and the access system 
% $\sim 100$~kCHF. Thus the identified extra
% costs amount to about 1~MCHF. Even allowing for the 
% inevitable additional costs that are likely to arise
% during a more detailed technical study, we conclude
% that the additional expense for installing a 100~m
% AI experiment at CERN should be below 2~MCHF, which is
The expected cost of the infrastructure is significantly smaller than the cost of the experiment
itself which is estimated to be in the range 30 to 50~MCHF.

\begin{figure}[htbp]
\centering % \begin{center}/\end{center} takes some additional vertical space
\includegraphics[width=1\textwidth]{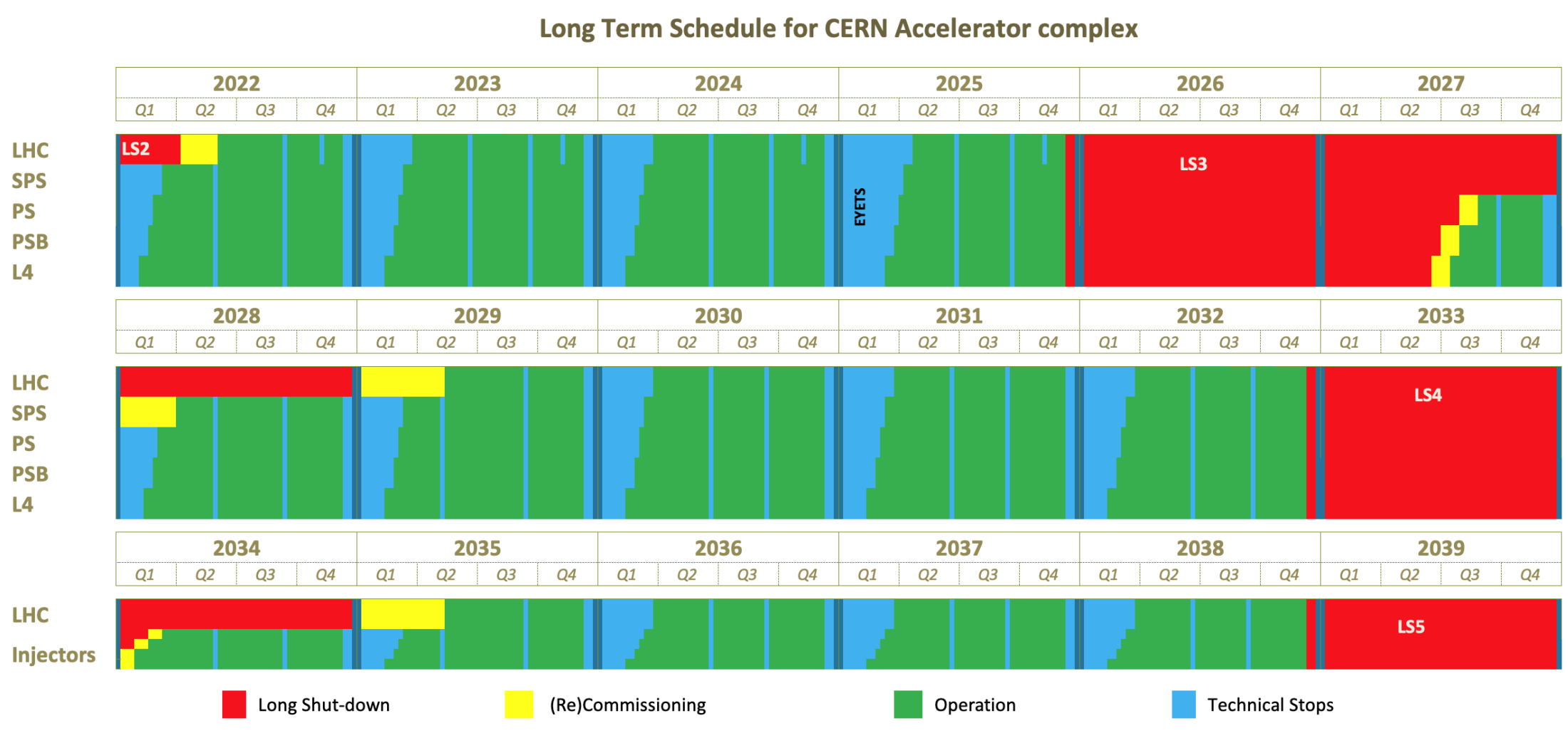}
\caption{\label{fig:LHCSchedule} The LHC and HL-LHC master schedules.}
\end{figure}

The feasible installation schedule of an atom interferometer at LHC Point~4 would be driven by the LHC and HL-LHC master schedule, whose current version is displayed in Figure~\ref{fig:LHCSchedule} \cite{Tock:2022}, as well as by the technical progress of forerunner projects and their results.
The major opportunities for extended installation work for the civil engineering and access and safety systems will be provided by the Long Shutdowns currently scheduled for 2026-2028 and 2033-2034, with the annual Technical Stops providing only limited opportunities.  Technical readiness for starting the construction a 100-m AI experiment is expected to align well with the earlier Long Shutdown. Once the shielding and access and safety systems are in place the rest of the installation of the experiment and of its ancillaries could continue during machine runs.

\vspace{-5mm}

%\section{Conclusions (G. Arduini, S. Calatroni, O. Buchmuller, J. Ellis)}
%\section{Conclusions (G. Arduini, S. Calatroni, O. Buchmuller, J. Ellis)}
\section{Conclusions} % (G. Arduini, S. Calatroni, O. Buchmuller, J. Ellis)
\label{sec:Conclusions}

% {\it 1 page: summary and outlook.}

This conceptual feasibility study has not identified any showstoppers
for the installation and operation of an $\sim 100$~m atom interferometer
experiment in the PX46 shaft at CERN. As discussed in Section~3, such
a detector would have unique capabilities to search for ultralight dark
matter, and would open the way to searches for gravitational waves and
other interesting phenomena in fundamental physics. It would therefore
complement the existing CERN programme in a novel way and
provide a very interesting addition to the Physics Beyond Colliders 
portfolio. Moreover, the technique of atom interferometry fits very 
naturally into CERN's Quantum Technology Initiative.

Section~4 of this report presented an overview of the experiment and of
the required infrastructure and environmental constraints. 

Section~5 reviewed exploratory studies how these requirements could be
accommodated at CERN without interfering with its core scientific
programme. Initial measurements indicated that the experimental
requirements for the spectra of vibrations, seismic and electromagnetic 
noise could all be met in PX46. Health and safety issues specific to CERN
such as radiation protection fire safety and helium release incidents have 
all been considered and suitable technical solutions found. Essential
aspects of the proposed new infrastructure were investigated, including
civil engineering, experimental access and the emergency evacuation of 
personnel, access control, ventilation, cooling and electricity supply.

As discussed in Section~6, the extra costs associated with installing
an $\sim 100$~m atom interferometer at CERN rather than elsewhere appear
to be small compared with the overall cost of such an experiment.

If the international community interested in future large-scale
atom interferometer experiments wishes to pursue the
possibility of siting a 100-m detector at CERN, in view of its
{\it prima facie} suitable infrastructure, a programme of relevant
future studies could be discussed. These could include, for example,
supplementary seismic measurements to understand better the diurnal 
variations that have been observed, extend the measurements to lower 
frequencies and investigate the possibility of mitigating the
gravity gradient noise via a network of sensors in the CERN shafts and
tunnels adjacent to the PX46 shaft. In the event of a specific 
experimental proposal, a programme of more detailed technical studies 
of the safety and infrastructure issues discussed here would also be 
needed, but we expect them to be surmountable.

\vspace{0.5cm}
\paragraph{Acknowledgements}

We would like to thank F.~Corsanego for extremely valuable discussions on  fire safety issues. The kind advice of P.~Berlinghi, J.~Bremer, F.~Cappelletti, L.~Colly, O.~Crespo-Lopez, E.~Dho, D.~Letant-Delrieux, O.~Pirotte, R.~Samoes and D.~Tshilumba is also gratefully acknowledged.
% \textbf{\color{red} Others to be acknowledged?}

\bigskip

%This is an example of citation \cite{Gorzawski:1697022}
%\clearpage
%\section*{Acknowledgments}
%Place for acknowledgements.\\
%The work of J.E. was supported in part by the United Kingdom STFC Grants
%ST/T00679X/1 and ST/T000759/1.

%\begin{thebibliography}{90}

%\bibitem{review} A reference

%\end{thebibliography}
\newpage

\bibliography{PBCreport}

\newpage
\appendix
\section{Definitions of acronyms}
\label{sec:Acronyms}

\textbf{}
\setlength{\parindent}{0pt}

\noindent
\textbf{AI}: Atom Interferometer/Interferometry

\textbf{AION}: Atom Interferometer Observatory and Network

\textbf{ASN}: Atom Shot Noise

\textbf{ATS}: Accelerators \& Technology Sector 

\textbf{AURIGA}: Antenna Ultracriogenica Risonante per l'Indagine Gravitazionale Astronomica~\footnote{Ultracryogenic Resonant Bar Gravitational Wave Detector}

\textbf{BSM}: Beyond the Standard Model

\textbf{CV}: Cooling and Ventilation

\textbf{DM}: Dark Matter

\textbf{EM}: ElectroMagnetic

\textbf{EMC}: Electromagnetic Compatibility

\textbf{EPPSU}: European Particle Physics Strategy Update

\textbf{GGN}: Gravity Gradient Noise

\textbf{GW}: Gravitational Wave

\textbf{HL-LHC}: High-Luminosity LHC

\textbf{HVAC}; Heating, Ventilation and Air Conditioning

\textbf{ISM} Industrial, Scientific, and Medical RF bands 

\textbf{KAGRA}: KAmioka GRAvitational wave detector

\textbf{LACS}: LHC Access Control System

\textbf{LASS}: LHC Access Safety System

\textbf{LHC}: Large Hadron Collider

\textbf{LIGO}: Laser Interferometer Gravitational Observatory experiment

\textbf{LISA}: Laser Interferometer Space Antenna

\textbf{LMT}: Large Momentum Transfer

\textbf{LS}: Long Shutdown

\textbf{LSBB}: Laboratoire Souterrain {\` a} Bas Bruit~\footnote{Low-Noise Underground Laboratory}

\textbf{MAD}: Material Access Device

\textbf{MAGIS}: Matter-wave Atomic Gradiometer Interferometric Sensor experiment

\textbf{MCI}: Maximum Credible Incident

\textbf{MICROSCOPE}: Micro-Satellite à traînée Compensée pour l'Observation du Principe d'Equivalence~\footnote{Micro-Satellite with Compensated Drag for Observing the Principle of Equivalence}

\textbf{MIGA}: Matter wave-laser based Interferometer Gravitation Antenna

\textbf{MOT} : Magneto-Optical Trap

\textbf{NHMN}: New High-Noise Model

\textbf{NLMN}: New Low-Noise Model

\textbf{ODH}: Oxygen Deficiency Hazard

\textbf{PAD}: Personal Access Device

\textbf{PBC}: Physics Beyond Colliders

\textbf{PM}: Puit Materiel~\footnote{Access shaft with stairs and lift used for the transfer of equipment}

\textbf{PPE}: Personal Protection Equipment

\textbf{PX}: Puit eXperience~\footnote{Access shaft to experimental cavern for (formerly) LEP or (currently) LHC detectors}

\textbf{PX46}: Access shaft at LHC Point 4

\textbf{QTI}: Quantum Technology Initiative

\textbf{RP}: Radiation Protection

\textbf{RF}: Radio Frequency

\textbf{SM}: Standard Model

\textbf{SUSI}: système de SUrveillance des SItes~\footnote{Site Surveillance system}

\textbf{SU4}: Surface building dedicated to the cooling and ventilation at Point 4

\textbf{SX4}: Surface building on top of the PX46 shaft

\textbf{TETRA}: Terrestrial Trunked Radio, formerly known as Trans-European Trunked Radio

\textbf{TX46}: Access gallery at LHC Point 4

\textbf{ULDM}: Ultra-Light Dark Matter

\textbf{UX45}: Experimental cavern at LHC Point 4

\textbf{WIMP}: Weakly Interacting Massive Particle

\textbf{YETS}: Year-End Technical Stop

\textbf{ZAIGA}: Zhaoshan Long-baseline Atom Interferometer Gravitation Antenna
\end{document}